\documentclass[11pt]{article}
\usepackage[utf8]{inputenc}
\usepackage{hyperref,url}

\usepackage{geometry}

\usepackage{graphicx} % support the \includegraphics command and options
\usepackage{adjustbox}

\usepackage{booktabs} % for much better looking tables
\usepackage{array} % for better arrays (eg matrices) in maths
\usepackage{paralist} % very flexible & customisable lists (eg. enumerate/itemize, etc.)
\usepackage{verbatim} % adds environment for commenting out blocks of text & for better verbatim
\usepackage{caption}
\usepackage{subcaption}
\usepackage{amsmath}
\usepackage{amsthm}
\usepackage{amssymb}
\usepackage{mathdots}
\usepackage[linesnumbered,ruled,vlined,noend]{algorithm2e}
\usepackage{bm}
\usepackage{xargs}
\usepackage{comment}
\usepackage{standalone}
\usepackage{tikz}
\usetikzlibrary{decorations.pathreplacing}
\usetikzlibrary{decorations.pathmorphing}
\usetikzlibrary{automata}
\usetikzlibrary{arrows.meta}
\tikzset{every state/.style={inner sep=.5pt,minimum size=6pt}}
\tikzset{edge/.style = {->, semithick, >=Latex}}
\tikzset{edgelabel/.style = {inner sep=1pt, fill=white}}
\tikzset{init/.style = {initial, initial text={}}}
\tikzset{>=Latex}
\tikzset{ball/.style={circle,draw,fill=black,inner sep=0pt, minimum width=4pt}}
\usepackage{tikz-cd}
\usepackage{tikz-network}
\usepackage{todonotes}
\usepackage{titling}

\theoremstyle{plain}
\newtheorem{thm}{Theorem}[section] % reset theorem numbering for each chapter

\theoremstyle{definition}
\newtheorem{defn}[thm]{Definition} % definition numbers are dependent on theorem numbers
\newtheorem{lem}[thm]{Lemma}

\newtheorem{prop}[thm]{Proposition}

\newcommand{\partto}{\rightharpoonup}

\DeclareMathOperator{\Ind}{Ind}

\DeclareMathOperator{\start}{start}
\DeclareMathOperator{\indeg}{in-degree}
\DeclareMathOperator{\core}{core}
\DeclareMathOperator{\cf}{cf}

\tikzset{commutative diagrams/.cd,
mysymbol/.style={start anchor=center,end anchor=center,draw=none}
}

\DeclareMathOperator{\repr}{repr}

\newcommand{\arc}[3][]{#2 \xrightarrow{#1} #3}
\newcommand{\pth}[3][]{#2 \stackrel{#1}{\leadsto} #3}

\title{Modular Decomposition of Hierarchical Finite State Machines}
\author{Oliver Biggar, Behzad Zamani, Iman Shames}
\date{}

\begin{document}
\maketitle
\begin{abstract}
Hierarchical Finite State Machines (HFSMs) are a standard software-modelling concept which extends the classical Finite State Machine (FSM) notion with the useful abstraction of \emph{hierarchical nesting}. That is, an HFSM is an FSM whose states can be other FSMs. The hierarchy in HFSMs is provided at design time, and can be removed by expanding nested states, allowing HFSMs to inherit the semantics of FSMs. However, because hierarchy is a useful modular representation of the structure of an FSM, we would like to be able to invert this operation: given an FSM, can we compute equivalent HFSMs? This is the topic of this paper. By adapting the analogous theory of `modular decomposition' from graph theory into automata theory, we are able to compute an efficient representation of the space of equivalent HFSMs to a given one. Specifically, we first define a \emph{module} of an FSM, which is a collection of nodes which can be treated as a nested FSM. Unlike modules in graphs, some modules in FSMs are lacking in algebraic structure. We identify a simple and natural restriction of the modules, called \emph{thin modules}, which regain many of the critical properties from modules in graphs. We then construct a linear-space directed graph which uniquely represents every thin module, and hence every equivalent (thin) HFSM. We call this graph the \emph{modular decomposition}. The modular decomposition makes clear the significant common structure underlying equivalent thin HFSMs. We provide an $O(n^2k)$ algorithm for constructing the modular decomposition of an $n$-state $k$-symbol FSM. We demonstrate the applicability of this theory on the following `bottleneck' problem: given an HFSM, find an equivalent one where the size of the largest component FSM is minimised. The modular decomposition gives a simple greedy algorithm for the bottleneck problem on thin HFSMs, which we demonstrate on a wristwatch HFSM example from Harel \cite{harel1987statecharts}.
%In this paper we develop an analogue of the graph-theoretic `modular decomposition' in automata theory. This decomposition allows us to identify \emph{hierarchical finite state machines} (HFSMs) equivalent to a given \emph{finite state machine} (FSM). We first define a \emph{module} of an FSM, which is a collection of nodes which can be treated as a nested FSM. We then identify a natural subset of FSM modules called \emph{thin modules}, which are algebraically well-behaved. We construct a linear-space directed graph, which uniquely represents every thin module, and hence every equivalent (thin) HFSM. We call this graph the \emph{modular decomposition}. The modular decomposition makes clear the significant common structure underlying equivalent HFSMs, and allows us to efficiently construct equivalent HFSMs. Finally, we provide an $O(n^2k)$ algorithm for constructing the modular decomposition of an $n$-state $k$-symbol FSM.
\end{abstract}

\section{Introduction} \label{sec:introduction}

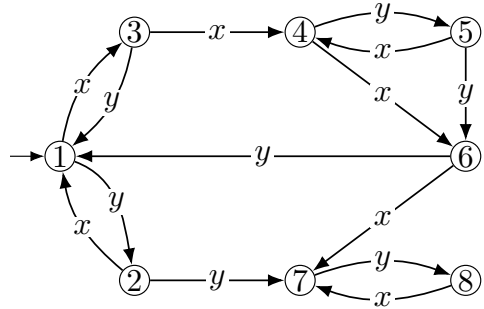
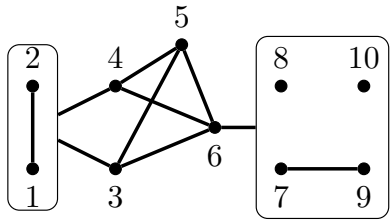
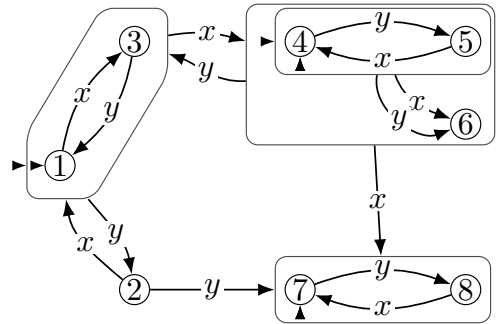
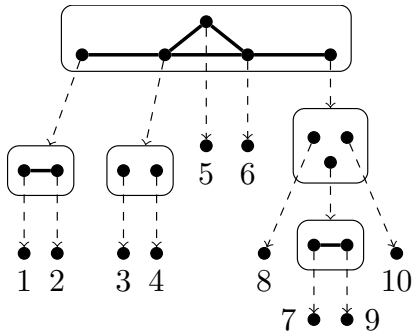
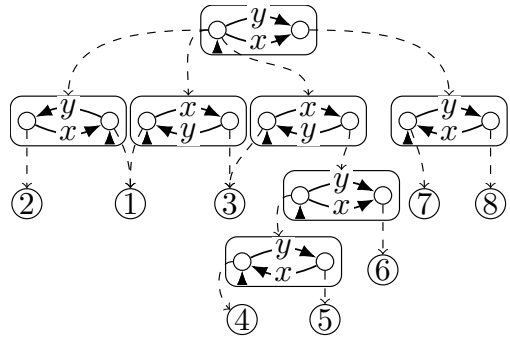
\begin{figure}
    \centering
    \begin{subfigure}{0.4\textwidth}
        \begin{tikzpicture}
\path [use as bounding box] (0,-0.4) rectangle (6,3.6);
\node (1) [label = below:{1},ball] at (1,1) {};
\node (2) [label = above:2,ball] at (1,2) {};
\node (3) [label = below:3,ball] at (2,1) {};
\node (4) [label = above:4,ball] at (2,2) {};
\node (5) [label = above:5,ball] at (2.8,2.5) {};
\node (6) [label = below:6,ball] at (3.6,1.5) {};
\node (7) [label = below:7,ball] at (4,1) {};
\node (8) [label = above:8,ball] at (4,2) {};
\node (9) [label = below:9,ball] at (5,1) {};
\node (10) [label = above:10,ball] at (5,2) {};

\draw[very thick] (1) to (2);
\draw[very thick] (2) to (3);
\draw[very thick] (1) to (3);
\draw[very thick] (1) to (4);
\draw[very thick] (2) to (4);
\draw[very thick] (3) to (5);
\draw[very thick] (4) to (5);
\draw[very thick] (3) to (6);
\draw[very thick] (4) to (6);
\draw[very thick] (5) to (6);
\draw[very thick] (6) to (8);
\draw[very thick] (6) to (7);
\draw[very thick] (6) to (9);
\draw[very thick] (6) to (10);
\draw[very thick] (7) to (9);
 \end{tikzpicture}
        \caption{A graph $G$}
        \label{fig:G}
    \end{subfigure}
    \hskip 1cm
    \begin{subfigure}{0.4\textwidth}
        \begin{tikzpicture}
\path [use as bounding box] (0,0) rectangle (6,4);
%\draw (0,0) rectangle (6,4);
%BASIS BOXES
% { 1, 2 }
%\draw[ opacity=0.2, rounded corners] (0,2.7) -- (0.8,2.7) -- (3.6,1) -- (3.6,0) -- (2.9,0) -- (0,1.9) -- cycle;
% { 1, 3 }
%\draw[ opacity=0.2, rounded corners] (0,2.7) -- (2.6,4.4) -- (3.6,4.4) -- (3.6,3.6) -- (0.8,1.9) -- (0,1.9) -- cycle;
% { 10, 11 }
%\draw[ opacity=0.2, rounded corners] (10.4,2.05) -- (14,2.05) -- (14,0.6) -- (10.4,0.6) -- cycle;
% { 4, 5 }
%\draw[ opacity=0.2, rounded corners] (5.2,4.75) -- (8.8,4.75) -- (8.8,3.3) -- (5.2,3.3) -- cycle;
% { 8, 9 }
%\draw[ opacity=0.2, rounded corners] (10.4,4.4) -- (13,5.5) -- (13.9,5.5) -- (13.9,4.8) -- (11.2,3.6) -- (10.4,3.6) -- cycle;
% { 4, 5, 6 }
%\draw[ opacity=0.2, rounded corners] (5.1,4.85) -- (8.9,4.85) -- (8.9,1.9) -- (8.1,1.9) -- (5.1,3.2) -- cycle;
% { 4, 5, 6, 8, 9 }
%\draw[ opacity=0.2, rounded corners] (5,4.95) -- (13,5.6) -- (14,5.6) -- (14,4.7) -- (9,1.8) -- (8,1.8) -- (5,3.1) -- cycle;
% { 3, 4, 5, 6, 8, 9 }
%\draw[ opacity=0.2, rounded corners] (2.7,4.95) -- (13,5.7) -- (14.1,5.7) -- (14.1,4.6) -- (9,1.7) -- (8,1.7) -- (2.7,3.5) -- cycle;

\node (1) [state,init] at (0.6,2) {1};
\node (2) [state] at (1.5,0.5) {2};
\node (3) [state] at (1.5,3.5) {3};
\node (4) [state] at (3.5,3.5) {4};
\node (5) [state] at (5.5,3.5) {5};
\node (6) [state] at (5.5,2) {6};
\node (7) [state] at (3.5,.5) {7};
\node (8) [state] at (5.5,.5) {8};
% 1 ---y--> 2
\draw[edge] (1) to[out=-20,in=100] node[edgelabel] {$y$} (2);
% 1 ---x--> 3
\draw[edge] (1) to[out=80,in=-140] node[edgelabel] {$x$} (3);
% 2 ---x--> 1
\draw[edge] (2) to[out=140,in=-80] node[edgelabel] {$x$} (1);
% 2 ---y--> 1
\draw[edge] (2) to node[edgelabel] {$y$} (7);
% 3 ---y--> 1
\draw[edge] (3) to[out=-100,in=40] node[edgelabel] {$y$} (1);
% 3 ---x--> 4
\draw[edge] (3) to[out=0,in=180] node[edgelabel] {$x$} (4);
% 4 ---y--> 5
\draw[edge] (4) to[out=20,in=160] node[edgelabel] {$y$} (5);
% 4 ---x--> 6
\draw[edge] (4) to node[edgelabel] {$x$} (6);
% 5 ---y--> 6
\draw[edge] (5) to node[edgelabel] {$y$} (6);
% 6 ---y--> 1
\draw[edge] (6) to[out=180.00,in=0.00] node[edgelabel] {$y$} (1);
% 6 ---x--> 7
\draw[edge] (6) to node[edgelabel] {$x$} (7);
% 5 ---x--> 4
\draw[edge] (5) to[out=-160,in=-20] node[edgelabel] {$x$} (4);
% 7 ---y--> 8
\draw[edge] (7) to[out=20,in=160] node[edgelabel] {$y$} (8);
% 8 ---x--> 7
\draw[edge] (8) to[out=-160.60,in=-20] node[edgelabel] {$x$} (7);
 \end{tikzpicture}
        \caption{An FSM $Z$}
        \label{fig:Z}
    \end{subfigure}
    \vskip 1cm
    \begin{subfigure}{0.4\textwidth}
        \begin{tikzpicture}
\path [use as bounding box] (0,-0.4) rectangle (6,3.6);
\node (1) [label = below:1,ball] at (1,1) {};
\node (2) [label = above:2,ball] at (1,2) {};
\node (3) [label = below:3,ball] at (2,1) {};
\node (4) [label = above:4,ball] at (2,2) {};
\node (5) [label = above:5,ball] at (2.8,2.5) {};
\node (6) [label = below:6,ball] at (3.2,1.5) {};
\node (7) [label = below:7,ball] at (4,1) {};
\node (8) [label = above:8,ball] at (4,2) {};
\node (9) [label = below:9,ball] at (5,1) {};
\node (10) [label = above:10,ball] at (5,2) {};

\node[draw,rounded corners,minimum height=2cm, minimum width=0.6cm] (12) at (1,1.5) {};
%\node[draw,rounded corners,minimum height=2cm, minimum width=0.6cm] (34) at (2,1.5) {};
\node[draw,rounded corners,minimum height=2.2cm, minimum width=1.6cm] (78910) at (4.5,1.5) {};
%\node[draw,rounded corners,minimum height=0.8cm, minimum width=1.4cm] (79) at (4.5,0.9) {};

\draw[very thick] (1) to (2);
\draw[very thick] (12) to (3);
\draw[very thick] (12) to (4);
\draw[very thick] (3) to (5);
\draw[very thick] (4) to (5);
\draw[very thick] (3) to (6);
\draw[very thick] (4) to (6);
\draw[very thick] (5) to (6);
\draw[very thick] (6) to (78910);
\draw[very thick] (7) to (9);

 \end{tikzpicture}
        \caption{Hierarchical graph equivalent to $G$}
        \label{fig:G nested}
    \end{subfigure}
    \hskip 1cm
    \begin{subfigure}{0.4\textwidth}
        \begin{tikzpicture}
\path [use as bounding box] (0,0) rectangle (6,4);
%\draw (0,0) rectangle (6,4);

\node[opacity=0.7,draw,rounded corners,minimum height=0.8cm, minimum width=2.6cm,initial, initial text={}, initial distance=1ex] (45) at (4.5,3.5) {};

\node[opacity=0.7,draw,rounded corners,minimum height=1.7cm, minimum width=3cm] (456) at (4.35,3.1) {};

\node[opacity=0.7,draw,rounded corners,minimum height=0.8cm, minimum width=2.6cm] (78) at (4.5,0.5) {};

% { 1, 2 }
%\draw[opacity=0.7, rounded corners] (0.2,2.4) -- (1,2.4) -- (1.9,.9) -- (1.9,0.1) -- (1.1,0.1) -- (0.2,1.6) -- cycle;
% { 1, 3 }
\draw[opacity=0.7, rounded corners] (0.2,1.6) -- (1,1.6) -- (1.9,3.1) -- (1.9,3.9) -- (1.1,3.9) -- (0.2,2.4) -- cycle;

% { 3, 4, 5, 6}
%\draw[opacity=0.7, rounded corners] (1.1,4) -- (1.1,3.1) -- (2.1,3.1) -- (2.9,2.2) -- (5.9,2.2) -- (5.9,4) --cycle;

\node[opacity=0.7,rounded corners,minimum height=0.8cm, minimum width=0.8cm,initial, initial text={}, initial distance= 1ex] (1f) at (0.6,2) {};
\node[opacity=0.7,rounded corners,minimum height=0.8cm, minimum width=0.8cm] (2f) at (1.5,0.5) {};
\node[opacity=0.7,rounded corners,minimum height=0.8cm, minimum width=0.8cm] (3f) at (1.5,3.5) {};

\node (1) [state,initial, initial text={}, initial distance= 1ex] at (0.6,2) {1};
\node (2) [state] at (1.5,0.5) {2};
\node (3) [state] at (1.5,3.5) {3};
\node (4) [state, initial, initial text={}, initial distance= 1ex, initial below] at (3.5,3.5) {4};
\node (5) [state] at (5.5,3.5) {5};
\node (6) [state] at (5.5,2.5) {6};
\node (7) [state, initial, initial text={}, initial distance= 1ex, initial below] at (3.5,.5) {7};
\node (8) [state] at (5.5,.5) {8};
% 1 ---y--> 2
\draw[edge] (1f) to[out=-50,in=100] node[edgelabel] {$y$} (2);
% 1 ---x--> 3
\draw[edge] (1) to[out=80,in=-140] node[edgelabel] {$x$} (3);
% 2 ---x--> 1
\draw[edge] (2) to[out=140,in=-80] node[edgelabel] {$x$} (1f);
% 2 ---y--> 7
\draw[edge] (2) to node[edgelabel] {$y$} (78);
% 3 ---y--> 1
\draw[edge] (3) to[out=-100,in=40] node[edgelabel] {$y$} (1);
% 3 ---x--> 4
\draw[edge] (3f) to[out=10,in=165] node[edgelabel] {$x$} (456);
% 4 ---y--> 5
\draw[edge] (4) to[out=20,in=160] node[edgelabel] {$y$} (5);
% 4 ---x--> 6
\draw[edge] (45) to[out=-70,in=160] node[edgelabel] {$x$} (6);
% 5 ---y--> 6
\draw[edge] (45) to[out=-100,in=-160] node[edgelabel] {$y$} (6);
% 6 ---y--> 1
\draw[edge] (456) to[out=-177,in=-20] node[edgelabel] {$y$} (3f);
% 6 ---x--> 7
\draw[edge] (456) to node[edgelabel] {$x$} (78);
% 5 ---x--> 4
\draw[edge] (5) to[out=-160,in=-20] node[edgelabel] {$x$} (4);
% 7 ---y--> 8
\draw[edge] (7) to[out=20,in=160] node[edgelabel] {$y$} (8);
% 8 ---x--> 7
\draw[edge] (8) to[out=-160.60,in=-20] node[edgelabel] {$x$} (7);
 \end{tikzpicture}
        \caption{Hierarchical FSM equivalent to $Z$}
        \label{fig:Z nested}
    \end{subfigure}
    \vskip 1cm
    \begin{subfigure}{0.4\textwidth}
        \begin{tikzpicture}
\path [use as bounding box] (0,0) rectangle (6,4);

\node (1) [label = below:1,ball] at (0.8,1) {};
\node (2) [label = below:2,ball] at (1.2,1) {};
\node (3) [label = below:3,ball] at (2,1) {};
\node (4) [label = below:4,ball] at (2.4,1) {};
\node (5) [label = below:5,ball] at (3,2.3) {};
\node (6) [label = below:6,ball] at (3.5,2.3) {};
\node (7) [label = left:7,ball] at (4.3,0.2) {};
\node (8) [label = below:8,ball] at (3.7,1) {};
\node (9) [label = right:9,ball] at (4.7,0.2) {};
\node (10) [label = below:10,ball] at (5.3,1) {};

\node (a) [ball] at (1.5,3.4) {};
\node (b) [ball] at (2.5,3.4) {};
\node (c) [ball] at (3,3.8) {};
\node (d) [ball] at (3.5,3.4) {};
\node (e) [ball] at (4.5,3.4) {};
\node[draw,rounded corners,minimum height=.8cm, minimum width=3.5cm] (abcde) at (3,3.6) {};
\draw[very thick] (a) to (b);
\draw[very thick] (b) to (c);
\draw[very thick] (b) to (d);
\draw[very thick] (c) to (d);
\draw[very thick] (d) to (e);

\draw[-to, dashed] (c) to (5);
\draw[-to, dashed] (d) to (6);

\node (x) [ball] at (4.3,2.4) {};
\node (y) [ball] at (4.5,2.1) {};
\node (z) [ball] at (4.7,2.4) {};
\node[draw,rounded corners,minimum height=.9cm, minimum width=.9cm] (xyz) at (4.5,2.3) {};
\draw[-to, dashed] (e) to (xyz);
\draw[-to, dashed] (x) to (8);
\draw[-to, dashed] (z) to (10);

\node (v) [ball] at (4.3,1.1) {};
\node (w) [ball] at (4.7,1.1) {};
\draw[very thick] (v) to (w);
\node[draw,rounded corners,minimum height=0.6cm, minimum width=.8cm] (vw) at (4.5,1.1) {};
\draw[-to, dashed] (y) to (vw);
\draw[-to, dashed] (v) to (7);
\draw[-to, dashed] (w) to (9);

\node (e) [ball] at (0.8,2) {};
\node (f) [ball] at (1.2,2) {};
\draw[very thick] (e) to (f);
\node[draw,rounded corners,minimum height=0.6cm, minimum width=.8cm] (12) at (1,2) {};
\draw[-to, dashed] (a) to (12);
\draw[-to, dashed] (e) to (1);
\draw[-to, dashed] (f) to (2);

\node (g) [ball] at (2,2) {};
\node (h) [ball] at (2.4,2) {};
\node[draw,rounded corners,minimum height=0.6cm, minimum width=.8cm] (34) at (2.2,2) {};
\draw[-to, dashed] (b) to (34);
\draw[-to, dashed] (g) to (3);
\draw[-to, dashed] (h) to (4);

 \end{tikzpicture}
        \caption{\textbf{Modular decomposition} of $G$ (represents all equivalent hierarchical graphs)}
        \label{fig:G tree}
    \end{subfigure}
    \hskip 1cm
    \begin{subfigure}{0.4\textwidth}
        \begin{tikzpicture}
\path [use as bounding box] (0,0) rectangle (6,4);

\newcommand{\locx}{3}
\newcommand{\locy}{3.7}
\node (123456o) [state,initial, initial text={}, initial distance=1ex,initial below] at (\locx - 0.5,\locy) {};
\node (78o) [state] at (\locx + 0.5,\locy) {};
\node[draw,rounded corners,minimum height=.6cm, minimum width=1.4cm] (all) at (\locx,\locy) {};
\draw[edge] (123456o) to[out=25,in=155] node[edgelabel] {$y$} (78o);
\draw[edge] (123456o) to[out=-25,in=-155] node[edgelabel] {$x$} (78o);

\renewcommand{\locx}{.7}
\renewcommand{\locy}{2.6}
\node (2o) [state] at (\locx - 0.5,\locy) {};
\node (1oL) [state,initial, initial text={}, initial distance=1ex,initial below] at (\locx + 0.5,\locy) {};
\node[draw,rounded corners,minimum height=.6cm, minimum width=1.4cm] (12) at (\locx,\locy) {};
\draw[edge] (1oL) to[out=155,in=25] node[edgelabel] {$y$} (2o);
\draw[edge] (2o) to[out=-25,in=-155] node[edgelabel] {$x$} (1oL);
\draw[-to, dashed] (123456o) to[out=180,in=90] (12);

\node (1) [state] at (\locx + 0.75,1.6) {1};
\draw[-to, dashed] (1oL) to[out=-60,in=90] (1);
\node (2) [state] at (\locx-0.5,1.6) {2};
\draw[-to, dashed] (2o) to[out=-90,in=90] (2);

\renewcommand{\locx}{2.15}
\node (1oR) [state,initial, initial text={}, initial distance=1ex,initial below] at (\locx - 0.5,\locy) {};
\node (3oL) [state] at (\locx + 0.5,\locy) {};
\node[draw,rounded corners,minimum height=.6cm, minimum width=1.4cm] (13) at (\locx,\locy) {};
\draw[edge] (1oR) to[in=155,out=25] node[edgelabel] {$x$} (3oL);
\draw[edge] (3oL) to[in=-25,out=-155] node[edgelabel] {$y$} (1oR);
\draw[-to, dashed] (123456o) to[out=180,in=90] (13);

\node (3) [state] at (\locx+0.5,1.6) {3};
\draw[-to, dashed] (1oR) to[out=-120,in=90] (1);
\draw[-to, dashed] (3oL) to[out=-90,in=90] (3);

\renewcommand{\locx}{3.6}
\node (3oR) [state,initial, initial text={}, initial distance=1ex,initial below] at (\locx - 0.5,\locy) {};
\node (456o) [state] at (\locx + 0.5,\locy) {};
\node[draw,rounded corners,minimum height=.6cm, minimum width=1.4cm] (3456) at (\locx,\locy) {};
\draw[edge] (3oR) to[in=155,out=25] node[edgelabel] {$x$} (456o);
\draw[edge] (456o) to[in=-25,out=-155] node[edgelabel] {$y$} (3oR);
\draw[-to, dashed] (123456o) to[out=-60,in=90] (3456);
\draw[-to, dashed] (3oR) to[out=-120,in=90] (3);

\renewcommand{\locx}{5.3}
\node (7o) [state,initial, initial text={}, initial distance=1ex,initial below] at (\locx - 0.5,\locy) {};
\node (8o) [state] at (\locx + 0.5,\locy) {};
\node[draw,rounded corners,minimum height=.6cm, minimum width=1.4cm] (78) at (\locx,\locy) {};
\draw[edge] (7o) to[in=155,out=25] node[edgelabel] {$y$} (8o);
\draw[edge] (8o) to[in=-25,out=-155] node[edgelabel] {$x$} (7o);
\draw[-to, dashed] (78o) to[out=0,in=90] (78);

\node (7) [state] at (\locx-0.3,1.6) {7};
\draw[-to, dashed] (7o) to[out=-70,in=90] (7);
\node (8) [state] at (\locx +.5,1.6) {8};
\draw[-to, dashed] (8o) to[out=-90,in=90] (8);

\renewcommand{\locx}{4}
\renewcommand{\locy}{1.7}
\node (45o) [state,initial, initial text={}, initial distance=1ex,initial below] at (\locx - 0.5,\locy) {};
\node (6o) [state] at (\locx + 0.5,\locy) {};
\node[draw,rounded corners,minimum height=.6cm, minimum width=1.4cm] (456) at (\locx,\locy) {};
\draw[edge] (45o) to[out=25,in=155] node[edgelabel] {$y$} (6o);
\draw[edge] (45o) to[out=-25,in=-155] node[edgelabel] {$x$} (6o);
\draw[-to, dashed] (456o) to[out=-90,in=90] (456);

\node (6) [state] at (\locx + 0.5,0.8) {6};
\draw[-to, dashed] (6o) to[out=-90,in=90] (6);

\renewcommand{\locx}{3.3}
\renewcommand{\locy}{0.9}
\node (4o) [state,initial, initial text={}, initial distance=1ex,initial below] at (\locx - 0.5,\locy) {};
\node (5o) [state] at (\locx + 0.5,\locy) {};
\node[draw,rounded corners,minimum height=.6cm, minimum width=1.4cm] (45) at (\locx,\locy) {};
\draw[edge] (4o) to[out=25,in=155] node[edgelabel] {$y$} (5o);
\draw[edge] (5o) to[in=-25,out=-155] node[edgelabel] {$x$} (4o);
\draw[-to, dashed] (45o) to[out=180,in=100] (45);

\node (4) [state] at (\locx - 0.5,0.2) {4};
\draw[-to, dashed] (4o) to[out=180,in=140] (4);
\node (5) [state] at (\locx + 0.5,0.2) {5};
\draw[-to, dashed] (5o) to[out=-90,in=90] (5);

 \end{tikzpicture}
        \caption{\textbf{Modular decomposition} of $Z$ (represents all equivalent HFSMs)}
        \label{fig:Z tree}
    \end{subfigure}
    %\begin{subfigure}{0.4\textwidth}
    %    \includestandalone[width=\textwidth]{figs/graph_grouped}
    %    \caption{A nested drawing of the modular decomposition of $G$}
    %    \label{fig:G mods}
    %\end{subfigure}
    %\hskip 1cm
    %\begin{subfigure}{0.4\textwidth}
    %    \includestandalone[width=\textwidth]{figs/fsm_grouped}
    %    \caption{A nested drawing of the modular decomposition of $Z$}
    %    \label{fig:Z mods}
    %\end{subfigure}
    \caption{A graphical depiction of the contribution of this paper. The left column shows the modular decomposition of a graph $G$, which constructs a hierarchical representation of $G$ which captures all other such representations~\cite{gallai1967transitiv}. The right column shows our analogous theory for FSMs. The main contribution is the modular decomposition of an FSM (Fig.~\ref{fig:Z tree}), a tree-like structure which represents all HFSMs equivalent to a given FSM.}
    \label{fig:graph modular decomposition}
\end{figure}
%maybe list the fields.
Finite State Machines (FSMs) are a fundamental model in theoretical computer science, with applications across numerous disciplines of science and engineering. One well-known extension of this model, originally introduced by Harel~\cite{harel1987statecharts}, is the notion of \emph{hierarchy}. Harel proposed allowing FSMs to be \emph{nested} with states of other FSMs, leading to a tree-like structure which nowadays is called a \emph{hierarchical finite state machine} (HFSM)~\cite{alur1998model}. See Figures~\ref{fig:Z} and \ref{fig:Z nested}. HFSMs manage complexity through modularity, separating independent areas of a complex system. Standard FSMs have no such way of being broken down, so understanding their behaviour can be difficult as they grow in size. HFSMs can provide a compact representation of FSMs, which can be exploited for model checking~\cite{alur1998model,alur2000efficient,alur2005analysis,laster1998modular}. Nowadays, HFSMs are a ubiquitous modelling tool, being a standardised part of the Unified Modelling Language (UML)~\cite{boochunified}.

The hierarchy of HFSMs is an optional design tool to improve clarity---from a semantic perspective, HFSMs are the same as FSMs. In fact, it is easy to remove the hierarchy from an HFSM by recursively \emph{expanding} nested states, transforming it into an equivalent FSM. This allows HFSMs to inherit the semantics of FSMs.%us to assign semantics to HFSMs and gives them a natural equivalence relation and partial ordering.%: an HFSM $Z$ \emph{refines} an HFSM $Y$ if we can obtain $Z$ by expanding nested states in $Y$. Two HFSMs are equivalent if they are both refinements of the same FSM.

However, because hierarchy is \emph{a good thing}, we generally don't want to expand HFSMs. We would prefer the opposite transformation: given an FSM, \emph{we want to find which hierarchical FSMs are equivalent to it}. We arrive at our central question: \textbf{how can we discover the innate hierarchical structure in FSMs?} This is the goal of this paper. This line of research is motivated in two ways: (1) it lends insight into FSM structure, %and (2) it gives us the means to explore equivalence classes of HFSMs, so that we can solve optimisation problems on HFSMs efficiently. A particular motivating example is the \emph{bottleneck problem} for HFSMs: given an HFSM, find an equivalent one with minimal \emph{width}, which is the size of the largest component FSM. Many optimisation problems on HFSMs (such as finding shortest paths) have complexity bounded by a superlinear function of the size of each component FSM, and so minimising the width of the HFSM bounds the complexity of such problems.
and (2) it allows us efficiently solve optimisation problems on HFSMs, by identifying \emph{structural bottlenecks} for these algorithms. As an example, the shortest path problem on HFSMs is an important algorithm for planning applications. Given an HFSM, we can find shortest paths in a recursive fashion, so the overall computational complexity is a superlinear function of the size of the \emph{largest nested FSM} in the HFSM, a parameter we call its \emph{width}. Like `treewidth' in graph theory, the width of an HFSM bounds the complexity of recursive optimisation problems\footnote{A different but similar notion is McCabe's \emph{cyclomatic complexity}, a graph-theoretic metric for software complexity~\cite{mccabe1976complexity}.}. This leads to the \emph{bottleneck problem} for HFSMs: given an HFSM, find an equivalent one with minimal width. However, to our knowledge, this important problem for HFSMs has received no formal study---all hierarchy in HFSMs is assumed to be constructed at design time. More generally, we would like a representation of the space of equivalent HFSMs, allowing us to explore this space efficiently.

Luckily, we have a direct source of inspiration for this problem: the \emph{modular decomposition}, from graph theory. The modular decomposition is the graph-theoretic analogue of our goal for FSMs: given a graph, it provides an efficient representation of the space of equivalent hierarchically nested graphs (Figures~\ref{fig:G tree} and~\ref{fig:G nested}). Originally developed by Gallai~\cite{gallai1967transitiv} for the purpose of constructing transitive orientations of comparability graphs, it has since been applied to a large number of problems in graph theory and combinatorics%. These include recognising and performing optimisation problems on a variety of classes of graphs and partial orders
~\cite{habib2010survey,mohring1984substitution}. The concept has been generalised to directed graphs, set systems, matroids, hypergraphs and other combinatorial objects~\cite{mohring1984substitution,mcconnell1999modular}. When we have a hierarchical graph, such as Figure~\ref{fig:G nested}, we can remove the hierarchy by recursively \emph{expanding} the nested graphs (see Figure~\ref{fig:expansion}), an operation which replaces a node $v$ containing a nested graph $H$ with the nodes of $H$, where all nodes previously adjacent to $v$ become adjacent to all nodes of $H$. The goal of modular decomposition theory is to invert this operation, finding the sets of nodes in a graph which could have resulted from an expansion. These sets are called \emph{modules}. Intuitively, modules are sets of nodes which can be
%, that of a \emph{module}, which is a set $M$ of nodes in a graph which all share a common neighbourhood among nodes not in $M$. %Precisely, if $V(G)$ is the set of nodes of $G$, then a set $M\subseteq V$ is a \emph{module} if every element of $V\setminus M$ is adjacent to all nodes of $M$ or none.
%So, for instance, in the above example, the nodes of $H$ always form a module in $G\cdot_v H$. The converse is always also true---every module $M$ can be considered as the image of an expansion, specifically the expansion of $G[M]$ at $M$ in $G/M$ (Theorem~\ref{concrete graphs}), where $G/M$ is the \emph{contraction} of $M$ (the graph obtained by identifying the nodes of $M$, Definition~\ref{contraction}) and $G[M]$ is the \emph{restriction} (the induced subgraph of $M$ in $G$, Definition~\ref{restriction}). That is, a set is a module precisely when it can be 
treated as a nested graph in an equivalent hierarchical representation. Further, modules of modules are themselves modules (Theorem~\ref{gallai},~\cite{gallai1967transitiv}), so we can construct equivalent hierarchical graphs by repeating the process of identifying and nesting modules. Modules can overlap, so graphs can have exponentially many modules, and hence there are a large number of equivalent hierarchical graphs. However, Gallai~\cite{gallai1967transitiv} constructed a unique tree which succinctly represents all equivalent hierarchical graphs, and this tree is what is known as the \textbf{modular decomposition} of the graph. The modular decomposition allows us to efficiently solve combinatorial optimisation problems on graphs. In this paper, we develop the theory of modular decomposition on HFSMs, and show how it can be used to solve the bottleneck problem efficiently.

%much of the richness of the graph theory
Our contributions are as follows. First, we identify the analogue of graph modules in FSMs. We do this by characterising the \emph{role} which modules play with respect to the operations of expansion, contraction and restriction in graph theory (Section~\ref{sec:modules}). By formalising these operations on FSMs, we identify FSM modules as the sets of states which play the same role. Here we discover a key fact which---at first---appears to be a significant obstacle to a modular decomposition theory for FSMs. FSM modules, unlike graph modules, lack algebraic structure. Overlapping graph modules are closed under intersection, union, difference and symmetric difference~\cite{bui2008tree} but FSM modules are not closed under any of these! See Figure~\ref{fig:fsm_counter}. Fortunately, we find a path around this obstacle: there is a subset of modules called \emph{thin modules}, which exist in all FSMs, and which are closed under overlapping intersection and union. Thin modules are motivated by the observation that, while a single $x$-input may not cause an $x$-transition in an HFSM, \emph{repeated} $x$-inputs should cause an $x$-transition to occur. From this point we restrict our attention to thin modules and thin HFSMs, which are HFSMs whose nested FSMs are thin modules. Thinness is a somewhat natural property; for instance, Harel's wristwatch model in his original statecharts paper is thin \cite{harel1987statecharts}. See Section~\ref{sec:thin modules}.

%\footnote{Modules in graphs satisfy Lemma~\ref{union and intersection}~\cite{bui2008tree}, as do modules in decision structures~\cite{biggar2022modularity}. Modules in graphs are additionally closed under set difference~\cite{bui2008tree}, which FSM modules are not (even when thin). As shown in~\cite{mohring1984substitution, bui2008tree}, set families where pairs of overlapping sets are closed under difference, union and intersection always have a tree representation. Our result is not a consequence of this because we do not require modules to be closed under set difference. Separately, it is proved in~\cite{gabow1995centroids} that set families closed under union and intersection of overlapping sets (such as the thin modules of an FSM) have a $O(n^2)$-size representation. Our representation is $O(nk)$ in the size of the input FSM.} lemma which holds for thin modules but not for modules in general.

%Unfortunately, FSM modules lack key algebraic properties which graph modules possess. However, we identify a natural restriction of this definition, called \emph{thin modules}, which, like graph modules, are closed under intersection and union (Section~\ref{sec:thin modules}). 

%The results of our paper from here apply only to thin modules and thin HFSMs, which are those whose nested FSMs are thin modules. 
We then define the \emph{modular decomposition} of a thin HFSM, which is a directed acyclic graph built from the non-singleton \emph{indecomposable} thin modules, which we call the \emph{basis modules} of the decomposition. Like in the graph case, we label each node in the modular decomposition by the \emph{contracted form} (Definition~\ref{contracted form}) of the associated basis module (see Figure~\ref{fig:Z tree}). Our main theorem (Theorem~\ref{basis theorem}) establishes that the modular decomposition is linear in size and uniquely represents all thin modules and hence all equivalent thin HFSMs. Two equivalent HFSMs (such as Figures~\ref{fig:W1} and \ref{fig:W2}) have a one-to-one correspondence between their bases which preserves contracted forms, so their modular decomposition always consists of the same component FSMs, possibly nested in different orders. We further show that this modular decomposition can be constructed efficiently, for which we describe an algorithm in Section~\ref{sec:computing}. This leads to a simple greedy solution to the bottleneck problem: by selecting and contracting modules in the modular decomposition, we can extend any thin HFSM to one with minimal width. As an example, $Z$'s modular decomposition (Figure~\ref{fig:Z tree}) has three overlapping basis modules $\{1,2\}$, $\{1,3\}$ and $\{3,4,5,6\}$. This can then be easily extended to (possibly multiple equivalent) minimal width HFSMs (Figure~\ref{fig:W1} and \ref{fig:W2}), but the bases of these HFSMs remain in a one-to-one correspondence. In Section~\ref{sec: applications} we demonstrate our technique on a larger real-world example: Harel's wristwatch model from \cite{harel1987statecharts}. Using the modular decomposition we efficiently construct an equivalent minimal-width model to Harel's, and in doing so demonstrate the subcomponent which acts as a bottleneck for decomposition. We conclude and discuss open problems in Section~\ref{sec:conclusions}. Missing proofs can be found in the Appendix.

\section{Related Work} \label{sec:relatedwork}

Decomposition of automata was extensively studied in the sixties~\cite{karp1964some,kohavi1965decomposition,hartmanis1966algebraic,weiner1967modular}, with the focus on a `cascading' composition, where the output of one machine is fed into the input of another. An important result was Krohn-Rhodes theory~\cite{krohn1965algebraic}, which presents a decomposition of an automaton and its associated \emph{transition monoid}~\cite{eilenberg1974automata}. Our theory is distinctly different, because our FSMs do not have output, and so the decomposition is hierarchical rather than cascading\footnote{Unfortunately, this decomposition is also sometimes called `modular decomposition'~\cite{weiner1967modular}, which can lead to confusion.}.

Our approach and naming follow the modular decomposition in graph theory, a parallel development beginning around the same period~\cite{gallai1967transitiv}. Later work generalised the modular decomposition to many other kinds of mathematical structure~\cite{mohring1984substitution,mcconnell1999modular}; a survey is given in~\cite{habib2010survey}. There are established criteria for when a family of sets has a tree representation analogous to that of the modular decomposition~\cite{mohring1984substitution,bui2008tree,habib2010survey}. For instance, families of sets closed under overlapping union, intersection, set difference and symmetric difference always have a linear representation~\cite{habib2010survey}. Further, families closed under only overlapping union and intersection have a quadratic-size representation~\cite{gabow1995centroids}. Thin FSM modules are only closed under overlapping union and intersection, so our linear-space representation improves on these results. While modular decomposition has been applied to directed graphs, which are similar to FSMs, the notions of `equivalence' and `nesting' on FSMs and graphs do not coincide, so the concept of a module is different. See Section~\ref{sec:modules}.

%In software engineering, `modularity' (in various senses) has been identified as a desirable property for many years. Several established studies focus on defining techniques for breaking software into smaller and more maintainable modules~\cite{parnas1972criteria,parnas1985modular}. More specifically, using hierarchy to manage complexity of FSMs seems to begin with the work of Harel~\cite{harel1987statecharts}, and has since been much discussed in the field of software modelling. However, to the best of our knowledge there has been no mathematical study identifying the possible hierarchical finite state machines equivalent to a given one.

In recent years, HFSMs have received interest in formal analysis of software, particularly in model checking~\cite{alur1998model,alur2000efficient,laster1998modular,clarke2000modular,alur2003hierarchical,alur2005analysis}. It is shown in~\cite{alur1998model} that HFSMs can be (when states can be equivalent to each other) significantly smaller than their equivalent FSMs, and model checking can be performed in time proportional to its size, thus providing a significant improvement in complexity when the HFSM is small. Our work complements this by showing how we can construct HFSMs from FSMs, suggesting that it may be possible to use modular decomposition as a preprocessing step for model checking to reduce the size of an FSM's representation. %compress an FSM into a smaller one
% BE PRECISE HERE, WE CAN'T EXACTLY DO THIS. TALK ABOUT NP-HARDNESS
In~\cite{biggar2022modularity}, the authors perform a modular decomposition on an acyclic FSM-like architecture called a \emph{decision structure}. The module definition presented there is a special case of the thin modules we define in this paper. However, FSMs are allowed to have cycles, which adds significant complexity to the theory.
%Most closely, the module definition we use here directly generalises our module definition in where we perform a similar decomposition for  This provided a convenient explanation of why popular decision-making architectures like Behavior Trees~\cite{btbook} and Decision Trees are `modular'. The algorithm we present here will also work to identify modules in that case.

In theoretical computer science, FSMs are often studied as \emph{deterministic finite automata}, which differ only in the addition of `accepting sets'. In this paper, we focus on the software motivations and so treat every state as distinct. However, our theory lays the foundations for this interesting problem to be tackled in future work. If we add an equivalence relation on states (such as membership in an accepting set), then HFSMs can be `compressed' by merging identical sub-HFSMs. As shown in \cite{alur1998model}, this allows HFSMs to be represented in logarithmic size, and model checking on such HFSMs can be done on this compressed representation. This task may, however, have significant complexity obstacles---finding the `most compressed' HFSM in this sense is likely NP-hard, via a reduction to the \emph{smallest grammar problem}~\cite{charikar2005smallest}.

%`Compressing' an HFSM consists of merging identically labelled subtrees in the nesting tree. By Theorem~\ref{basis theorem}, each basis module is associated with a contracted form FSM, so the modular decomposition could provide a starting point for compressing HFSMs by merging contracted forms. While we suspect (via a reduction to the \emph{smallest grammar problem}~\cite{charikar2005smallest}) that finding the `most compressed' HFSM is NP-hard, heuristic approaches may be useful in practice.

%\item (Compressing HFSMs) In this paper, we assumed all states in an FSM were distinct. If we have an equivalence relation on states (such as membership in an accepting set, like that on DFAs) then HFSMs can be `compressed' by merging identical nested HFSMs. 

\section{Finite State Machines}

% \subsection{Some preliminaries}
First, some preliminaries. We use the term \emph{graph} to mean a simple undirected graph, and \emph{digraphs} for directed graphs, where we allow parallel arcs and loops~\cite{bondy1976graph}. In digraphs, we write $\arc{u}{v}$ to indicate that there is an arc from the node $u$ to $v$. A \emph{directed path} (simply a \emph{path} when there is no ambiguity) is a sequence of nodes $\begin{tikzcd}v_1 \ar[r] & v_2 \ar[r] &\dots \ar[r] & v_n \end{tikzcd}$ in a digraph where each $v_i$ is distinct and there is an arc from each to the subsequent node in the sequence. We write $\pth{v_1}{v_n}$ to mean that there is a path from $v_1$ to $v_n$, in which case we say that $v_n$ is \emph{reachable} from $v_1$. If there is also an arc $\arc{v_n}{v_1}$, we call this a \emph{cycle}. 
In FSMs, the digraphs are \emph{labelled} by a set of symbols $X$, meaning each arc is assigned a symbol from $X$. We call an arc $\arc{u}{v}$ labelled by $x\in X$ an \emph{$x$-arc}, denoted by $\arc[x]{u}{v}$. Similarly, a path or cycle made up of $x$-arcs we call an $x$-path or $x$-cycle respectively, and denote an $x$-path from $u$ to $v$ by $\pth[x]{u}{v}$. A digraph with no cycles is called \emph{acyclic}, and we call it a \emph{directed acyclic graph}, which we contract to \emph{dag}. %Given a node $v$ in a dag we write $\uparrow v$ to the denote the \emph{ancestors of $v$}, which are those nodes from which $v$ is reachable, and $\downarrow\! v$ the \emph{descendants of $v$}, which are those nodes reachable from $v$. 
A \emph{tree} is a dag where there is a single node called the \emph{root} which has precisely one path to every other node. A digraph is \emph{strongly connected} if there exists a directed path between every pair of nodes. It is \emph{connected} if there exists an undirected path between every pair of nodes. We denote a partial function $f$ from $X$ to $Y$ by $f:X\partto Y$. Given $f:X\partto Y$ and $x\in X$ we write $f(x) = \emptyset$ to mean that $f$ is not defined on $x$, and adopt the convention that for any function $g$, partial or otherwise, $g(\emptyset) = \emptyset$. Given a set $X$, we write $X^*$ for the set of all finite sequences of elements of $X$.

\begin{defn}[Overlapping Sets] \label{overlapping sets}
A pair of sets $X$ and $Y$ is \emph{overlapping} if $X\cap Y \neq \emptyset$, $X\not\subseteq Y$, and $Y\not\subseteq X$. We say a collection of sets $X = \{X_1,\dots,X_n\}$ is \emph{overlapping} if for any $X_i,X_j\in X$ there exists a sequence $X_i,X_{a_1},X_{a_2},\dots X_j$ where each adjacent pair in the sequence is overlapping\footnote{If we view these sets as edges in a hypergraph, the collection being overlapping is equivalent to this hypergraph being connected.}%This can be understood using the `overlap graph' of the sets, whose whose nodes are the sets $\{X_1,X_2,\dots,X_n\}$, where two sets are adjacent if they overlap pairwise. A collection of sets is overlapping if the overlap graph is connected.}.
%if any partition of $X$ into two nonempty blocks $A=\{X_{a_1}, \dots, X_{a_m}\}$ and $B= \{X_{b_1},\dots, X_{b_k}\}$ with $A\cap B = \emptyset$ and $A\cup B = X$, there exist a $X_a\in A$ and $X_b\in B$ that are overlapping.
\end{defn}
 %The above definition is the natural extension of an overlapping pair of sets to a family of sets.
%To begin, we provide a definition of a Finite State Machine, and discuss how such objects operate.
\begin{defn}[Finite state machine, \cite{eilenberg1974automata}]\label{def:fsm}
 A \emph{finite state machine (FSM)} is a 4-tuple $(Q,\Sigma,\delta, s)$, where $Q$ is a finite set of \emph{states}; $\Sigma$ is a finite set called the \emph{alphabet}, whose elements we call \emph{symbols}; $\delta:Q\times \Sigma\partto Q$ is the \emph{transition function}, which can be a partial function%\footnote{An FSM is called \emph{complete} if the transition function is total. In this paper we allow but do not restrict ourselves to complete FSMs, because nesting a complete FSM inside another FSM has no effect.}
 ; and $s\in Q$ is a state we call the \emph{start state}\footnote{This definition of an FSM is very similar to that of a Deterministic Finite Automaton (DFA)~\cite{eilenberg1974automata}. The difference is that FSMs lack `accept' or `reject' states. Consequently, FSMs have a different notion of equivalence than DFAs, as we discuss in Section~\ref{sec:relatedwork}.}.
%When given an FSM $Z$, we will often write $Q(Z)$ to mean the state set of $Z$.
\end{defn}
FSMs are equivalently thought of as labelled digraphs~\cite{eilenberg1974automata}, where an arc $\arc[x]{u}{v}$ means that $\delta(u,x)=v$, where $x\in\Sigma$ and $u,v\in Q$. This is how FSMs are normally depicted, as in Figure~\ref{fig:Z}.
%That is, $\delta$ naturally defines a set $A\subseteq Q\times Q$ of arcs and a labelling
%The following alternative definition makes this clearer.
%\begin{defn}[Finite State Machine (FSM)]
%A \emph{finite state machine (FSM)} is a 5-tuple $(Q,A,\Sigma,\ell,s)$, where 
%\begin{itemize}
%    \item $Q$, $\Sigma$ and $s\in Q$ are defined as above;
%    \item $A \subseteq Q\times Q$ is a set of \emph{arcs};
%    \item $\ell:A \to \Sigma$ is an arc-labelling function, with the property that each arc out of a given node has a distinct label;
%\end{itemize}
%\end{defn}
 %Here, $Q$ and $A$ together have the data of a pseudodigraph. The equivalence between these definitions can be observed by noting that $A$ and $\ell$ together store precisely the data of $\delta$. 
 Consequently, we can apply graph-theoretic concepts to FSMs, such as connectedness and reachability. %. We will also refer to $Z$ as a digraph when we mean the underlying digraph defined by $\delta$,%FSM execution begins in the start state $s$ and follows transitions corresponding to an input sequence of symbols. At each point there is a single \emph{current state}
We will impose one further requirement on FSMs (and HFSMs), which is that they be \emph{accessible}~\cite{eilenberg1974automata}, meaning that all states are reachable from the start state. This will cause no loss of generality, because all FSMs are equivalent to an accessible one~\cite{eilenberg1974automata}, by ignoring the unreachable states. %The following is notation we will use for FSMs.
%Finally, we must establish what we mean when we say two FSMs are equivalent.
 We define FSM `execution' as follows.

%\begin{defn}
%Let $Z = (Q,A,\Sigma,\ell,s)$ be an FSM. We define the \emph{output function} $\varphi_Z$ as a partial function $\varphi_Z:\Sigma^*\partto Q^*$ which maps sequences of symbols to sequences of states, defined as follows. Let $x_1x_2\dots x_n\in\Sigma^*$ be a finite sequence of symbols. Then $\varphi_Z(x_1x_2\dots x_n) = q_0q_1\dots q_n$ where $q_0 = s$ and for each $i$ there is an arc $\arc[x_i]{q_{i-1}}{q_i}$. If for any $i$, $q_{i-1}$ has no arc labelled $x_i$ out of it, then $\varphi_Z(x_1x_2\dots x_n)$ does not exist.
%\end{defn}
\begin{defn}[Output Function of FSMs] \label{fsm output function}
Let $Z = (Q,\Sigma,\delta,s)$ be an FSM. The \emph{output function} of $Z$ is the partial function $\varphi_Z:\Sigma^*\partto Q$ which maps sequences of symbols to a state, defined by $\varphi_Z(x_1x_2\dots x_n) = \delta(\dots\delta(\delta(s,x_1),x_2),\dots,x_n)$.%\footnote{In functional programming terminology, this is a \emph{fold}, and we might write $\varphi_Z = \mathrm{fold}(\delta,s)$}. 
\end{defn}

The output function takes a sequence of symbols and returns the state reached after the associated sequence of transitions has been performed in the FSM, starting from the start state. If at any point there is no transition matching the next symbol, the output function returns $\emptyset$. Formally, if the $i$-th symbol $x_i$ has no matching transition (that is, $\delta(\dots\delta(\delta(s,x_1),x_2),\dots,x_i) = \emptyset$) then $\varphi_Z(x_1x_2\dots x_n) = \emptyset$, because $\delta(\emptyset,x)=\emptyset$. %This is why we allow $\varphi_Z$ to be a partial function. 
This is the standard way in which FSMs execution is defined, when $\delta$ is allowed to be partial. We now extend these definitions to HFSMs.

%\begin{defn}
%A \emph{hierarchical finite state machine (HFSM)} is a triple $(X,T,\mu)$, where $X$ is a set of FSMs, $T$ is a tree whose node set is $X$, and $\mu$ is a function which takes an FSM $f\in X$ and produces a partial function $\eta_f:Q(f)\partto X$ which takes an FSM in $X$ and maps its nodes to the successors of $f$ in $T$.
%\end{defn}

\begin{defn}[Hierarchical Finite State Machine]\label{def:HFSM}
A \emph{hierarchical finite state machine (HFSM)} is a pair $Z = (X,T)$, where $X=\{X_1,\dots,X_n\}$ is a set of FSMs with alphabet $\Sigma$ and $T$ is a tree that we call $Z$'s \emph{nesting tree}\footnote{A more general definition would allow subHFSMs to be reused, so this tree would become a dag, as in \cite{alur1998model}. This could allow for much smaller HFSM representations, but for simplicity we leave this for future work (Section~\ref{sec:conclusions}).}. The non-leaf nodes of $T$ are the FSMs $X_i$. Arcs out of $X_i$ in $T$ are labelled by the states $v\in Q(X_i)$, and each $v$ labels exactly one arc. If an FSM $X_j$ is nested at state $v$ in $X_i$, there is an arc $\arc[v]{X_i}{X_j}$ in $T$. Otherwise, there is an arc $\arc[v]{X_i}{v}$, where $v$ is a leaf of $T$. % The nodes of $T$ are labelled by FSMs in $X$, and the arcs out of a tree-node labelled by $X_i$ are labelled by nodes $v_i\in Q(X_i)$.
The \emph{states} of $Z$, written $Q(Z)$, are the states in $\bigcup_{X_i\in X} Q(X_i)$ which are leaves of $T$, that is, they do not contain a nested FSM. %The cardinality of $X$ we call the \emph{order of $Z$} which we write $\order(Z) := |X|$. \todo{this needs to be fixed}
\end{defn}

As is standard for HFSMs~\cite{harel1987statecharts,alur1998model}, we will depict them as FSMs where FSMs can be nested in the states of other FSMs, as in Figure~\ref{fig:Z nested}. %, and its nesting tree is shown in Figure~\ref{fig:nesting tree Y}. 
An HFSM $Z=(X,T)$ where $X=\{Y\}$ we call a \emph{flat} HFSM. Flat HFSMs have exactly the same data as FSMs in the normal sense (as defined in Definition~\ref{def:fsm}) and we will generally not distinguish between them.%If there is an arc $\arc[v]{Y}{X}$ in $T$ we say that $X$ is nested in $v$. Because both $X$ and $v$ are uniquely determined by each other, we call $v$ \emph{the} node which $X$ is nested in and $X$ \emph{the} FSM which is nested in $v$.

%Intuitively, an HFSM is a tree whose node set is $X$, and whose root is $R$. The successors of an FSM $f$ in this tree are the FSMs that are the image of a state of $f$ under $\eta$ (we will draw this by labelling an arc from $f$ to $g$ by the unique state $v$ of $f$ such that $\eta(v) = g$, if this exists). The fact that $\eta$ is injective and that $R$ is not in its range ensures that this has a tree structure. $\eta$ is never a total function because $X \setminus\{R\}$ is always smaller than $\bigcup_{f\in X} Q(f)$. We assume for simplicity that the node sets of all the FSMs in $X$ are disjoint.
%\begin{rmk}
%This definition allows us to define HFSMs inductively, making use of their tree structure. If $Y$ is the set of FSMs in a subtree of an HFSM $(X,R,\eta)$, and that subtree has root $y$, then $(Y,y,\eta|_Y)$ is also an HFSM.
%\end{rmk}

\begin{defn} \label{def:nested start state}
Let $Z = (X,T)$ be an HFSM. We define the function $\start: X \to Q(Z)$, by
\[
\start(X_i) = \begin{cases}\start(X_j), & \text{there is an arc}\ \arc[s_i]{X_i}{X_j}\ \text{in $T$}\\
s_i, & \text{otherwise}
\end{cases}
\] where $s_i$ is the start state of the FSM $X_i$.
\end{defn}

%\begin{defn}
%Let $Z = (X,R,\eta)$ be an HFSM. The cardinality of $X$ we call the \emph{order} of $Z$ (written $\order(Z)$), and the height of this tree we call the \emph{height} of $Z$. $Z$ is \emph{flat} if the order of $Z$ is 1 (equivalently if the height is 0). Finally, define $\start: \bigcup_{f\in X} Q(f) \to \bigcup_{f\in X} Q(f)$ recursively by 
%\[
%\start(v) = \begin{cases}
%v, & \text{$\eta(v)$ is undefined} \\
%\start(s(\eta(v))), & \text{otherwise}
%\end{cases}
%\]
%\end{defn}
The `start' function identifies the `start state' of the HFSM, found by following the unique path in $T$ where all arcs are labelled by start states in their respective FSMs.
This recursive definition works because HFSMs have a finite tree structure. 
\begin{defn}[Hierarchical Transition Function] \label{def hierarchical transition}
Let $Z=(X,T)$ be an HFSM, $q\in Q(Z)$ a state and $x\in\Sigma$ a symbol. Suppose $q\in Q(X_j)$, $X_j\in X$. The \emph{hierarchical transition function} of $Z$ is defined recursively as
\[
\psi(q,x) = \begin{cases}
\start(Y), &\delta_j(q,x) = v,\arc[v]{X_j}{Y}\in T,\ Y\in X \\ %\text{ $Y$ is nested in $v$}\\
\psi(w,x), &\delta_j(q,x) =\emptyset, \arc[w]{W}{X_j}\in T,\ W\in X \\%\text{ $X_j$ is nested in $w$} \\
\delta_j(q,x), &\text{ otherwise}\\
\end{cases}
\]
\end{defn}
%\begin{defn}
%Let $q$ be a node in an HFSM $Z = (X, T)$, where $q\in Q(P_1)$, with $P_1\in X$. Let $P_1,P_2,\dots,P_k$ be the sequence of nodes of $T$ on the unique path from $P_1$ to the root of $T$. We write $V_i$ for the unique node of $P_{i+1}$ labelling the arc to $P_i$, and define $V_0 := q$. Given $x$ in $\Sigma$, we let 
%\[j = \arg\min_{0\leq i\leq k} \{V_i|\ V_i\ \text{has an $x$-arc in $P_{i+1}$}\}\] and let $v = \delta_{P_{j+1}}(P_{j},x)$. Finally, we define the \emph{hierarchical transition function of $Z$} as the partial function $\psi: Q_Z\times\Sigma \partto Q_Z$,
%\[
%\psi (q, x) = \begin{cases} \start(Y), &\text{there is an arc}\ \arc[v]{P_{j+1}}{Y}\\
%v, &\text{otherwise}
%\end{cases}
%\]
%In general, $\{V_i|\ V_i\ \text{has an $x$-arc in $P_{i+1}$}\}$ may be empty, and $j$ and $v$ do not exist. In this case, $\psi$ is undefined. Figure~\ref{fig:nested} visualises the definition of $\psi$. Note that if $Z$ is flat then this is the same as the transition function of its sole FSM.
%\end{defn} 

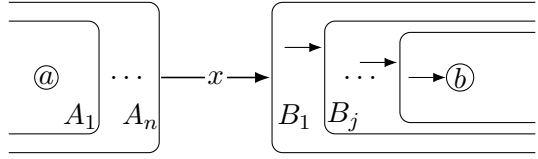
\begin{figure}
    \centering
    \begin{tikzpicture}
\node (a) [state] at (-2.5,2) {$a$};
\node (A1) at (-2.05,1.48) {$A_1$};
\node (An) at (-1.25,1.48) {$A_n$};
\node (B1) at (0.8,1.48) {$B_1$};
\node (Bj) at (1.45,1.48) {$B_j$};
\node (b) [state,init] at (3,2) {$b$};
\node [init] (inouter) at (1.3,2.4) {};
\node [init] (ininner) at (2.3,2.2) {};
\node (rc) at (0.61,2) {};
\node (lc) at (-1.11,2) {};
\node at (-1.4,2) {\dots};
\node at (1.7,2) {\dots};
%\node[draw,minimum height=2cm,minimum width=2cm,rounded corners] (ll) at (-2,2) {};
%\node[draw,minimum height=1.5cm,minimum width=1.2cm,rounded corners] (l) at (-2.4,2) {};
%\node[draw,minimum height=1.2cm,minimum width=2cm,rounded corners] (lll) at (3.2,2) {};
% 1, 2, 3, 4, 5, 6, 7, 8, 9
% \node[draw,minimum height=2cm,minimum width=3.7cm,rounded corners] (rr) at (2.4,2) {};
%\node[draw,minimum height=1.5cm,minimum width=3cm,rounded corners,initial left, initial text={}] (r) at (2.7,2) {};
\draw[edge] (lc) to node[edgelabel] {$x$} (rc);

\draw[rounded corners] (4,3) -- (0.5,3) -- (0.5,1) -- (4,1);
\draw[rounded corners] (4,2.75) -- (1.2,2.75) -- (1.2,1.25) -- (4,1.25);
\draw[rounded corners] (4,2.6) -- (2.2,2.6) -- (2.2,1.4) -- (4,1.4);
\draw[rounded corners] (-3,3) -- (-1,3) -- (-1,1) -- (-3,1);
\draw[rounded corners] (-3,2.75) -- (-1.8,2.75) -- (-1.8,1.25) -- (-3,1.25);
 \end{tikzpicture}
    \caption{A transition $\psi(a,x) = b$ in an HFSM, defined by the hierarchical transition function (Definition~\ref{def hierarchical transition}). Here, $a$ has no $x$-arc, and nor do any of the containing FSMs $A_1,\dots,A_{n-1}$. An $x$-arc exists out of $A_n$, which transitions to the superstate $B_1$. The state $b$ is $\start(B_1)$ (Definition~\ref{def:nested start state}), the nested start state of $B_1$.}
    \label{fig:nested}
\end{figure}

This definition formalises the functioning of HFSMs in software engineering~\cite{harel1987statecharts}. There, the symbols in the input sequence $x_1\dots x_n$ are thought of as `events' to which the machine responds, arriving one-by-one, with the `current state' $q$ of the machine stored in-between. A symbol $x_i$ has been `processed' once the appropriate transition has been performed, updating the current state to $q' \in Q(Z)$. In the hierarchical case, this is done in a depth-first manner. As an example, suppose the current state $q$ is contained in an FSM $X_j \in X$, and $X_j$ is nested as a state $w$ in an FSM $W\in X$. When the `input event' $x_i$ is received, the FSM checks first if a transition $\delta_{X_j}(q,x_i)$ exists in $X_j$; if it does, this transition is performed, and the current state is updated to $\delta_{X_j}(q,x_i)$. If we perform a transition $\arc[x]{q}{q'}$, and the state $q'$ contains a nested FSM, then the current state becomes the start state of this FSM, unless that too is nested; this process is repeated until a state is reached that does not contain a nested FSM (following Definition~\ref{def:nested start state}). \emph{Otherwise} if the transition $\delta_{X_j}(q,x_i)$ does not exists in $X_j$, the program checks whether a transition out of $w$ on symbol $x_i$ exists in the `parent' FSM $W$ (that is, if $\delta_W(w,x_i)$ exists); if so, this transition is performed, with the current state updated to $\delta_W(w,x_i)$. This process is repeated for each parent until either a transition is performed, or the state has no parent, in which case $\psi(q,x_i) = \emptyset$. This is visualised in Figure~\ref{fig:nested}. Conceptually, each FSM attempts to `handle' events that its nested FSMs have not handled. %When $Z$ is flat, the hierarchical transition function reduces to the transition function of its sole FSM. %We now have sufficient setup to define the output function of an HFSM, analogously to Definition~\ref{fsm output function}.
\begin{defn}[Output Function of HFSMs]
Let $Z = (X,T)$ be an HFSM. Let $R\in X$ be the FSM labelling the root of $T$. Then the \emph{output function} of $Z$ is the partial function $\varphi_Z:\Sigma^*\partto Q_Z$ where $\varphi_Z(x_1x_2\dots x_n) = \psi(\dots\psi(\psi(\psi(\start(R),x_1),x_2),\dots,x_n)$.\footnote{Again recall that $\varphi_Z(x_1x_2\dots x_n) = \emptyset$ if $\psi$ is not defined on some input in this expression.}
\end{defn}

%\begin{defn}
%Let $Z = (X,R,\eta)$ be an HFSM. Define $\hat Q := \bigcup_{f\in X} Q(f)$. For each $q\in\hat Q$, we write $\delta_q$ for the transition function of the FSM containing $q$. We define the \emph{hierarchical transition function} $\psi:\hat Q\times\Sigma \partto \hat Q$ recursively by \[
%\psi(q,x) = \begin{cases}
%\psi(s(\eta(q)),x), & \text{if defined and $\eta(q)$ is defined} \\
%\start(\eta(\delta_q(q,x))), & \text{otherwise}
%\end{cases}
%\]
%Finally, we define the \emph{output function} $\varphi_Z$ as a partial function $\varphi_Z:\Sigma^*\partto \hat Q$ defined by $\varphi_Z(x_1x_2\dots x_n) = \psi(\dots\psi(\psi(\psi(\start(s(R)),x_1),x_2),\dots,x_n)$. If $\psi$ is undefined for any $i$ in this expression, then $\varphi_Z(x_1x_2\dots x_n)$ is undefined.
%end{defn}

\begin{defn}[(H)FSM Equivalence] \label{def: equivalence}
Two (H)FSMs $Y$ and $Z$ are \emph{equivalent}, written $Y\cong Z$, if their output functions $\psi_Y$ and $\psi_Z$ are equal.%\footnote{All the theory up until this point applies to both DFAs and FSMs, because DFAs are FSMs with the addition of a set $F\subseteq Q$ of \emph{accept} states~\cite{eilenberg1974automata}. However, DFAs have a different notion of equivalence. DFAs $Y$ and $Z$ are equivalent if given any input sequence $x_1\dots x_n\in\Sigma^*$, $\varphi_Y(x_1\dots x_n)\in F_Y$ iff $\varphi_Z(x_1\dots x_n)\in F_Z$, so $Y$ and $Z$ accept the same \emph{language}. As a result, every FSM module is a DFA module but not conversely.}
\end{defn}

It is straightforward to show that two FSMs are equivalent if and only if they are equal (have the same states, transition functions and start states). This is not true of HFSMs. However, by repeatedly expanding (Definition~\ref{defn:expansion}) nested FSMs we can convert any HFSM $Z$ into a unique equivalent (by Definition~\ref{def: equivalence}) flat HFSM, which we call $Z^F$~\cite{alur1998model}. It follows that two HFSMs $Y$ and $Z$ are equivalent if and only if $Y^F$ and $Z^F$ are equal. %The proof of this theorem as well as other proofs for the rest of the results in this section can be found in \ref{app_proof:modules}. 
\begin{thm} \label{flat equivalence} 
If $Z$ and $Y$ are accessible HFSMs, $Z^F$ and $Y^F$ are unique and $Z\cong Y$ if and only if $Z^F = Y^F$.
\end{thm}

\section{Modules in Graphs and FSMs} \label{sec:modules}

\begin{defn}[$x$-Exit and $x$-Entrance] \label{def:entrance and exit}
Let $Z$ be an FSM, $M \subseteq Q(Z)$ a set of states and $x\in\Sigma$ a symbol. We call a state $v\not\in M$ an \emph{$x$-exit} of $M$ if there is an arc $\arc[x]{u}{v}$, with $u\in M$. We call a state $u\in M$ an \emph{$x$-entrance} of $M$ if there is an arc $\arc[x]{v}{u}$ with $v\not\in M$, \emph{or} $v$ is the start state of $Z$. By an \emph{entrance} or \emph{exit}, we mean an $x$-entrance or $x$-exit for any $x$.
\end{defn}
\begin{defn}[FSM Modules] \label{concrete modules}
Let $Z$ be an FSM. A non-empty subset $M\subseteq Q(Z)$ is a \emph{module} if and only if it has one entrance, and for each $x\in \Sigma$, if $M$ has an $x$-exit then (1) that $x$-exit is unique and (2) every state in $M$ has an $x$-arc.
\end{defn}
\begin{defn}[Graph Modules,~{\cite{mcconnell1999modular,habib2010survey}}] \label{concrete graphs}
Let $G$ be a graph. A non-empty\footnote{We exclude the trivial empty module for simplicity.} subset $M\subseteq N(G)$ is a \emph{module} if and only if for every $v\not\in M$, $v$ is adjacent to all of $M$ or none of $M$.
\end{defn}
For hierarchical FSMs, modules correspond to modules in the constituent FSMs.
\begin{defn}[HFSM Modules] \label{HFSM module}
Let $Z=(X,T)$ be an HFSM. A set $M\subseteq Q(Z)$ is a \emph{module} if there is an FSM $X_i\in X$ and module $H\subseteq Q(X_i)$ where $M$ is the set of HFSM states recursively nested within states in $H$.
\end{defn}
The definition of an FSM module (Definition~\ref{concrete modules}) bears little resemblance to that of a graph module (Definition~\ref{concrete graphs}). The connection between them only becomes apparent by considering the `independence' property of graph modules. Specifically, given a graph $G$ and module $M$, the original $G$ can be recovered uniquely from the \emph{contraction} $G/M$ (Definition~\ref{contraction}) and the \emph{restriction} $G[M]$ (Definition~\ref{restriction}), which are each well-defined graphs.

%is the correct definition of a module in an FSM. In particular, it is not obviously related to modules in graphs . However, in graphs, modules are sets of nodes which satisfy an `independence' property,  
We view this as the defining property of modules, and we arrive at Definition~\ref{concrete modules} by defining contraction and restriction of FSMs. Theorem~\ref{module theorem} establishes that graph and FSM modules are characterised by these operations. %To begin, we define $G[M]$ and $G/M$ for both graphs and FSMs.

%ModulesConcretely, in the graph case a subset $M$ is a module of a graph $G$ if the graph can be recovered from the quotient $G/M$ (which we call the \emph{outer structure}) and the induced subgraph $G[M]$ (the \emph{inner structure}). We will define both of these formally for graphs and FSMs and then define modules.

\begin{defn}[Contraction] \label{contraction}
Let $Z$ be an FSM/graph and $M$ a set of its states/nodes. The \emph{contraction} of $M$ in $Z$, written $Z/M$, is the FSM/graph whose state/node set is $\{\overbrace{v_1,v_2,\dots}^{Z\setminus M},M\}$. For FSMs, given any $v_i,v_j\not\in M$ and $x\in\Sigma$, there is a transition $\delta_{Z/M}(v_i,x)=v_j$ if and only if $\delta_{Z}(v_i,x)=v_j$. There is a transition $\delta_{Z/M}(M,x)=v_j$ if and only if $\delta_{Z}(u,x)=v_j$ for some $u$ in $M$, and there is a transition $\delta_{Z/M}(v_i,x)=M$ if and only if $\delta_{Z}(v_i,x)=u$ for some $u$ in $M$. For graphs, there is an edge $(v_i,v_j)$ in $G/M$ if and only if $(v_i,v_j)$ is an edge in $G$ and there is an edge $(M,v_i)$ if and only if there is some $u\in M$ where $(u,v_i)$ is an edge in $G$.
\end{defn}

Contractions in graphs is defined by the graph quotient, and the FSM definition is the analogue on the underlying digraph of the FSM. The astute reader will notice that the FSM definition is not always well-defined---there may be multiple $x$-transitions out of $M$, so $\delta_{Z/M}$ may not be a function. This motivates the requirement in Definition~\ref{concrete modules} that the $x$-exit of a module is unique, because then $Z/M$ is a well-defined FSM. Given multiple disjoint sets $M_1,M_2,\dots$, we can perform multiple contractions at once, which we write $Z/M_1,M_2,\dots$. Note that the order does not matter.

%\begin{defn} [Quotient Graph]
%Let $G$ be a graph and $P$ a partition of its node set. The \emph{quotient graph} $G/P$ is the graph whose node set is $P$ and where there is an edge $(p_i,p_j)$ iff $p_i\neq p_j$ and there is an edge $(u,v)$ in $E(G)$ with $u\in p_i$ and $v\in p_j$.\todo{cite this}
%\end{defn}

%This is the usual quotient of simple graphs, and the definition for FSMs is the graph quotient on the underlying labelled pseudodigraph.

%\begin{defn}[Quotient FSM]
%Let $Z=(Q,\Sigma,\delta,s)$ be an FSM and $P$ a partition of $Q$. The \emph{quotient FSM} $Z/P$ is the FSM $(Q/P,\Sigma,\delta',[s]_P)$, where 
%$\delta'([q]_P,a)=[r]_P\ \text{iff}\ \delta(q,a)=r$, and $[q]_P$ denotes the block of the partition $P$ which contains $q$. 
%\end{defn}

\begin{defn}[Restriction] \label{restriction}
If $Z$ is an FSM/graph and $M$ is a subset of its states/nodes, then the \emph{restriction} of $Z$ to $M$, written $Z[M]$, is the FSM/graph defined on the subgraph induced by $M$ in $Z$. For FSMs, the start state of $Z[M]$ is any entrance of $M$ (Definition~\ref{def:entrance and exit}).
\end{defn}
Restriction is not well-defined for abitrary sets $M$, because if $M$ has multiple entrances the start state of $Z[M]$ isn't uniquely determined. However, modules require that the entrance be unique (Definition~\ref{concrete modules}) which ensures that $Z[M]$ is always a well-defined FSM. In fact, we will refer to the unique entrance of a module $M$ as its \emph{start state}, where we overload the term because this state is exactly the start state of the restriction $Z[M]$. Finally, we define the \emph{expansion} of nested FSMs or graphs. On graphs this operation has previously been called \emph{X-join} or \emph{substitution}~\cite{mcconnell1999modular}. We demonstrate FSM expansion (Definition~\ref{defn:expansion}) in Figure~\ref{fig:expansion}.
%Again, for an FSM this induced subgraph may not be fully defined, because it lacks a start node. For simplicity we define the start node of $G[M]$ as either the start node of $G$, if that is in $M$, or otherwise as any node with a maximal number (in $M$) of in-arcs from $G\setminus M$.
%\begin{rmk}
%Both contraction and restriction are defined identically for graphs as they are for FSMs in Definition~\ref{contraction} and~\ref{restriction}, ignoring start states and using the graph definition of quotient.
%\end{rmk}

\begin{defn}[Graph Expansion,~\cite{mcconnell1999modular}]
Let $G$ and $H$ be graphs, with $v\in G$. The \emph{expansion of $H$ at $v$}, written $G\cdot_v H$, is the graph whose node set is $(N(G)\setminus \{v\})\cup N(H)$ where edges within $G$ or $H$ are unchanged and there is an edge $(g,h)$ between $g\in G$ and $h\in H$ if and only if there is an edge $(g,v)$ in $G$.
\end{defn}
\begin{defn}[FSM expansion] \label{defn:expansion}
Let $G$ and $H$ be FSMs, and let $v$ be a state of $G$. Then, the \emph{expansion of $H$ at $v$}, written $G\cdot_v H$ is the FSM $\left((Q(G)\setminus \{v\})\cup Q(H), \Sigma(G)\cup \Sigma(H), \delta',\begin{cases} s_H, & v = s_G \\ s_G, &v \neq s_G\end{cases}\right)
$ where \[
\delta'(q,a) = \begin{cases}\delta_G(q,a), & q\in Q(G) \land \delta_G(q,a)\neq v \\ s_H, &q\in Q(G)\land \delta_G(q,a)=v \\ \delta_H(q,a), &q\in Q(H) \land \delta_H(q,a)\neq \emptyset\\ \delta_G(v,a), &q\in Q(H) \land \delta_H(q,a)= \emptyset
\end{cases}
\]
\end{defn}
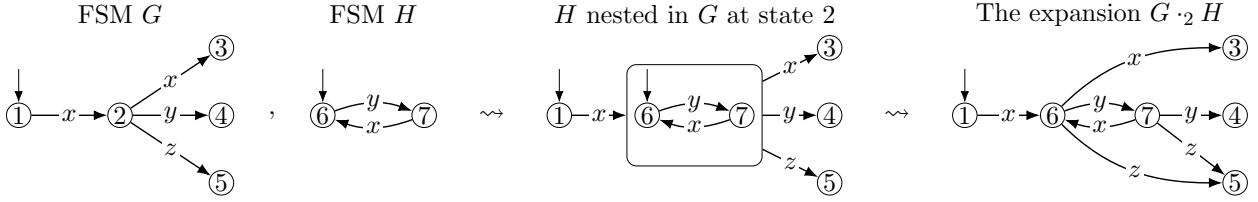
\begin{figure}
    \centering
    \begin{tikzpicture}
\node (G) at (-2.5,1.5) {FSM $G$};
\node (H) at (1.25,1.5) {FSM $H$};
\node (H) at (12,1.5) {The expansion $G\cdot_2 H$};

% G
\node (1) [state,initial above, initial text={}] at (-4,0) {1};
\node (2) [state] at (-2.5,0) {2};
\node (3) [state] at (-1,1) {3};
\node (4) [state] at (-1,0) {4};
\node (5) [state] at (-1,-1) {5};
\draw[edge] (1) to node[edgelabel] {$x$}(2);
\draw[edge] (2) to node[edgelabel] {$x$}(3);
\draw[edge] (2) to node[edgelabel] {$y$}(4);
\draw[edge] (2) to node[edgelabel] {$z$}(5);

\node (comma) at (-0.25,0) {,};

% H
\node (6) [state,initial above, initial text={}] at (0.5,0) {6};
\node (7) [state] at (2,0) {7};
\draw[edge] (6) to[out=20,in=160] node[edgelabel] {$y$}(7);
\draw[edge] (7) to[in=-20,out=-160] node[edgelabel] {$x$}(6);

\node at (3,0) {$\leadsto$};

\node at (6,1.5) {$H$ nested in $G$ at state 2};

% H in G
\node (13) [state,initial above, initial text={}] at (4,0) {1};
\node (8) [state] at (8,1) {3};
\node (9) [state] at (8,0) {4};
\node (10) [state] at (8,-1) {5};
\node[draw,minimum height=1.5cm,minimum width=2cm,rounded corners] (67) at (6,0) {};
\node (11) [state,initial above, initial text={}] at (5.3,0) {6};
\node (12) [state] at (6.7,0) {7};
\draw[edge] (11) to[out=20,in=160] node[edgelabel] {$y$}(12);
\draw[edge] (12) to[in=-20,out=-160] node[edgelabel] {$x$}(11);
\draw[edge] (13) to node[edgelabel] {$x$}(67);
\draw[edge] (67) to node[edgelabel] {$x$}(8);
\draw[edge] (67) to node[edgelabel] {$y$}(9);
\draw[edge] (67) to node[edgelabel] {$z$}(10);

\node at (9,0) {$\leadsto$};

% G.H
\node (14) [state,initial above, initial text={}] at (10,0) {1};
\node (15) [state] at (14,1) {3};
\node (16) [state] at (14,0) {4};
\node (17) [state] at (14,-1) {5};
\node (18) [state] at (11.3,0) {6};
\node (19) [state] at (12.7,0) {7};
\draw[edge] (18) to[out=20,in=160] node[edgelabel] {$y$}(19);
\draw[edge] (19) to[in=-20,out=-160] node[edgelabel] {$x$}(18);
\draw[edge] (14) to node[edgelabel] {$x$}(18);
\draw[edge] (18) to[in=180,out=45] node[edgelabel] {$x$}(15);
\draw[edge] (19) to node[edgelabel] {$y$}(16);
\draw[edge] (18) to[out=-45,in=180] node[edgelabel] {$z$}(17);
\draw[edge] (19) to node[edgelabel] {$z$}(17);

 \end{tikzpicture}
    \caption{An example of FSM expansion (Definition~\ref{defn:expansion})}
    \label{fig:expansion}
\end{figure}

%We also point out the following theorem, which highlights some inherent non-reversibility in the expansion operation.
%\begin{thm}
%Let $Z$ and $Y$ be FSMs, and $v$ a node in $Z$. If $v$ has no self-loops, and for each $x$-arc out of $v$, $Y$ is not $x$-sealed, then $(Z\cdot_v Y)[N(Y)] \cong Y$ and $(Z\cdot_v Y)/N(Y)\cong Z$.
%\end{thm}
Having defined expansion, restriction, contraction and equivalence on FSMs, we can justify Definition~\ref{concrete modules} by showing that modules in graphs and FSMs are characterised by these operations. This provides an alternative abstract definition of modules which is similar to a universal property in algebra, in that it characterises the \emph{role} of a module with respect to the operations of contraction, restriction and expansion, independent of the definitions of these operations for each class of objects.

\begin{thm} \label{module theorem}
 Let $Z$ be a graph/FSM. A non-empty set $M\subseteq Q(Z)$ is a \emph{module} if and only if $Z/M\cdot_M Z[M] \cong Z$.
\end{thm}
 % and by the definition a module is a set that can be `equivalently' considered as an induced FSM. It will usually be clear from context whether we are discussing an FSM's start node or a module's start node.
Modules in FSMs share important properties with modules of graphs. First, in any FSM the singleton sets and the whole state set $Q$ are modules, which in graphs are called the \emph{trivial modules} (Lemma~\ref{trivial modules}). An FSM with only trivial modules we call \emph{prime}, following the convention for graphs~\cite{mcconnell1999modular}. Second, we can reason hierarchically about modules because `modules of modules are modules' (Theorem~\ref{gallai}). This is the FSM analogue of an important theorem on graph modules~\cite{gallai1967transitiv}.

\begin{lem} \label{trivial modules}
For any FSM $Z$, $Q(Z)$ is a module and $\{v\}$ is a module for any state $v\in Q(Z)$.
\end{lem}

%\begin{defn}[Trivial Modules and Prime FSMs]
%The modules characterised in Lemma~\ref{trivial modules} are called \emph{trivial modules}. 
%\end{defn}
%For graphs, modules are generalisations of connected components. That is also true for FSMs.

%\begin{lem} \label{components are modules}
%Every connected component (in the graph sense) of an FSM is a module. Every collection of components is a module.
%\end{lem}

%In general, there are families of graphs and FSMs which have exponentially many modules\footnote{This is easy to see from Lemma~\ref{components are modules}, though this is not the exclusive case, because there are also families of single-component graphs and FSMs with exponentially many modules.} in the number of states in the FSM. In order to reason algorithmically about FSM modules we would like a compact representation of them, and this is what we derive (for \emph{thin} modules) in the next section. We will need the following important theorem.

\begin{thm} \label{gallai}
Let $Z$ be an FSM, $X$ a module, and $Y\subseteq X$ a subset of $X$. Then $Y$ is a module of $Z$ if and only if it is a module of $Z[X]$.
\end{thm}
%\begin{rmk}
%Lemma~\ref{trivial modules} and Theorem~\ref{gallai} were first proved for graphs in~\cite{gallai1967transitiv}.
%\end{rmk}

% analogue of a famous theorem

\subsection{Thin modules} \label{sec:thin modules}

\begin{figure}
    \centering
    \begin{tikzpicture}
\node (1) [state,init] at (1,3) {1};
\node (2) [state] at (4,3) {2};
\node (5) [state] at (1,1) {5};
\node (3) [state] at (7,3) {3};
\node (4) [state] at (10,3) {4};
\node (6) [state] at (10,1) {6};
\node[draw,opacity=0.2,shape=rectangle,minimum width=7cm,minimum height=1.7cm,rounded corners] (a) at (4,3) {};
\node[draw,opacity=0.2,shape=rectangle,minimum width=7cm,minimum height=1.9cm,rounded corners] (a) at (7,3) {};
% 1 ---x--> 2
\draw[edge] (1) to[out=30, in=150] node[edgelabel] {$y$} (2);
% 1 ---y--> 5
\draw[edge] (1) to node[edgelabel] {$x$} (5);
% 2 ---x--> 1
\draw[edge] (2) to[out=-150, in =-30] node[edgelabel] {$y$} (1);
% 2 ---y--> 3
\draw[edge] (2) to[out=30, in=150] node[edgelabel] {$x$} (3);
% 3 ---y--> 2
\draw[edge] (3) to[out=-150, in =-30] node[edgelabel] {$x$} (2);
% 3 ---x--> 4
\draw[edge] (3) to node[edgelabel] {$y$} (4);
% 4 ---x--> 1
\draw[edge] (4) to[out=-140,in=-40] node[edgelabel] {$y$} (1);
% 4 ---y--> 6
\draw[edge] (4) to node[edgelabel] {$x$} (6);
 \end{tikzpicture}
    \caption{This FSM has two non-trivial modules, $\{1,2,3\}$ and $\{2,3,4\}$ (represented by grey rectangles), neither of which are thin; they both contain an $x$-cycle and have an $x$-exit. These modules overlap, but neither their union $\{1,2,3,4\}$ nor their intersection $\{2,3\}$ are modules.}
    \label{fig:fsm_counter}
\end{figure}
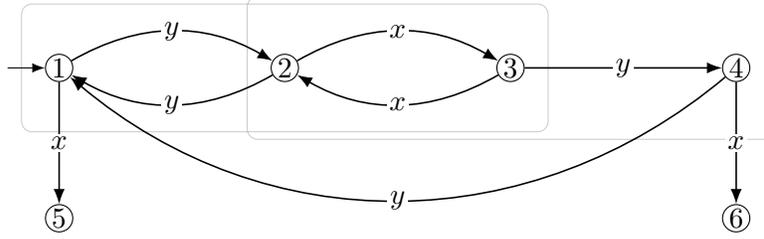

However, as we discussed in the introduction, FSM modules are not as algebraically well-behaved as we would hope. Figure~\ref{fig:fsm_counter} shows how, unlike graph modules, ovelapping FSM modules need not be closed under intersection and union. Similarly, while `modules of modules are modules' (Theorem~\ref{gallai}), the same doesn't hold for contractions---if $X$ and $Y$ are modules of $Z$ with $X\subseteq Y$, $Y/X$ may not be a module of $Z/X$. Graph modules, on the other hand, do possess this property. Fortunately, a simple restriction of the definition of a module---called a \emph{thin module}---suffices to obtain all of these critical algebraic properties (Theorem~\ref{reverse gt} and Lemma~\ref{union and intersection}).
% This seems, at first, to be an insurmountable obstacle for a modular decomposition theory for HFSMs. But fear not!
%We now introduce a stricter notion of a module, called a \emph{thin module}, which satisfies the above property (Theorem~\ref{reverse gt}) and is closed under overlapping unions and intersections (Lemma~\ref{union and intersection}).
% I should be more precise about this

\begin{defn}[Thin Modules of FSMs] \label{def: thin module}
%A module $M$ of an FSM $Z$ is \emph{thin} if for every $x\in \Sigma$, either $M$ has no $x$-exit or $M$ contains no $x$-cycles.
Let $Z$ be an FSM. A non-empty subset $M\subseteq Q(Z)$ is a \emph{thin module} if and only if it has one entrance, and for each $x\in \Sigma$, if $M$ has an $x$-exit $v$, then every state in $M$ has an $x$-path to $v$ with all states other than $v$ inside $M$.
\end{defn}
Definition~\ref{def: thin module} differs from Definition~\ref{concrete modules} by requiring that, for each $x\in\Sigma$ with an $x$-exit $v$, \emph{not only} must each state have an $x$-arc, but also an $x$-path which leads to $v$ in $M$. Note that uniqueness of the $x$-exit is implied by the requirement of it being reachable from each state in $M$. Thinness extends naturally to HFSMs.
\begin{defn}\label{thin hfsms}
An HFSM $Z = (X,T)$ is \emph{thin} if $Q(X_i)$ is a thin module of $Z^F$ for each $X_i$.
\end{defn}
The definition of thinness can be opaque at first, so it is worth examining more deeply. %By contrast non-thin modules require the existence of both an $x$-cycle and an $x$-exit for some symbol $x$, which is what allows for constructions like Figure~\ref{fig:fsm_counter}. 
Definition~\ref{concrete modules} require that all states in a module behave the same under any input symbol $x$. Definition~\ref{def: thin module} extends this by requiring that all states in a \emph{thin} module behave the same under \emph{repeated application} of any symbol $x$. That is, either repeated $x$-inputs cause us to stay in the module forever, or they cause an eventual transition to the $x$-exit, but the outcome does not depend on the starting state within the module. Note that thinness is not too restrictive a property---for instance, all trivial modules are thin.

%Conceptually, thinness ensures that \emph{$x$-transitions in an HFSM remain consistent with reachability}. In an FSM, an $x$-arc from $v$ to $w$ means that a single $x$-event moves the current state from $v$ to $w$. In an HFSM, an $x$-event could instead cause a transition inside an FSM nested at $v$. However, thinness ensures that all states in a module behave the same way under \emph{repeated} $x$-inputs: 
%In general, no amount of $x$-events can ensure the transition to $w$ actually takes place. However, if the FSM nested at $v$ is thin (as a module of the equivalent flat FSM), we can guarantee that \emph{some finite number} of repeated $x$-events causes the current state to transition to $w$.

When viewed on a concrete example, thinness has a somewhat intuitive UI interpretation. In Section~\ref{sec: applications} we apply our results to an HFSM model of a wristwatch, originally from Harel~\cite{harel1987statecharts}, which we present in Figure~\ref{fig:harel}. A surprising feature of this well-studied HFSM is that it is thin, demonstrating that thinness can arise naturally as a design feature of HFSMs. For a user, thinness allows for navigation: repeatedly pressing any one button either cycles within a subsystem ($d$ in the `chime' subsystem), or eventually leaves a subsystem ($c$ in the `update' subsystem), but never both.

Fortunately, thin modules allow us to obtain much stronger algebraic properties. From here, we will restrict our attention to thin HFSMs.
\begin{lem} \label{union and intersection}
If $A$ and $B$ are overlapping thin modules, then $A\cup B$ and $A\cap B$ are both thin modules.
\end{lem}
\begin{thm} \label{reverse gt}
Let $Z$ be an FSM and $X$ a module, and $Y$ a superset of $X$. If $X$ is thin, then $Y$ is a thin module of $Z$ if and only if $Y/X$ is a thin module of $Z/X$.
\end{thm}

\section{The Modular Decomposition of FSMs} \label{sec:modular decomp}

Because thin modules can overlap, there can be exponentially many, and so an FSM may have many decompositions into thin HFSMs, depending on choices of modules. In this section we define a structure called the \emph{modular decomposition}, and prove the main theorem of our paper, which shows that the modular decomposition represents all thin modules, and hence allows us to efficiently construct and search the space of equivalent (thin) HFSMs.
%graph theory, the main result of modular decomposition theory is a tree which provides an $O(n)$-size representation of the set of all possible modules~\cite{gallai1967transitiv}. In this section  %From here we will assume all FSMs we mention are accessible, because as established every FSM is equivalent to a unique accessible one, where unreachable nodes are ignored. 
%This section is devoted to investigating the properties of the decomposition tree, and we give an efficient algorithm for computing it in Section~\ref{sec:computing}.
Our representation is built on overlapping unions of modules, using Lemma~\ref{union and intersection}. Specifically, we identify the thin modules which are \emph{indecomposable} under overlapping unions, and use these to generate all other thin modules.% The modular decomposition is then the transitive reduction of the inclusion order on these modules.
\begin{defn}[Decomposable and Indecomposable Sets]\label{def:dec_indec_modules}
Let $F$ be a family of subsets of a finite set $X$. An element $M\in F$ is \emph{decomposable} if there exists an overlapping (Def.~\ref{overlapping sets}) collection $D_1,\dots,D_n$ ($n\geq 2$) of sets in $F$ such that $M=D_1\cup\dots\cup D_n$. Otherwise, $M$ is \emph{indecomposable}. %We denote the indecomposable elements of $F$ by $\Ind(F)$.
\end{defn}
\begin{defn}[Modular Decomposition of an HFSM]\label{def:FSM_decom_tree}
Let $Z$ be an HFSM. Let $D$ be the dag whose nodes are the indecomposable thin modules of $Z$, with an arc from modules $K$ to $M$ in $D$ if and only if $M\subseteq K$. Then the \emph{modular decomposition} %\footnote{This dag is a tree in the digraph sense iff all indecomposable modules are \emph{strong}~\cite{habib2010survey}. See the Appendix.}
$\mathcal{T}_Z$ of $Z$ is the transitive reduction \footnote{The transitive reduction of a dag is the unique subgraph with the minimal number of arcs and the same reachability relation~\cite{aho1972transitive}.} of $D$. We call the non-singleton indecomposable modules the \emph{basis} of the modular decomposition.
\end{defn}
Some properties of the modular decomposition are easy to establish. Firstly, it is unique because the transitive reduction of a dag is unique. Secondly, the sinks of the modular decomposition are the singletons (which are trivially indecomposable), and the non-sink nodes are basis modules. The descendants of a given node define the module assigned to that node\footnote{Consequently, we don't need to explicitly label non-sink nodes by the associated module. This ensures the tree is linear-space. The same technique is used for the graph modular decomposition~\cite{mcconnell1999modular}.} (see Figure~\ref{fig:Z tree}). Every module is an overlapping union of basis modules, and so is defined by an overlapping set of descendants of basis modules in the modular decomposition. The name `basis' is by analogy with linear algebra: basis modules cannot be formed as overlapping unions of other modules  (they are `independent') and every thin module can be uniquely formed as an overlapping union of indecomposable ones (they `span' the set of thin modules). We refer to the size of the basis as the \emph{dimension} of an HFSM. Like the dimension of a vector space, it is an invariant of equivalent HFSMs.
%If the decomposition tree is a tree in the digraph sense, then modules are subtrees. In fact, more generally, this is how the decomposition tree stores all modules: a set is a thin module if and only if it is a union of connected subtrees---connected meaning that they share descendants, that is, they overlap. 
We have one crucial tool for proving results about the basis, which will also prove useful when computing it (Theorem~\ref{algorithm correct}). This tool is that each basis module has a `representative' state.
\begin{thm}[Representative Theorem] \label{Kq theorem}
Let $q$ be a state in an FSM $Z$ that is not the start state. Define $\repr_Z(q)$ as the intersection of all thin modules which contain $q$ but where $q$ is not the start state.
%\[
%\repr_Z(q) := \bigcap_{\substack{M\text{ thin module} \\q\in M\\q\text{ not start node of $M$}}} M
%\]
Then $\repr_Z(q)$ is a basis module, and each basis module $H$ has a $q$ such that $\repr_Z(q)=H$.
\end{thm}
Showing that $\repr_Z(q)$ is a module, and is indecomposable, follows from closure under overlapping intersections (Lemma~\ref{union and intersection}). The main novelty of Theorem~\ref{Kq theorem} is that each basis module can be represented this way. It follows easily that $\mathcal{T}$ has a linear number of nodes, because there are at most $n-1$ distinct representatives for basis modules\footnote{More precisely, every accessible $n$-state FSM with $n > 1$ has between $n+1$ and $2n-1$ indecomposable modules (Lemma~\ref{linear growth}). These bounds are tight. Any prime FSM has $n+1$ trivial indecomposable modules, and one FSM with $2n-1$ indecomposable modules is that consisting of a single $x$-path of length $n$ beginning at the start state.}.

There is one more ingredient we must add to the modular decomposition. In the graph modular decomposition, each non-sink node is labelled by a graph (Figure~\ref{fig:G tree}), and this is what allows us to reconstruct the original graph from the modular decomposition. Each module $M$ is labelled by the restriction $G[M]$, with smaller modules $K_1,K_2,\dots\subseteq M$ contracted. We call this the \emph{contracted form} of $M$. We do the same in our modular decomposition, labelling each basis module by its contracted form, leading to Figure~\ref{fig:Z tree}. Formally:
\begin{defn}[Contracted Form] \label{contracted form}
Let $Z$ be an FSM and $M$ a module. The \emph{contracted form} of $M$ is the FSM $Z[M]/K_1,\dots,K_n$, where $K_i$ are the \emph{maximal} thin modules contained in $M$, that is for each $i$ there is no thin module $H$ such that $K_i\subset H\subset M$.
\end{defn}
This is well-defined and unique because each maximal thin module is disjoint from all others; this follows from Lemma~\ref{union and intersection}. As a result, the order of contraction does not matter.
%More precisely, Theorem~\ref{basis theorem}, our main theorem, lays out the properties of the decomposition tree. 

 %When drawing decomposition trees we will therefore label sinks by the element in their respective singleton and all other nodes will remain unlabelled, because the modules corresponding to any node can be recovered from its descendants. This assumption is important to ensure the tree is linear-space\footnote{There are at most a linear number of indecomposable modules, but storing the contents of each may take quadratic space, so we cannot label each node in the decomposition tree by its module without incurring a significant increase in space. The decomposition tree of a graph is constructed in the same way.}.

\begin{thm}[Properties of the Modular Decomposition] \label{basis theorem}
Let $Z$ be an accessible HFSM, and $\mathcal{T}$ its modular decomposition. Then,%(Definition~\ref{def:FSM_decom_tree}). Given $t\in\mathcal{T}$, we write $\downarrow\! t$ to denote the set of nodes of $Z$ labelling sinks reachable from $t$. Then, 
\begin{enumerate}
    \item \textbf{$\mathcal{T}$ is small}: $\mathcal{T}$ has a linear number of nodes and arcs compared to $Z$;
    \item \textbf{$\mathcal{T}$ represents all thin modules}: a set $M\subseteq Q(Z)$ is a thin module of $Z$ if and only if it is a union of overlapping basis modules, and each thin module is an overlapping union of a unique smallest set of basis modules;
    %each thin module has a unique smallestthere are nodes $t_1,\dots,t_n\in \mathcal{T}$ where the sets $\downarrow\! t_i$ are overlapping and $M = \bigcup_{i=1}^n \downarrow\!t_i$. Further, each thin module $M$ has a unique minimal set $\{t_1,\dots,t_k\}$ where the sets $\downarrow\! t_i$ are overlapping and $M = \bigcup_{i=1}^k\downarrow\! t_i$.\todo{change this notation?}
    \item \textbf{$\mathcal{T}$ represents HFSM equivalence}: if $Z$ and $Y$ are equivalent HFSMs then $\repr_Z(q)\mapsto \repr_Y(q)$ is a one-to-one correspondence between basis modules, and the contracted forms of $\repr_Z(q)$ and $\repr_Y(q)$ are equal up to state relabelling.
\end{enumerate}
\end{thm}
We will sketch the proof here---the full version is in the Appendix. First, Theorem~\ref{Kq theorem} ensures there are a linear number of basis modules\footnote{A dag with $O(n)$ nodes could still potentially have a quadratic number of arcs. However, we show that if a node $t_M$ in $\mathcal{T}$ representing a basis module $M$ has many arcs out of it in $\mathcal{T}$, then these modules overlap $M$ in a specific branching structure in $Z$, and this implies the existence of a proportional number of arcs in $Z$ (Proposition~\ref{linear arcs})}, establishing the first claim. Second, we show that in any family of sets that is closed under overlapping unions, every set can be constructed as an overlapping union of indecomposable ones (Proposition~\ref{set bases}). Using the fact that they are also closed under intersection, we further prove that each thin module is a union of a \emph{unique smallest} set of overlapping basis modules (Proposition~\ref{decomposition theorem}), proving the second claim. For the third claim we use induction, showing that for two HFSMs $Z$ and $Y$ which differ by a single nesting, $\repr_Z(q)\mapsto\repr_Y(q)$ is well-defined in both directions.

%\subsection{Application: maximising HFSMs} \label{sec:maximising hfsms}

\begin{figure}
    \centering
    \begin{subfigure}{0.4\textwidth}
        \centering
        \begin{tikzpicture}
\path [use as bounding box] (0,-2) rectangle (6.2,4);

\newcommand{\locx}{3}
\newcommand{\locy}{3.7}
\node (123456o) [state,initial, initial text={}, initial distance=1ex,initial below] at (\locx - 0.5,\locy) {};
\node (78o) [state] at (\locx + 0.5,\locy) {};
\node[draw,rounded corners,minimum height=.6cm, minimum width=1.4cm] (all) at (\locx,\locy) {};
\draw[edge] (123456o) to[out=25,in=155] node[edgelabel] {$y$} (78o);
\draw[edge] (123456o) to[out=-25,in=-155] node[edgelabel] {$x$} (78o);

\renewcommand{\locx}{2.15}
\renewcommand{\locy}{2.6}
\node (1oR) [state,initial, initial text={}, initial distance=1ex,initial below] at (\locx - 0.5,\locy) {};
\node (3oL) [state] at (\locx + 0.5,\locy) {};
\node[draw,rounded corners,minimum height=.6cm, minimum width=1.4cm] (13) at (\locx,\locy) {};
\draw[edge] (1oR) to[in=155,out=25] node[edgelabel] {$x$} (3oL);
\draw[edge] (3oL) to[in=-25,out=-155] node[edgelabel] {$y$} (1oR);
\draw[-to, dashed] (123456o) to[out=180,in=90] (13);

\node (3) [state] at (\locx+0.5,.5) {3};

\renewcommand{\locx}{5.5}
\node (7o) [state,initial, initial text={}, initial distance=1ex,initial below] at (\locx - 0.5,\locy) {};
\node (8o) [state] at (\locx + 0.5,\locy) {};
\node[draw,rounded corners,minimum height=.6cm, minimum width=1.4cm] (78) at (\locx,\locy) {};
\draw[edge] (7o) to[in=155,out=25] node[edgelabel] {$y$} (8o);
\draw[edge] (8o) to[in=-25,out=-155] node[edgelabel] {$x$} (7o);
\draw[-to, dashed] (78o) to[out=0,in=90] (78);

\node (7) [state] at (\locx-0.9,1.6) {7};
\draw[-to, dashed] (7o) to[out=180,in=90] (7);
\node (8) [state] at (\locx +.5,1.6) {8};
\draw[-to, dashed] (8o) to[out=-90,in=90] (8);

\renewcommand{\locx}{.7}
\renewcommand{\locy}{1.5}
\node (2o) [state] at (\locx - 0.5,\locy) {};
\node (1oL) [state,initial, initial text={}, initial distance=1ex,initial below] at (\locx + 0.5,\locy) {};
\node[draw,rounded corners,minimum height=.6cm, minimum width=1.4cm] (12) at (\locx,\locy) {};
\draw[edge] (1oL) to[out=155,in=25] node[edgelabel] {$y$} (2o);
\draw[edge] (2o) to[out=-25,in=-155] node[edgelabel] {$x$} (1oL);
%\draw[-to, dashed] (123456o) to[out=180,in=90] (12);

\node (1) [state] at (\locx + 0.9,.5) {1};
\draw[-to, dashed] (1oL) to[out=0,in=90] (1);
\node (2) [state] at (\locx-0.5,.5) {2};
\draw[-to, dashed] (2o) to[out=-90,in=90] (2);

\renewcommand{\locx}{3.6}
\node (3oR) [state,initial, initial text={}, initial distance=1ex,initial below] at (\locx - 0.5,\locy) {};
\node (456o) [state] at (\locx + 0.5,\locy) {};
\node[draw,rounded corners,minimum height=.6cm, minimum width=1.4cm] (3456) at (\locx,\locy) {};
\draw[edge] (3oR) to[in=155,out=25] node[edgelabel] {$x$} (456o);
\draw[edge] (456o) to[in=-25,out=-155] node[edgelabel] {$y$} (3oR);
%\draw[-to, dashed] (123456o) to[out=-60,in=90] (3456);
\draw[-to, dashed] (3oR) to[out=180,in=90] (3);

\draw[-to, dashed] (1oR) to[out=180,in=90] (12);
\draw[-to, dashed] (3oL) to[out=0,in=90] (3456);

\renewcommand{\locx}{4.1}
\renewcommand{\locy}{0.5}
\node (45o) [state,initial, initial text={}, initial distance=1ex,initial below] at (\locx - 0.5,\locy) {};
\node (6o) [state] at (\locx + 0.5,\locy) {};
\node[draw,rounded corners,minimum height=.6cm, minimum width=1.4cm] (456) at (\locx,\locy) {};
\draw[edge] (45o) to[out=25,in=155] node[edgelabel] {$y$} (6o);
\draw[edge] (45o) to[out=-25,in=-155] node[edgelabel] {$x$} (6o);
\draw[-to, dashed] (456o) to[out=-90,in=90] (456);

\node (6) [state] at (\locx + 0.5,-0.5) {6};
\draw[-to, dashed] (6o) to[out=-90,in=90] (6);

\renewcommand{\locx}{3.3}
\renewcommand{\locy}{-0.5}
\node (4o) [state,initial, initial text={}, initial distance=1ex,initial below] at (\locx - 0.5,\locy) {};
\node (5o) [state] at (\locx + 0.5,\locy) {};
\node[draw,rounded corners,minimum height=.6cm, minimum width=1.4cm] (45) at (\locx,\locy) {};
\draw[edge] (4o) to[out=25,in=155] node[edgelabel] {$y$} (5o);
\draw[edge] (5o) to[in=-25,out=-155] node[edgelabel] {$x$} (4o);
\draw[-to, dashed] (45o) to[out=180,in=100] (45);

\node (4) [state] at (\locx - 0.9,-1.3) {4};
\draw[-to, dashed] (4o) to[out=180,in=90] (4);
\node (5) [state] at (\locx + 0.5,-1.3) {5};
\draw[-to, dashed] (5o) to[out=-90,in=90] (5);

 \end{tikzpicture}
        \caption{HFSM $W_1$}
        \label{fig:W1}
    \end{subfigure}
    \hskip 1cm
    \begin{subfigure}{0.4\textwidth}
        \centering
        \begin{tikzpicture}
\path [use as bounding box] (0,-2) rectangle (6.2,4);

\newcommand{\locx}{3}
\newcommand{\locy}{3.7}
\node (123456o) [state,initial, initial text={}, initial distance=1ex,initial below] at (\locx - 0.5,\locy) {};
\node (78o) [state] at (\locx + 0.5,\locy) {};
\node[draw,rounded corners,minimum height=.6cm, minimum width=1.4cm] (all) at (\locx,\locy) {};
\draw[edge] (123456o) to[out=25,in=155] node[edgelabel] {$y$} (78o);
\draw[edge] (123456o) to[out=-25,in=-155] node[edgelabel] {$x$} (78o);

\renewcommand{\locx}{5.5}
\renewcommand{\locy}{2.6}
\node (7o) [state,initial, initial text={}, initial distance=1ex,initial below] at (\locx - 0.5,\locy) {};
\node (8o) [state] at (\locx + 0.5,\locy) {};
\node[draw,rounded corners,minimum height=.6cm, minimum width=1.4cm] (78) at (\locx,\locy) {};
\draw[edge] (7o) to[in=155,out=25] node[edgelabel] {$y$} (8o);
\draw[edge] (8o) to[in=-25,out=-155] node[edgelabel] {$x$} (7o);
\draw[-to, dashed] (78o) to[out=0,in=90] (78);

\node (7) [state] at (\locx-0.9,1.6) {7};
\draw[-to, dashed] (7o) to[out=180,in=90] (7);
\node (8) [state] at (\locx +.5,1.6) {8};
\draw[-to, dashed] (8o) to[out=-90,in=90] (8);

\renewcommand{\locx}{3}
\node (3oR) [state,initial, initial text={}, initial distance=1ex,initial below] at (\locx - 0.5,\locy) {};
\node (456o) [state] at (\locx + 0.5,\locy) {};
\node[draw,rounded corners,minimum height=.6cm, minimum width=1.4cm] (3456) at (\locx,\locy) {};
\draw[edge] (3oR) to[in=155,out=25] node[edgelabel] {$x$} (456o);
\draw[edge] (456o) to[in=-25,out=-155] node[edgelabel] {$y$} (3oR);
%\draw[-to, dashed] (123456o) to[out=-60,in=90] (3456);

\renewcommand{\locx}{.7}
\renewcommand{\locy}{1.6}
\node (2o) [state] at (\locx - 0.5,\locy) {};
\node (1oL) [state,initial, initial text={}, initial distance=1ex,initial below] at (\locx + 0.5,\locy) {};
\node[draw,rounded corners,minimum height=.6cm, minimum width=1.4cm] (12) at (\locx,\locy) {};
\draw[edge] (1oL) to[out=155,in=25] node[edgelabel] {$y$} (2o);
\draw[edge] (2o) to[out=-25,in=-155] node[edgelabel] {$x$} (1oL);
%\draw[-to, dashed] (123456o) to[out=180,in=90] (12);
\draw[-to, dashed] (3oR) to[out=180,in=90] (12);

\node (2) [state] at (\locx-0.5,.5) {2};
\draw[-to, dashed] (2o) to[out=-90,in=90] (2);

\renewcommand{\locx}{4.1}
\renewcommand{\locy}{0.5}
\node (45o) [state,initial, initial text={}, initial distance=1ex,initial below] at (\locx - 0.5,\locy) {};
\node (6o) [state] at (\locx + 0.5,\locy) {};
\node[draw,rounded corners,minimum height=.6cm, minimum width=1.4cm] (456) at (\locx,\locy) {};
\draw[edge] (45o) to[out=25,in=155] node[edgelabel] {$y$} (6o);
\draw[edge] (45o) to[out=-25,in=-155] node[edgelabel] {$x$} (6o);
\draw[-to, dashed] (456o) to[out=-90,in=90] (456);

\node (6) [state] at (\locx + 0.5,-0.5) {6};
\draw[-to, dashed] (6o) to[out=-90,in=90] (6);

\renewcommand{\locx}{2.15}
\node (1oR) [state,initial, initial text={}, initial distance=1ex,initial below] at (\locx - 0.5,\locy) {};
\node (3oL) [state] at (\locx + 0.5,\locy) {};
\node[draw,rounded corners,minimum height=.6cm, minimum width=1.4cm] (13) at (\locx,\locy) {};
\draw[edge] (1oR) to[in=155,out=25] node[edgelabel] {$x$} (3oL);
\draw[edge] (3oL) to[in=-25,out=-155] node[edgelabel] {$y$} (1oR);
\draw[-to, dashed] (123456o) to[out=-160,in=160] (3456);

\node (1) [state] at (\locx - 0.9,-.4) {1};
\draw[-to, dashed] (1oL) to[out=0,in=90] (13);
\node (3) [state] at (\locx+0.5,-0.4) {3};

\renewcommand{\locx}{3.3}
\renewcommand{\locy}{-1}
\node (4o) [state,initial, initial text={}, initial distance=1ex,initial below] at (\locx - 0.5,\locy) {};
\node (5o) [state] at (\locx + 0.5,\locy) {};
\node[draw,rounded corners,minimum height=.6cm, minimum width=1.4cm] (45) at (\locx,\locy) {};
\draw[edge] (4o) to[out=25,in=155] node[edgelabel] {$y$} (5o);
\draw[edge] (5o) to[in=-25,out=-155] node[edgelabel] {$x$} (4o);
\draw[-to, dashed] (45o) to[out=180,in=100] (45);

\node (4) [state] at (\locx - 0.9,-1.8) {4};
\draw[-to, dashed] (4o) to[out=180,in=90] (4);
\node (5) [state] at (\locx + 0.5,-1.8) {5};
\draw[-to, dashed] (5o) to[out=-90,in=90] (5);

\draw[-to, dashed] (1oR) to[out=180,in=90] (1);
\draw[-to, dashed] (3oL) to[out=-90,in=90] (3);

 \end{tikzpicture}
        \caption{HFSM $W_2$}
        \label{fig:W2}
    \end{subfigure}
\caption{Two HFSMs equivalent to $Z$ (Figure~\ref{fig:Z}), constructed by contracting different choices of modules from $Z$. Comparing to Figure~\ref{fig:Z tree}, the modular decomposition of $Z$, we see that $W_1$ and $W_2$ represent different ways to extend the modular decomposition to a tree form.}
    \label{fig:W}
\end{figure}
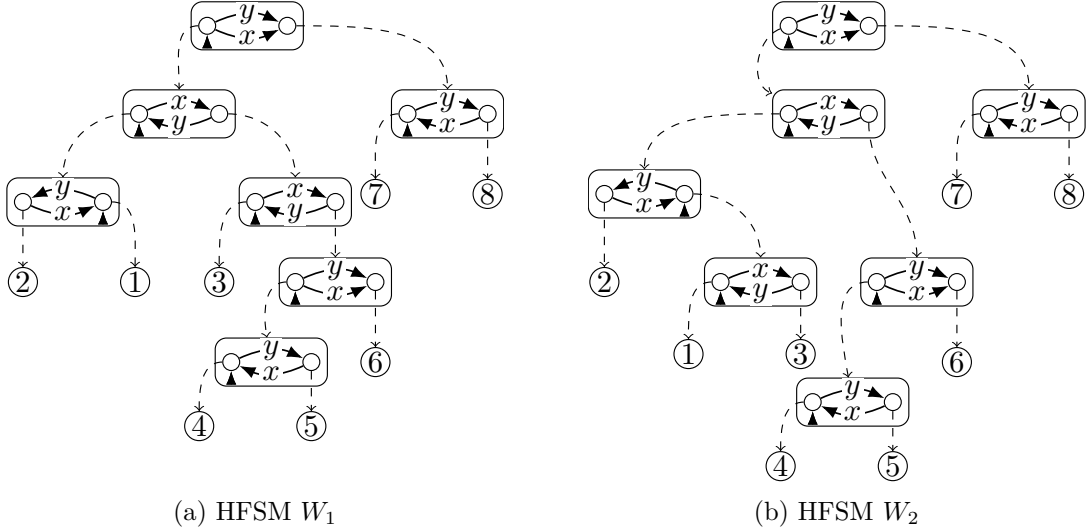

To understand Theorem~\ref{basis theorem}, let's consider the concrete example of $Z$ (Figure~\ref{fig:Z}), and its modular decomposition (Figure~\ref{fig:Z tree}). We can contruct an equivalent HFSM by repeatedly selecting a thin module $M$ and nesting $Z[M]$ at $M$ in $Z/M$. Where $Z$'s modules are properly nested, such as $\{4,5,6\}$ and $\{4,5\}$, there is no choice for how to decompose these modules into nested FSMs. However, where $Z$'s modules overlap, such as $\{1,2\}$, $\{1,3\}$ and $\{3,4,5,6\}$, selecting a module to nest removes the overlap, and so different choices of module can lead to HFSMs with different nesting trees. For instance, Figure~\ref{fig:W} shows two HFSMs $W_1$ and $W_2$, both equivalent to $Z$ from Figure~\ref{fig:Z}.  However, because basis modules are in a one-to-one correspondence, equivalent HFSMs have the same dimension. Repeating this process, we find that the number of times we can recursively nest FSMs is always \emph{exactly} the dimension, which in the case of $Z$ is seven. At that point, regardless of choice of modules, we arrive at an HFSM whose modular decomposition is \emph{exactly the same} as its nesting tree (such as $W_1$ and $W_2$ in Figure~\ref{fig:W}, which have seven component FSMs). These HFSMs are maximal, in that they can't be further decomposed. Theorem~\ref{basis theorem} implies that all maximal HFSMs have the same amount of nesting, and so finding maximal HFSMs can be done greedily. Further, all maximal HFSMs have the same set of component FSMs, which are \emph{exactly the contracted forms of the basis modules}, possibly in different orders. We can verify this by comparing Figure~\ref{fig:W} to Figure~\ref{fig:Z tree}. Looking at a specific state, such as 3, we see it is contained in different nested FSMs in $W_1$ and $W_2$, with $\repr_{W_1}(3) = \{1,2,3,4,5,6\}$ and $\repr_{W_2}(3) = \{1,3\}$. However, the contracted forms of $\repr_{W_1}(3)$ and $\repr_{W_2}(3)$ are equal up to state labels.

This gives a greedy solution to the bottleneck problem. Every thin HFSM must have width at least that of the maximum size of the contracted form of any basis module. However, this means that every maximal HFSM has minimal width, and we can construct maximal HFSMs greedily. For example, $Z$ can be decomposed into $W_1$ or $W_2$ both with width 2, which is minimal.

\section{Computing the Modular Decomposition} \label{sec:computing}

We have defined the modular decomposition, but how do we compute it? In this section we give an algorithm which constructs the modular decomposition of a $n$-state $k$-symbol accessible FSM in $O(n^2k)$ time. The modular decomposition of an HFSM can then be constructed from its component FSMs.
%The main idea behind the algorithm proposed in this section (Algorithm~\ref{alg:findmodules}) is constructing a preorder on $Q(Z)$ for each state $v$, and identifying the indecomposable thin module as the ancestor sets of certain nodes in this order, using a characterisation described in Theorem~\ref{module characterisation}. Finally, a node is added to the decomposition tree for each indecomposable module computed, using Algorithm~\ref{alg:merge module}.
The algorithm has two main steps. The first involves computing the basis modules (Algorithms~\ref{alg:Gv} and~\ref{alg:findmodules}), and the second constructs the modular decomposition by ordering the basis by containment (Algorithm~\ref{alg:merge module}). The latter part is comparatively straightforward\footnote{Algorithm~\ref{alg:merge module} uses a variant of breadth-first-search on the modular decomposition which, given a basis module $M$ to add to $\mathcal{T}$, locates the nodes $t_H$ representing modules $H$ which are immediate successors of $K$ in the inclusion order on basis modules.} and we defer it to the Appendix. Here we focus on computing the basis modules. Given that every module has a unique start state, we begin our search by fixing a state $v$, and identifying the modules for which this is the start state. In the following, we will call a module with start state $v$ a \emph{$v$-module}. The next step is to formalise the following relationship: if a given state $u$ is in a $v$-module, then which states $w$ must also be in this module? This relationship is a preorder on $Q(Z)$ for each fixed $v$. Algorithm~\ref{alg:Gv} computes this preorder by constructing a graph $\mathcal{G}_v$ where (1) for any states $u$ and $w$, there is a path $\pth{u}{w}$ in $\mathcal{G}_v$ if and only if $u$ is contained in every $v$-module containing $w$; and (2) every state in $\mathcal{G}_v$ is contained in some $v$-module.

\begin{algorithm}
\SetAlgoLined
\SetKwInOut{Input}{Input}
\SetKwInOut{Output}{Output}
\Input{\ Accessible FSM $Z$, state $\bm{v}\in Q(Z)$}
\Output{\ A digraph $\mathcal{G}_{\bm{v}}$}
 Create an empty digraph $\mathcal{G}_{\bm{v}}$ with state set $Q(Z)$\;
 Given $x\in\Sigma$, let $q_x$ be the state with the longest $x$-path $\pth[x]{\bm{v}}{q_x}$\;
 \For{each state $u$ in $Q(Z)$}{
    \For{each symbol $x$ in $\Sigma$}{
        \eIf{there is an arc $\arc[x]{u}{w}$ in $Z$}{
            \If{$w\neq \bm{v}$}{
                ($\mathbf{a}$) Add $\arc{u}{w}$ to $\mathcal{G}_v$\;
                \eIf{there is a path \begin{tikzcd} \bm{v} \ar[r,squiggly,"x"] \pgfmatrixnextcell t \ar[r,"x"] \pgfmatrixnextcell w\end{tikzcd} in $Z$}{
                    ($\mathbf{c}$) Add $\arc{t}{u}$ to $\mathcal{G}_v$ (if $t\neq u$)\;
                }{
                    ($\mathbf{b}$) Add $\arc{w}{u}$ to $\mathcal{G}_v$\;
                }
            }
        }{
            ($\mathbf{d}$) Add $\arc{q_x}{u}$ to $\mathcal{G}_v$ (if $q_x\neq u)$\;
        }
    }
 }
 If $v\neq \bm{s}$ ($Z$'s start state), remove from $\mathcal{G}_v$ any state which is reachable from $s$ in $\mathcal{G}_v$\;
 \Return $\mathcal{G}_v$\;
 \caption{Constructing $\mathcal{G}_v$}
 \label{alg:Gv}
\end{algorithm}

Algorithm~\ref{alg:Gv} begins with an empty graph whose node set is $Q(Z)$. We then iterate through each arc of $Z$, adding arcs to $\mathcal{G}_v$ in four cases, called (\textbf{a}), (\textbf{b}), (\textbf{c}) and (\textbf{d}). Let $\arc[x]{u}{w}$ be some transition in $Z$. First, if $w$ is contained in some $v$-module $H$, and $w\neq v$, then $w$ is not an entrance of $H$, so $u$ is also in $H$, and we add the arc $\arc{u}{w}$ to $\mathcal{G}_v$ (\textbf{a}) (line~7 of Alg.~\ref{alg:Gv}). Now suppose $u$ is in a $v$-module $H$ but \emph{not} also $w$. For this to be true, $w$ must be the $x$-exit of $H$. By thinness, $w$ is on the unique $x$-path out of $v$, so \begin{tikzcd} v \ar[r,squiggly,"x"] & t \ar[r,"x"] & w \end{tikzcd}. Now, if $u$ is on this path it must be $t$, but if $u$ is not on this path then the entire subpath $\pth[x]{v}{t}$ must be in $H$. We enforce this by the arc $\arc{t}{u}$ ((\textbf{c}), line~9 of Alg.~\ref{alg:Gv}). Otherwise, $u$ and $w$ must be contained in all the same $v$-modules, so we also add an $\arc{w}{u}$ to $\mathcal{G}_v$ ((\textbf{b}), line~11).  %If $w\in H$, then all other elements of the path are also in $H$ by the first part of Rule~2. Algorithmically, we can represent this by a single added arc $\arc{w}{a}$ in $\mathcal{G}_v$.
Finally, if a state $u$ in a $v$-module $H$ has no $x$-arc, then $H$ must have no $x$-exits, and so all states on the unique $x$-path out of $v$ must be in $H$. This is (\textbf{d}) (line~13 of Alg.~\ref{alg:Gv}). By the end of the outer loop of Algorithm~\ref{alg:Gv}, property (1) is satisfied. % that is, there is a path $\pth{u}{w}$ in $\mathcal{G}_v$ if and only if $u$ is contained in every $v$-module containing $w$. 
However, we must still remove states which are not in any $v$-module. If $v=s$ (the start state of $Z$), then every state is in at least one $v$-module (the trivial module $Q(Z)$). If $v\neq s$, then $s$ is in no $v$-module, and if there is a path $\pth{s}{u}$ in $\mathcal{G}_v$ then every $v$-module containing $u$ contains $s$, and so $u$ can't be contained in a $v$-module. As a result, we remove all such states from $\mathcal{G}_v$ (line~14 of Alg.~\ref{alg:Gv}). Theorem~\ref{module characterisation} proves that this is sufficient for $\mathcal{G}_v$ to satisfy our criteria.
\begin{thm} \label{module characterisation}
For $q\in Q(Z)$, we write $\uparrow_v\!\! q$ to denote the ancestors of $q$ in $\mathcal{G}_v$. Then $\uparrow_v\!\! q$ is a thin $v$-module, and a subset $H \subset Z$ is a thin $v$-module if and only if there exist states $q_1,\dots,q_m\in Q(Z)$ where $H = \bigcup_{i\in 1,\dots,m}\uparrow_v\!\!q_i$.
\end{thm}
Having constructed $\mathcal{G}_v$, finding the basis modules can be done using Algorithm~\ref{alg:findmodules}.% and makes use of Algorithms~\ref{alg:Gv} and \ref{alg:merge module} as subroutines.
\begin{algorithm}
\SetAlgoLined
\SetKwInOut{Input}{Input}
\SetKwInOut{Output}{Output}
\SetKwData{TREE}{TREE}
\SetKwData{MOD}{MOD}
\SetKwData{used}{used}
\Input{\ Accessible FSM $Z$, with start state $s$}
\Output{\ $Z$'s modular decomposition $\mathcal{T}$}
 Create a digraph $\mathcal{T}$ with node set $Q(Z)$\;
 $\used \gets \emptyset$\;
 \For{state $v$ in any reversed breadth-first-search of $Z$ from $s$}{
 Construct $\mathcal{G}_v$, using Algorithm~\ref{alg:Gv}\;
 \For{each strongly connected component $M\neq \{v\}$ of $\mathcal{G}_v$ in topological order}{
        Choose an arbitrary $q \in M \setminus \used $\;
        Add the module $\uparrow_v\!\!q$ to $\mathcal{T}$ using Algorithm~\ref{alg:merge module}\;
        $\used\gets \used \cup (\uparrow_v\!\!q\setminus\{v\})$\;
    }
 }
 \Return $\mathcal{T}$
 \caption{Constructing the Modular Decomposition}
 \label{alg:findmodules}
\end{algorithm}

\begin{thm} \label{algorithm correct}
Algorithm~\ref{alg:findmodules} works, i.e. the sets $\uparrow_v\!\!q$ added to $\mathcal{T}$ are exactly the basis modules of $Z$.
\end{thm}
\begin{proof}[Proof sketch; full proof in Appendix]
%Algorithm~\ref{alg:findmodules} works by: iterating through each node $v$ in a reverse breadth-first-search order from $s$; constructing $\mathcal{G}_v$; topologically ordering $\mathcal{G}_v$; constructing $\uparrow_v\!\!q$ for some $q\in M$ that is not yet in $used$; adding $\uparrow_v\!\!q$ to the modular decomposition using Algorithm~\ref{alg:merge module}; then adding $\uparrow_v\!\!q\setminus\{v\}$ to $used$. We will show that the sets $\uparrow_v\!\!q$ on which Algorithm~\ref{alg:merge module} is called are the basis modules.

First, observe that indecomposable $v$-modules must have the form $\uparrow_v\!\!q$ for some $q$. This is because each of these is a module (Theorem~\ref{module characterisation}), and \emph{indecomposable} modules cannot be a union of multiple overlapping modules. Unfortunately, not all $\uparrow_v\!\!q$ are indecomposable. To find those that are, we use Theorem~\ref{Kq theorem}. It turns out that $\uparrow_v\!\!q$ is indecomposable if and only if $\repr_Z(q) = \uparrow_v\!\!q$. We can find only the desired $q$ by choosing the states $v$ in a specific order---this is where the reverse breadth-first-search comes in. If $\uparrow_v\!\!q$ is decomposable, then $\repr_Z(q) \subset \uparrow_v\!\!q$, and so there is a state $w\in \uparrow_v\!\!q$ with $\repr_Z(q) = \uparrow_w\!\!q$. Because $\uparrow_w\!\!q \subset \uparrow_v\!\!q$, we show that all paths from $s$ to $w$ must pass through $v$. Equivalently, $w$ follows $v$ on any breadth-first-search order from $s$; so $w$ precedes $v$ in the reverse order. Using induction in reverse breadth-first-search order, we show that each $q$ is added to $used$ precisely as $\repr_Z(q)$ is added to $\mathcal{T}$.% Fix some $v$ and $q$. Either $\repr_Z(q) \subset \uparrow_v\!\!q$, in which case $\repr_Z(q)=\uparrow_w\!\!q$, and since $w$ has already been visited, we conclude $q$ is in $used$ by the inductive hypothesis. Otherwise, $\repr_Z(q) = \uparrow_v\!\!q$, so $\uparrow_v\!\!q$ is indecomposable, and we add it to the modular decomposition while adding $q$ to $used$. By searching through $\mathcal{G}_v$ in topological order for fixed $v$ we ensure that $q$ is not in $used$ until we construct $\uparrow_v\!\!q$. Consequently, by induction, we add precisely the basis modules to $\mathcal{T}$.
\end{proof}

\begin{thm}
Algorithm~\ref{alg:findmodules} constructs the modular decomposition of an FSM $Z$ in $O(n^2k)$ time.
\end{thm}
\begin{proof}
Firstly, observe that $Z$ has $O(nk)$ arcs. To begin, we perform a BFS of $Z$, and then reverse this order; BFS takes $O(nk)$ steps. Algorithm~\ref{alg:Gv} constructs $\mathcal{G}_v$ with an outer loop of size $n$ and inner loop of size $k$ (with some additional pre- and post-processing), and so overall takes $O(nk)$. At most two arcs are added to $\mathcal{G}_v$ on each inner loop iteration, so $|A(\mathcal{G}_v)| \in O(nk)$. Topologically ordering $\mathcal{G}_v$ takes $O(n+nk)= O(nk)$~\cite{tarjan1972depth}. Now, we add states to $used$ every time we add a module to $\mathcal{T}$. By Theorem~\ref{algorithm correct}, we add precisely the basis modules to $\mathcal{T}$, and by Lemma~\ref{linear growth} there are at most $n-1$ of these. Since Algorithm~\ref{alg:merge module} visits every arc of $\mathcal{T}$ at most once, and $|A(\mathcal{T})|\in O(|A(Z)|)\in O(nk)$ by Theorem~\ref{basis theorem}, it takes $O(nk)$ operations to add a module to the modular decomposition. Constructing $\uparrow_v\!\!q$ takes also $O(nk)$ time, but this is performed at most $n-1$ times, so the algorithm runs in $O(n^2k)$ time.
\end{proof}

\section{Applications} \label{sec: applications}

We will demonstrate our results on a famous HFSM: the wristwatch example used in Harel's original paper on statecharts (\cite{harel1987statecharts}, Figure 31). In Figure~\ref{fig:harel} we present (a somewhat simplified version of\footnote{Statecharts, as originally presented, allow a broader range of modelling techniques than we discuss here, notably timed transitions, transitions between layers in an HFSM, and orthogonal states. For the purposes of this example, we have constructed a slightly simplified version of the `displays' subcomponent of Harel's example (see Fig. 31 of \cite{harel1987statecharts}), which does not use these additional techniques.}) this example. This HFSM models the behaviour of a wristwatch (the Citizen Quartz Multi-Alarm III). The HFSM has four symbols $a,b,c$ and $d$, representing each of the four buttons on the watch\footnote{Harel's example also includes extra symbols $\hat a, 
\hat b, \hat c$ and $\hat d$ for when these buttons are \emph{held} rather than pressed---we ignore these for simplicity.}. This model demonstrates the features and complexity of real-world systems. Note that, despite its real-world nature, we did not need to modify this example to make it thin---this demonstrates the naturalness of the thinness property.

\begin{figure}
    \centering
    \begin{tikzpicture}

% out
\node (displays) at (0,10) {displays};
\node[draw,minimum height=14cm,minimum width=15.5cm,rounded corners] (displaybox) at (7,3.5) {};

% stopwatch
\node (stopwatch) at (.9,9.2) {stopwatch};
\node[draw,minimum height=5cm,minimum width=5cm,rounded corners] (stop) at (2.5,7) {};
\node (zero) [state,initial right, initial text={}] at (2.5,9) {zero};
\node (son) [state] at (1.5,7.5) {on};
\node (soff) [state] at (3.5,7.5) {off};
\node (lon) [state] at (1.5,5.5) {lap on};
\node (loff) [state] at (3.5,5.5) {lap off};
\draw[edge] (son) to[out=40,in=140]  node[edgelabel] {$b$}(soff);
\draw[edge] (soff) to[out=-140,in=-40]  node[edgelabel] {$b$}(son);
\draw[edge] (lon) to[out=40,in=140]  node[edgelabel] {$b$}(loff);
\draw[edge] (loff) to[out=-140,in=-40]  node[edgelabel] {$b$}(lon);
\draw[edge] (son) to[out=-60,in=65]  node[edgelabel] {$d$}(lon);
\draw[edge] (lon) to[out=115,in=-120]  node[edgelabel] {$d$}(son);
\draw[edge] (zero) to  node[edgelabel] {$b$}(son);
\draw[edge] (soff) to  node[edgelabel] {$d$}(zero);
\draw[edge] (loff) to  node[edgelabel] {$d$}(soff);

% regular
%\node (regular) at (6.7,10.75) {regular};
%\node[draw,minimum height=6cm,minimum width=8cm,rounded corners,initial right,initial text={}] (reg) at (10,8) {};
\node (time) [state,initial right, initial text={}] at (12.5,6.5) {time};
\node (date) [state] at (12.5,9.5) {date};
\draw[edge] (date) to[out=-60,in=60]  node[edgelabel] {$d$}(time);
\draw[edge] (time) to[out=120,in=-120]  node[edgelabel] {$d$}(date);

% update
\node (up) at (7.25,10.25) {update};
\node[draw,minimum height=5cm,minimum width=4cm,rounded corners] (update) at (8.5,8) {};
\node (c3) [state] at (7.2,9.5) {mon};
\node (c2) [state] at (8.5,9.5) {hr};
\node (c1) [state, initial above, initial text={}] at (9.8,9.5) {sec};
\node (c6) [state] at (7.2,8) {date};
\node (c5) [state] at (8.5,8) {10mn};
\node (c4) [state] at (9.8,8) {1mn};
\node (c9) [state] at (7.2,6.5) {day};
\node (c8) [state] at (8.5,6.5) {year};
\node (c7) [state] at (9.8,6.5) {mode};
\draw[edge] (c1) to  node[edgelabel] {$c$}(c4);
\draw[edge] (c4) to  node[edgelabel] {$c$}(c5);
\draw[edge] (c5) to  node[edgelabel] {$c$}(c2);
\draw[edge] (c2) to  node[edgelabel] {$c$}(c3);
\draw[edge] (c3) to  node[edgelabel] {$c$}(c6);
\draw[edge] (c6) to  node[edgelabel] {$c$}(c9);
\draw[edge] (c9) to  node[edgelabel] {$c$}(c8);
\draw[edge] (c8) to  node[edgelabel] {$c$}(c7);

\draw[edge] (time) to[out=120, in =0]  node[edgelabel] {$c$}(update);
\draw[edge] (update) to[out=-40,in=-150]  node[edgelabel] {$c$}(time);
\draw[edge] (update) to[out=-45,in=-130]  node[edgelabel] {$d$}(time);

% out
\node (o) at (0,3.25) {out};
\node[draw,minimum height=7cm,minimum width=15cm,rounded corners] (out) at (7,0) {};

% alarm 1 + update1
\node (a1) at (4.2,2.25) {alarm 2};
\node[draw,minimum height=5cm,minimum width=2cm,rounded corners] (alarm1) at (4.5,0) {};
\node (off1) [state,init] at (4.5,1) {off};
\node (on1) [state] at (4.5,-1) {on};
\draw[edge] (off1) to[out=-60,in=60]  node[edgelabel] {$d$}(on1);
\draw[edge] (on1) to[out=120,in=-120]  node[edgelabel] {$d$}(off1);
\node (u1) at (6.8,2.25) {update 2};
\node[draw,minimum height=5cm,minimum width=2cm,rounded corners] (update1) at (7,0) {};
\node (x1) [state,init] at (7,1.5) {hr};
\node (x2) [state] at (7.5,0) {1mn};
\node (x3) [state] at (7,-1.5) {10mn};
\draw[edge] (x1) to  node[edgelabel] {$c$}(x2);
\draw[edge] (x2) to  node[edgelabel] {$c$}(x3);
\draw[edge] (alarm1) to[out=80,in=100]  node[edgelabel] {$c$}(update1);
\draw[edge] (update1) to[out=-105,in=-75]  node[edgelabel] {$c$}(alarm1);
\draw[edge] (update1) to[out=-95,in=-85]  node[edgelabel] {$b$}(alarm1);

% alarm 2 + update2
\node (a2) at (10.2,2.25) {alarm 1};
\node[draw,minimum height=5cm,minimum width=2cm,rounded corners, initial above, initial text={}] (alarm2) at (10.5,0) {};
\node (off2) [state,init] at (10.5,1) {off};
\node (on2) [state] at (10.5,-1) {on};
\draw[edge] (off2) to[out=-60,in=60]  node[edgelabel] {$d$}(on2);
\draw[edge] (on2) to[out=120,in=-120]  node[edgelabel] {$d$}(off2);
\node (u2) at (12.8,2.25) {update 1};
\node[draw,minimum height=5cm,minimum width=2cm,rounded corners] (update2) at (13,0) {};
\node (y1) [state,init] at (13,1.5) {hr};
\node (y2) [state] at (13.5,0) {1mn};
\node (y3) [state] at (13,-1.5) {10mn};
\draw[edge] (y1) to  node[edgelabel] {$c$}(y2);
\draw[edge] (y2) to  node[edgelabel] {$c$}(y3);
\draw[edge] (alarm2) to[out=80,in=100]  node[edgelabel] {$c$}(update2);
\draw[edge] (update2) to[out=-105,in=-75]  node[edgelabel] {$c$}(alarm2);
\draw[edge] (update2) to[out=-95,in=-85]  node[edgelabel] {$b$}(alarm2);

% chime
\node (c) at (0.55,2.75) {chime};
\node[draw,minimum height=6cm,minimum width=2cm,rounded corners] (chime) at (1,0) {};
\node (off3) [state,init] at (1,1.5) {off};
\node (on3) [state] at (1,-1.5) {on};
\draw[edge] (off3) to[out=-60,in=60]  node[edgelabel] {$d$}(on3);
\draw[edge] (on3) to[out=120,in=-120]  node[edgelabel] {$d$}(off3);

\draw[edge] (time) to[out=-80,in=40]  node[edgelabel] {$a$}(out);
\draw[edge] (alarm2) to[out=115,in=65]  node[edgelabel] {$a$}(alarm1);
\draw[edge] (alarm1) to[out=115,in=65]  node[edgelabel] {$a$}(chime);
\draw[edge] (out) to[out=115,in=-90]  node[edgelabel] {$a$}(stop);
\draw[edge] (stop) to [out=-25,in=-115] node[edgelabel] {$a$}(time);

\end{tikzpicture}
    \caption{The `displays' component of the wristwatch HFSM example from Harel~\cite{harel1987statecharts}.}
    \label{fig:harel}
\end{figure}

As a modeller, we wish to understand the conceptual bottleneck of this system---which subystem is the most complex? Further, can this design be improved to a `more modular' one? This is an instance of the bottleneck problem. We can calculate the width of this HFSM by counting sizes of each component FSM, which informs us that the `update' FSM is the most complex, with 9 states, followed by the `stopwatch', `out' and `displays' subcomponents, each with 5 states. The latter three appear to be more complex than `update', which has few arcs and a simple path structure.

Using our results, we can approach this problem algorithmically. Running the modular decomposition on this HFSM allows us to construct a new simpler HFSM which is equivalent to the original (Figure~\ref{fig:better harel}). The `out' HFSM can be reduced to width 3 by grouping `chime', `alarm 2' and `update 2' into a single nested HFSM which we call `alarm2+chime'. The `displays' FSM can be reduced in width from 5 to 4, as the `out' and `stopwatch' FSMs together form a module. With this change, the `displays' subcomponent gains a very simple structure, consisting of a start node `time' and three HFSMs (`date', `update' and `out+stopwatch') with arcs to and from `time', accessed by the symbols $d$, $c$ and $a$ respectively. The `update' FSM, as a path, can be decomposed into nested FSMs of size 2 each, giving it width 2, though we omit this for clarity. Only the `stopwatch' component has no non-trivial modules, and retains a width of five. Thus, we have obtained a new \emph{strictly more modular} model of this system, which identifies the `stopwatch' FSM as the conceptual bottleneck for modularity. If we were to redesign this wristwatch, the  `stopwatch' component would be the obvious target for refactoring.

% names are up to the modeller

% gives a new, stritly better model of Harel's original wristwatch

\begin{figure}
    \centering
    \begin{tikzpicture}

% disaplys
\node (displays) at (0,10) {displays};
\node[draw,minimum height=14cm,minimum width=15.5cm,rounded corners] (displaybox) at (7,3.5) {};

% stopwatch
\node (stopwatch) at (.9,9.2) {stopwatch};
\node[draw,minimum height=5cm,minimum width=5cm,rounded corners] (stop) at (2.5,7) {};
\node (zero) [state,initial right, initial text={}] at (2.5,9) {zero};
\node (son) [state] at (1.5,7.5) {on};
\node (soff) [state] at (3.5,7.5) {off};
\node (lon) [state] at (1.5,5.5) {lap on};
\node (loff) [state] at (3.5,5.5) {lap off};
\draw[edge] (son) to[out=40,in=140]  node[edgelabel] {$b$}(soff);
\draw[edge] (soff) to[out=-140,in=-40]  node[edgelabel] {$b$}(son);
\draw[edge] (lon) to[out=40,in=140]  node[edgelabel] {$b$}(loff);
\draw[edge] (loff) to[out=-140,in=-40]  node[edgelabel] {$b$}(lon);
\draw[edge] (son) to[out=-60,in=65]  node[edgelabel] {$d$}(lon);
\draw[edge] (lon) to[out=115,in=-120]  node[edgelabel] {$d$}(son);
\draw[edge] (zero) to  node[edgelabel] {$b$}(son);
\draw[edge] (soff) to  node[edgelabel] {$d$}(zero);
\draw[edge] (loff) to  node[edgelabel] {$d$}(soff);

% regular
%\node (regular) at (6.7,10.75) {regular};
%\node[draw,minimum height=6cm,minimum width=8cm,rounded corners,initial right,initial text={}] (reg) at (10,8) {};
\node (time) [state,initial right, initial text={}] at (12.5,6.5) {time};
\node (date) [state] at (12.5,9.5) {date};
\draw[edge] (date) to[out=-60,in=60]  node[edgelabel] {$d$}(time);
\draw[edge] (time) to[out=120,in=-120]  node[edgelabel] {$d$}(date);

% update
\node (up) at (7.25,10.25) {update};
\node[draw,minimum height=5cm,minimum width=4cm,rounded corners] (update) at (8.5,8) {};
\node (c3) [state] at (7.2,9.5) {mon};
\node (c2) [state] at (8.5,9.5) {hr};
\node (c1) [state, initial above, initial text={}] at (9.8,9.5) {sec};
\node (c6) [state] at (7.2,8) {date};
\node (c5) [state] at (8.5,8) {10mn};
\node (c4) [state] at (9.8,8) {1mn};
\node (c9) [state] at (7.2,6.5) {day};
\node (c8) [state] at (8.5,6.5) {year};
\node (c7) [state] at (9.8,6.5) {mode};
\draw[edge] (c1) to  node[edgelabel] {$c$}(c4);
\draw[edge] (c4) to  node[edgelabel] {$c$}(c5);
\draw[edge] (c5) to  node[edgelabel] {$c$}(c2);
\draw[edge] (c2) to  node[edgelabel] {$c$}(c3);
\draw[edge] (c3) to  node[edgelabel] {$c$}(c6);
\draw[edge] (c6) to  node[edgelabel] {$c$}(c9);
\draw[edge] (c9) to  node[edgelabel] {$c$}(c8);
\draw[edge] (c8) to  node[edgelabel] {$c$}(c7);

\draw[edge] (time) to[out=120, in =0]  node[edgelabel] {$c$}(update);
\draw[edge] (update) to[out=-40,in=-150]  node[edgelabel] {$c$}(time);
\draw[edge] (update) to[out=-45,in=-130]  node[edgelabel] {$d$}(time);

% out
\node (o) at (9,3) {out};
\node[draw,minimum height=6.8cm,minimum width=14.5cm,rounded corners] (out) at (7,0) {};

\node (newout) at (8.15,2.25) {};
\node[draw,minimum height=6.4cm,minimum width=8.5cm,rounded corners] (smallout) at (4,0) {};

% stopwatch + out
\draw[rounded corners] (-0.5,-3.5) -- (-.5,9.8) -- (5.5,9.8) -- (5.5,4) -- (14.5,4) -- (14.5,-3.5) -- cycle;
\node (stopout) at (12.5,3.9) {};

\node at (1,4) {out + stopwatch};
\node at (3.4,-2.8) {alarm2+chime};

% alarm 1 + update1
\node (a1) at (4.2,2.25) {alarm 2};
\node[draw,initial above, initial text={},minimum height=5cm,minimum width=2cm,rounded corners] (alarm1) at (4.5,0) {};
\node (off1) [state,init] at (4.5,1) {off};
\node (on1) [state] at (4.5,-1) {on};
\draw[edge] (off1) to[out=-60,in=60]  node[edgelabel] {$d$}(on1);
\draw[edge] (on1) to[out=120,in=-120]  node[edgelabel] {$d$}(off1);
\node (u1) at (6.8,2.25) {update 2};
\node[draw,minimum height=5cm,minimum width=2cm,rounded corners] (update1) at (7,0) {};
\node (x1) [state,init] at (7,1.5) {hr};
\node (x2) [state] at (7.5,0) {1mn};
\node (x3) [state] at (7,-1.5) {10mn};
\draw[edge] (x1) to  node[edgelabel] {$c$}(x2);
\draw[edge] (x2) to  node[edgelabel] {$c$}(x3);
\draw[edge] (alarm1) to[out=80,in=100]  node[edgelabel] {$c$}(update1);
\draw[edge] (update1) to[out=-105,in=-75]  node[edgelabel] {$c$}(alarm1);
\draw[edge] (update1) to[out=-95,in=-85]  node[edgelabel] {$b$}(alarm1);

% alarm 2 + update2
\node (a2) at (10.2,2.25) {alarm 1};
\node[draw,minimum height=5cm,minimum width=2cm,rounded corners, initial above, initial text={}] (alarm2) at (10.5,0) {};
\node (off2) [state,init] at (10.5,1) {off};
\node (on2) [state] at (10.5,-1) {on};
\draw[edge] (off2) to[out=-60,in=60]  node[edgelabel] {$d$}(on2);
\draw[edge] (on2) to[out=120,in=-120]  node[edgelabel] {$d$}(off2);
\node (u2) at (12.8,2.25) {update 1};
\node[draw,minimum height=5cm,minimum width=2cm,rounded corners] (update2) at (13,0) {};
\node (y1) [state,init] at (13,1.5) {hr};
\node (y2) [state] at (13.5,0) {1mn};
\node (y3) [state] at (13,-1.5) {10mn};
\draw[edge] (y1) to  node[edgelabel] {$c$}(y2);
\draw[edge] (y2) to  node[edgelabel] {$c$}(y3);
\draw[edge] (alarm2) to[out=80,in=100]  node[edgelabel] {$c$}(update2);
\draw[edge] (update2) to[out=-105,in=-75]  node[edgelabel] {$c$}(alarm2);
\draw[edge] (update2) to[out=-95,in=-85]  node[edgelabel] {$b$}(alarm2);

% chime
\node (c) at (0.55,2.75) {chime};
\node[draw,minimum height=6cm,minimum width=2cm,rounded corners] (chime) at (1,0) {};
\node (off3) [state,init] at (1,1.5) {off};
\node (on3) [state] at (1,-1.5) {on};
\draw[edge] (off3) to[out=-60,in=60]  node[edgelabel] {$d$}(on3);
\draw[edge] (on3) to[out=120,in=-120]  node[edgelabel] {$d$}(off3);

\draw[edge] (time) to  node[edgelabel] {$a$}(stopout);
\draw[edge] (alarm2) to  node[edgelabel] {$a$}(newout);
\draw[edge] (alarm1) to[out=115,in=65]  node[edgelabel] {$a$}(chime);
\draw[edge] (out) to[out=115,in=-90]  node[edgelabel] {$a$}(stop);

\node (newstop) at (5.4,4.5) {};
\draw[edge] (newstop) to [out=0,in=-115] node[edgelabel] {$a$}(time);

\end{tikzpicture}
    \caption{After applying the modular decomposition, we can simplify Harel's model.}
    \label{fig:better harel}
\end{figure}

\section{Conclusions and Open Problems} \label{sec:conclusions}
In this paper we laid the foundations of the modular decomposition theory of HFSMs, and demonstrated its applicability to optimisation problems like minimising bottleneck of an HFSM. Given the minimal existing literature on this problem, our new concepts and results raise a number of interesting questions, which are deserving of future work. Two particular examples are:
\begin{enumerate}
    %\item (Thinness) Almost all results in this paper required the modules to be thin. How much can be recovered when modules are not thin? Is there another refinement of the module concept other than thinness which broadens the applicability of these results?
    %\item (Do modules occur ``in the wild"?) In random FSMs, non-trivial modules may well be rare. Are they still rare in FSMs that are used in practice, such as in model checking? We expect that the principle of divide and conquer in engineering design would naturally lead to modules, and so modules may be common in engineered FSMs.
    \item (DFAs) Because we did not consider `accepting sets' in this paper, we did not answer the question of how the modular decomposition relates to the language recognised by an automaton. This would be interesting to explore.
    %\item (The complexity of the modular decomposition.) We conjecture that the running time of the modular decomposition can be improved from $O(n^2k)$ to $O(nk)$.
    \item (Compressing HFSMs) As we discussed in Section~\ref{sec:relatedwork}, in this paper we we treated all states as distinct, and so the nesting in HFSMs was a tree. If we allowed subHFSMs to be equivalent, we would instead require only that this relationship define a dag. This allows us to construct potentially exponentially smaller HFSMs, but the trade-off is a significant increase in computational complexity.%increases the potential to compress the HFSM's representation, but comes at the cost of a significant increase in complexity. However it may be that there are special cases or heuristics which are useful in practice.
    %In this paper, we assumed all states in an FSM were distinct. If we have an equivalence relation on states (such as membership in an accepting set, like that on DFAs) then HFSMs can be `compressed' by merging identical nested HFSMs. As shown in \cite{alur1998model}, this allows HFSMs to be represented in logarithmic size, and model checking on such HFSMs can be done on the compressed representation. `Compressing' an HFSM consists of merging identically labelled subtrees in the nesting tree. By Theorem~\ref{basis theorem}, each basis module is associated with a contracted form FSM, so the modular decomposition could provide a starting point for compressing HFSMs by merging contracted forms. While we suspect (via a reduction to the \emph{smallest grammar problem}~\cite{charikar2005smallest}) that finding the `most compressed' HFSM is NP-hard, heuristic approaches may be useful in practice.
\end{enumerate}
Both of these tasks are potentially very difficult---as we mentioned in Section~\ref{sec:relatedwork}, compressing HFSMs is very likely NP-hard. However, this might be overly pessimistic; the graphs used in practice may have underlying structure which allows for effective heuristic approaches. Our results allow both of these questions to be tackled in future work.
%\section*{Acknowledgements}This work is partially supported by Defence Science and Technology Group, through agreement MyIP: ID10266 entitled ``Hierarchical Verification of Autonomy Architectures'' and the Australian Government, via grant AUSMURIB000001 associated with ONR MURI grant N00014-19-1-2571.

\bibliographystyle{plain}
\bibliography{references}

\appendix
\section{Proofs}\label{app_proof:modules}

\begin{thm}[Theorem~\ref{flat equivalence}]
If $Z$ and $Y$ are accessible HFSMs, $Z^F$ and $Y^F$ are unique and $Z\cong Y$ if and only if $Z^F = Y^F$.
\end{thm}
\begin{proof}
\noindent\emph{Claim: for accessible FSMs $U$ and $V$, $U\cong V \iff U = V$.} 

If $U=V$, then clearly $U\cong V$. Now suppose $U\neq V$. There must exist a state $q$ and symbol $x\in \Sigma$ such that $\delta_U(q,x)\neq \delta_V(q,x)$. Let $w\in \Sigma^*$ be a word such that $\varphi_U(w) = q$ or $\varphi_V(w) = q$. Such a $w$ exists because both $U$ and $V$ are accessible. Assume w.l.o.g that $\varphi_U(w) = q$. If $\varphi_V(w)\neq q$, then we are done and $U\not\cong V$. Otherwise $\varphi_U(wx)=\delta_U(q,x)\neq\delta_V(q,x)=\varphi_V(wx)$, so $U\not\cong V$.

\noindent\emph{Claim: let $Z$ be an HFSM with an FSM $Y$ nested in it. Then the HFSM $\hat Z$ given by expanding $Y$ in $Z$ is equivalent to $Z$.}

Specifically, if $Z=(X,T)$, $\hat Z := ((X\setminus \{Y\})\cup \{W\cdot_w Y\},T')$, where $\arc[w]{W}{Y}$ is the arc to $Y$ in $T$. The nesting tree $T'$ of $\hat Z$ differs from $T$ by identifying nodes $W$ and $Y$ into $W\cdot_w Y$, with all children of $Y$ now children of $W\cdot_w Y$.
%is defined by
%\[
%\arc[v]{a}{b}\in T' \iff \begin{cases}
%a=W\cdot_w Y, \arc[v]{W}{b}\in A(T) \\
%a=W\cdot_w Y, \arc[v]{Y}{b}\in A(T) \\
%a\neq W\cdot_w Y, \arc[v]{a}{b}\in A(T)
%\end{cases}
%\]
We show that the hierarchical transition function $\psi_Z$ and $\psi_{\hat Z}$ are equal. Let $x$ be a symbol and $q$ a state of $Z$ and $\hat Z$ (noting first that they have the same state set). First, suppose $q\in Q(Y)$. There exists a sequence of FSMs $K_1,\dots,K_n$ in $Z$ and $\hat Z$ such that $q,Y,W,K_1,\dots,K_n$ and $q,W\cdot_w Y,K_1,\dots,K_n$ is the sequence of FSMs from $q$ to the root in $T$ and $T'$ respectively. If $\delta_Y(q,x)$ exists, then $\delta_{W\cdot_wY}(q,x) = \delta_Y(q,x)$ by Def.~\ref{defn:expansion} and so $\psi_Z(q,x) = \start(\delta_Y(q,x)) = \start(\delta_{W\cdot_wY}(q,x)) = \psi_{\hat Z}(q,x)$. If $\delta_Y(q,x)$ does not exist but $\delta_W(w,x)$ does, then by Def.~\ref{defn:expansion} $\delta_{W\cdot_wY}(q,x) = \delta_W(w,x)$ and so again $\psi_Z(q,x) = \start(\delta_W(w,x)) = \psi_{\hat Z}(q,x)$. If neither exists, then $\delta_Y(q,x)$ and $\delta_{W\cdot_wY}(q,x)$ do not exist and so $\psi_Z(q,x) = \psi_{\hat Z}(q,x)$.

Now suppose that $q\not\in Q(Y)$. If $W$ is not on the path from the root to $q$ in $Z$ then $\psi_Z(q,x) = \psi_{\hat Z}(q,x)$. Assume it is, and we find that again there are FSMs $K_1,\dots,K_n$ in $Z$ and $\hat Z$ such that \[
\begin{tikzcd}
q & \ar[l,"q"] K_1 & \ar[l,"k_1"] \dots & \ar[l,"\mu"] W\text{ or }W\cdot_wY & \ar[l] \dots & \ar[l,"k_n"] K_n
\end{tikzcd}\] where the only difference in these paths is that the $j$-th item on this path is either $W$ or $W\cdot_w Y$ for $Z$ and $\hat Z$ respectively. If $\delta_{K_i}(k_{i-1},x)$ exists for any $i<j$, then $\psi_Z(q,x) = \psi_{\hat Z}(q,x)$. Otherwise, observe that by Def.~\ref{defn:expansion} $\delta_W(\mu,x)$ exists if and only if $\delta_{W\cdot_wY}(\mu,x)$ exists, and if neither exists then $\psi_Z(q,x)$ cannot differ from $\psi_{\hat Z}(q,x)$. If they exist, then either
$\delta_W(\mu,x) \neq v \implies \delta_{W\cdot_wY}(\mu,x)=\delta_W(\mu,x) \implies \psi_Z(q,x) = \start(\delta_W(\mu,x)) = \psi_{\hat Z}$ \emph{or} $\delta_W(\mu,x) = v \implies \delta_{W\cdot_wY}(\mu,x)=s_Y$.  But then $\psi_Z(q,x) = \start(v) = \start(s_Y) = \start(\delta_{W\cdot_wY}(\mu,x)) = \psi_{\hat Z}(q,x)$ by the recursive definition of the start function.

\noindent\emph{Claim: For an HFSM $Z$, $Z^F\cong Z$, and $Z^F$ is unique.}

There exists a sequence of expansions of FSMs in $Z$ which take $Z$ to $Z^F$. By the previous claim, it follows that $Z^F\cong Z$. Then because flat FSMs are unique up to equivalence (the first claim) $Z^F$ is unique.

\noindent\emph{Claim: For HFSMs $Z$ and $Y$, $Z\cong Y\iff Z^F=Y^F$.}

Since $Z\cong Z^F$ and $Y\cong Y^F$, then $Z\cong Y\implies Z^F\cong Z\cong Y\cong Y^F \implies Z^F=Y^F$. Likewise, $Z\not \cong Y\implies Z^F\cong Z\not \cong Y\cong Y^F \implies Z^F\neq Y^F$.
\end{proof}

\begin{thm}[Theorem~\ref{module theorem}]
 Let $Z$ be a graph/FSM. A non-empty set $M\subseteq Z$ is a \emph{module} if and only if $Z/M\cdot_M Z[M] \cong Z$.
\end{thm}
\begin{proof}
First, we prove this for graphs. Observe first that the node sets of $Z$ and $Z/M\cdot_M Z[M]$ are the same. Suppose $M$ is a module. We show that for any node $v\in Z$, its neighbourhood in $Z$ and $Z/M\cdot_M Z[M]$ is the same, so these graphs are isomorphic. For any $v$, let $N_Z(v)$ be its neighbourhood in $Z$, $N_M(v)$ be its neighbourhood in $Z[M]$, and let $N_{M^C}(v)$ be its neighbourhood in $Z\setminus M$. If $v\not\in M$ is not adjacent to any nodes in $M$, then its neighbourhood is the same in $Z$ and $Z/M\cdot_M Z[M]$. Now suppose $v\in M$. By Definition~\ref{concrete graphs}, $N_{Z/M}(M) = N_{M^C}(v)$, and $N_{Z/M\cdot_M Z[M]}(v) = N_{Z/M}(M)\cup N_{Z[M]}(v)$ by definition of expansion, so $N_{Z/M\cdot_M Z[M]}(v) = N_{M^C}(v)\cup N_{Z[M]}(v) = N_Z(v)$ as required. Finally, suppose $v$ is adjacent to a node in $M$. Then by Definition~\ref{concrete graphs}, $v$ is adjacent to all of $M$. Now, $N_Z(v) = N_{M^C}(v)\cup N_M(v)$. Then $N_{Z/M}(v) = N_{M^C}(v)\cup \{M\}$ and $N_{Z/M\cdot_M Z[M]}(v) = N_{M^C}(v)\cup M = N_Z(v)$, again by definition of expansion. Now we show that if $M$ is not a module, then these graphs are not isomorphic. If $M$ is not a module, then there exists a $k$ which is adjacent to $u\in M$ and nonadjacent to $v\in M$. However, as $k$ is adjacent to some of $u$ is it adjacent to $M$ in $Z/M$, and so is adjacent to all of $M$ in $Z/M\cdot_M Z[M]$, and thus $Z/M\cdot_M Z[M]\not \cong Z$.

Second, we prove this for FSMs.
We call this property (\ddag). First observe that the state sets $Q(Z)$ and $Q(Z/M\cdot_M Z[M])$ are the same. 

\noindent ($M$ is a module $\implies$ $Z/M\cdot_M Z[M]\cong Z$)

Let $\alpha$ be the start state of $M$.
\begin{itemize}
    \item $u, v\not\in M$: $\arc[x]{u}{v}$ is an arc in $Z/M\cdot_M Z[M]$ iff $\arc[x]{u}{v}$ is an arc in $Z/M$ iff $\arc[x]{u}{v}$ is an arc in $Z$.
    \item $u\not\in M$, $v\in M$: Then $v=\alpha$ as $M$ is a module. Thus, if $\arc[x]{u}{v}$ is an arc in $Z$ then $\arc[x]{u}{M}$ is an arc in $Z/M$ and thus $\arc[x]{u}{\alpha}=\arc[x]{u}{v}$ is an arc in $Z/M\cdot_M Z[M]$. Similarly, $\arc[x]{u}{v}=\arc[x]{u}{\alpha}$ in $Z/M\cdot_M Z[M]$ means that $\arc[x]{u}{M}$ in $Z/M$, which implies there is an $x$-arc from $u$ into $M$ in $Z$, but all such arcs go to $\alpha=v$.
    \item $u\in M$, $v\not\in M$: Let $\arc[x]{u}{v}$ be in $Z$. The state $u$ has only one $x$-arc out of it in $Z$, so it has no $x$-arc in $Z[M]$ as $v\not\in M$. Since $\arc[x]{M}{v}$ is an arc in $Z/M$, we know in $Z/M\cdot_M Z[M]$ all states in $Z[M]$ without $x$-arcs now have arcs to $v$ (definition of expansion), so there is an arc $\arc[x]{u}{v}$. If $\arc[x]{u}{v}$ is in $Z/M\cdot_M Z[M]$, then there is an arc $\arc[x]{M}{v}$ in $Z/M$ and $u$ cannot have an $x$-arc in $Z[M]$. Now $\arc[x]{M}{v}$ implies that there is a state $k\in M$ with $\arc[x]{k}{v}$ in $Z$. If so, by Definition~\ref{concrete modules} every state in $M$ has an $x$-arc, including $u$, which is either internal to $M$ or to $v$. $u$ has no $x$-arc in $Z[M]$ so its $x$-arc in $Z$ must be outside of $M$, so there is an arc $\arc[x]{u}{v}$ in $Z$.
    \item $u\in M$, $v\in M$: If $\arc[x]{u}{v}$ in $Z$, then $\arc[x]{u}{v}$ in $Z[M]$, and so $\arc[x]{u}{v}$ in $Z/M\cdot_M Z[M]$. If $\arc[x]{u}{v}$ in $Z/M\cdot_M Z[M]$, then $\arc[x]{u}{v}$ in $Z[M]$, and so $\arc[x]{u}{v}$ in $Z$.
\end{itemize}

\noindent($M$ is not a module $\implies$ $Z/M\cdot_M Z[M]\not\cong Z$)

Let $M$ be a set with two distinct entrances. These cannot both be the start state of $Z[M]$, so assume w.l.o.g. that $v$ is not the start state of $Z[M]$. No arcs into $M$ in $Z/M\cdot_M Z[M]$ from outside $M$ can go to $v$, as by definition of expansion they all go to the start state of $Z[M]$. Thus the arc $\arc[x]{u}{v}$ cannot exist in $Z/M\cdot_M Z[M]$ and thus $Z/M\cdot_M Z[M]\not\cong Z$. Similarly suppose $M$ has two distinct $x$-exits $v$ and $w$. But then in $Z/M$ the state $M$ has two $x$-arcs, so this is not an FSM, and certainly $Z/M\cdot_M Z[M]\not\cong Z$.
\end{proof}

\begin{thm}[Theorem~\ref{gallai}]
Let $Z$ be an FSM, $X$ a module, and $Y\subseteq X$ a subset of $X$. Then $Y$ is a module of $Z$ if and only if it is a module of $Z[X]$.
\end{thm}
\begin{proof}
Suppose $Y$ is a module of $Z$. In $Z[X]$ there is still one start state. Let $\arc[z]{a}{b}$ be an arc in $Z[X]$ with $a\in Y$ and $b\in X\setminus Y$. Then this arc exists in $Z$ and so as $Y$ is a module we conclude every state in $Y$ has a $z$-arc within $Y$ or to $b$. All of these arcs exist in $Z[X]$, so $Y$ is a module of $Z[X]$. For the converse, observe first that any arc into $X$ from $Z\setminus X$ goes to $X$'s start state. If this is also the start state of $Y$ in $Z[X]$, then all arcs into $Y$ in $Z$ go to a single state. Otherwise, all arcs into $Y$ must come from $X\setminus Y$, and all such arcs go to $Y$'s start state in $Z[X]$, so $Y$ must have one start state in $Z$. Now consider an arc $\arc[z]{a}{c}$ out of $Y$ in $Z$. If $c\in X\setminus Y$, then all states in $Y$ have $z$-arcs to $x$ or within $Y$, as $Y$ is a module of $Z[X]$. If $x\in Z\setminus X$, then as $X$ is a module every state in $X$ has an $z$-arc either within $X$ or to $c$. Thus all states in $Y$ have an $z$-arc either to $c$ or within $Y$, and so $Y$ is a module of $Z$.
\end{proof}

\begin{thm}[Theorem~\ref{reverse gt}]
Let $Z$ be an FSM and $X$ a module, and $Y$ a superset of $X$. If $X$ is thin, then $Y$ is a thin module of $Z$ if and only if $Y/X$ is a thin module of $Z/X$.
\end{thm}
\begin{proof}
Suppose $Y$ is a thin module of $Z$. $Y/X$ must still contain one entrance. Similarly, suppose $Y$ has a $z$-exit. Then $Y/X$ also has exactly one $z$-exit. In this case every state in $Y$ has a $z$-path to the exit, by thinness. Thus $X$ every state has an $z$-path to its $z$-exit, and so $Y/X$ also has every state with a $z$-path to its $z$-exit. Thus $Y/X$ is a thin module. Now suppose $Y/X$ is a thin module of $Z/X$. Again $Y$ must have a single entrance, because $Y/X$ and $X$ both have a single entrance. Now, any $z$-exit of $Y$ is a $z$-exit of $Y/X$, so $Y$ has at most one. If there is a $z$-exit in $Y/X$, then every state has a $z$-path to that exit, including the state $X$. As a thin module, all states in $X$ have a $z$-path to $X$'s $z$-exit, which is either equal to $Y$'s $z$-exit or has a $z$-path to it. Thus all states in $Y$ have $z$-paths to the exit, because that is true of $Y/X$ and $X$. Thus $Y$ is a thin module of $Z$.
\end{proof}

%\section{Proofs: Section~\ref{sec:modular decomp}}\label{app_proof:modular_decomp}

%We will build the results required to prove Theorem~\ref{basis theorem}. Firstly, we state the following definition. We will write $\Ind(Z)$ for the set of indecomposable thin modules of an FSM $Z$.

%\begin{defn} 
%Let $F$ be a family of subsets of a finite set $X$. An element $M\in F$ is \emph{strong}~\cite{habib2010survey} if no other element overlaps it.\todo{remove}
%\end{defn}

\begin{lem} \label{paths through start node}
Let $M$ be a module with start state $s$ in a FSM $Z$, with $m\in M$ and $v\not\in M$. Every path $\pth{v}{m}$ contains $s$.
\end{lem}
\begin{proof}
Consider any path $\pth{v}{m}$. $v\not\in M$, $m\in M$, so there exists a pair of states $\arc{u}{w}$ on this path with $u\not\in M$ and $w\in M$. All arcs into $M$ go to $s$ by Theorem~\ref{concrete modules}, so $w = s$.
\end{proof}

\begin{lem} \label{modules connected}
If $M$ is a module of an accessible FSM $Z$. All states in $M$ are reachable from $M$'s start state $s$.
\end{lem}
\begin{proof}
$Z$ is accessible, so there is a path $\pth{g}{m}$ from the start state $g$ of $Z$ to any $m\in M$. If $g\in M$, then $g=s$ and we are done. If $g\not\in M$, then $s$ is on any path $\pth{g}{m}$ by Lemma~\ref{paths through start node}, so we have a subpath $\pth{s}{m}$.
\end{proof}

%\begin{lem} \label{paths out of modules}
%If $M$ is a thin module in an FSM $Z$ with $v$ an $x$-exit of $M$, then every state in $M$ has an $x$-path to $v$.
%\end{lem}
%\begin{proof}
%Every state in $M$ has an $x$-arc, and $M$ has a unique $x$-exit. The $x$-path out of any state is unique. Start at a state $q\in M$. Let $q_x$ be the $x$-successor of $q$. If this is outside $M$, then $q_x=v$ as there is only one $x$-exit. If $q_x$ is in $M$, then repeat the argument. Because there are no $x$-cycles in $M$, the $x$-path out of every state must take finitely many steps to reach an $x$-exit of $M$.
%\end{proof}

%\begin{lem} \label{start node in intersection}
%Let $Z$ be accessible, and let $A$ and $B$ be modules with non-empty intersection, and $\alpha$, $\beta$ their start nodes. Then either $\alpha=\beta$ or exactly one of $\alpha$ or $\beta$ is in $A\cap B$.
%\end{lem}
%\begin{proof}
%Suppose $\alpha,\beta\not\in A\cap B$. But then there are non-repeating paths $\pth{\alpha}{v}$ and $\pth{\beta}{v}$ entirely within $A$ and $B$ respectively by Lemma~\ref{modules connected}, but as $v\in B$, $\beta$ is on $\pth{\alpha}{v}$ by Lemma~\ref{paths through start node}, but then the subpath $\pth{\beta}{v}$ must contain $\alpha$, which leads to a contradiction as both paths cannot contain each other as $\alpha\neq\beta$. Thus at least one start node is in the intersection. If both are in the intersection then they must be identical, as at least one is either the global start node or receives an arc from outside the module, by accessibility.
%\end{proof}

\begin{lem} \label{unions}
Let $Z$ be an FSM, and $A$ and $B$ overlapping thin modules. Then $A\cup B$ is a thin module.
\end{lem}
\begin{proof}
Let $\alpha$ and $\beta$ be the start states of $A$ and $B$ respectively, and let $v$ be in $A\cap B$. Then there is a path $\pth{\alpha}{v}$ within $A$. If $\alpha\not\in B$, then by Lemma~\ref{paths through start node} this path passes through $\beta$, so $\beta\in A$. Likewise, if $\beta\not\in A$ then every path into $A\cap B$ passes through $\alpha$, and so we conclude that either $\alpha$ or $\beta$ is in $A\cap B$. One of these must receive an arc from outside $A\cup B$ or must be the start state of $Z$, by accessibility, and so if both $\alpha$ and $\beta$ are in $A\cap B$ then $\alpha=\beta$. Now assume w.l.o.g. that $\beta$ is the start state of $A\cap B$. Any arc into $A\cup B$ goes to $\alpha$ or $\beta$, and since $\beta\in A$ all arcs must go to $\alpha$ ($\alpha$ and $\beta$ may be the same state). Now consider an arc $\arc[x]{u}{w}$, with $u\in A\cup B$ and $w\not\in A\cup B$. If $u\in A\cap B$ then by thinness all $x$-arcs out of both $A$ and $B$ go to $w$. If $u\in A\setminus B$ then all $x$-arcs out of $A$ go to $w$ by thinness. The state $v$ is in $A$, so there is a path $\pth[x]{v}{w}$ within $A$. However $v\in B$, so again by thinness every state $b\in B$ has a path $\pth[x]{b}{w}$ which is within $A\cup B$, and so all $x$-arcs out of $A\cup B$ must go to $w$. The $B\setminus A$ case is similar. Thus $A\cup B$ is a module.
\end{proof}
\begin{lem} \label{start node in intersection}
    Let $A_1,A_2,\dots,A_n$ be an overlapping (Def.~\ref{overlapping sets}) collection of thin modules. Then $H = A_1\cup\dots\cup A_n$ is a thin module and the start state of $H$ is the unique state in $H$ that is the start state of every $A_i$ which contains it.
\end{lem}
\begin{proof}
    Suppose $A$ overlaps $B$. By Lemma~\ref{unions}, $A\cup B$ is a module and its start state is either the start state of both modules or is only contained in one. This is the base case. Now assume that the overlapping union $A_1\cup \dots \cup A_m$ is a thin module, and $B$ is a thin module overlapping $A_i$ in this union. Let $\alpha$ and $\beta$ be the start states of $A_1\cup \dots \cup A_m$ and $B$ respectively. By the inductive hypothesis, $\alpha$ is the only state in $A_1\cup \dots \cup A_m$ which is the start state of all $A_i$ which contain it in this union. Suppose that $\beta\not\in A_1\cup \dots \cup A_m$. Then by Lemma~\ref{unions}, $B$ overlaps this module, and $B\cup A_1\cup \dots \cup A_m$ is a thin module whose start state is $\beta$, which is contained in only $B$, while $\alpha$ must be in $B$ but not the start state, so now $\beta$ is the only state that is the start state of all modules which contain it. Otherwise, $\beta\in B\cup A_1\cup \dots \cup A_m$. In this case, if $\alpha\in B$ then $\alpha=\beta$, and otherwise $\beta$ is contained in some $A_i$ where it is not the start state, by the inductive hypothesis. In both cases, $B\cup A_1\cup \dots \cup A_m$ because if $B$ doesn't overlap $A_1\cup \dots \cup A_m$ it is contained within it. $\alpha$ is the start state of $B\cup A_1\cup \dots \cup A_m$ and is the only state that is the start state of all modules which contain it in this union. By the definition of overlapping sets, for any $A_i$ and $A_j$ there is a path $A_{a_1},A_{a_2},\dots,A_{a_n}$ where each overlaps the next pairwise, so we can iteratively add all modules to this union, completing the proof by induction.
\end{proof}
%\begin{lem} 
%Let $Z$ be accessible, $A_1,A_2,\dots,A_n$ overlapping thin modules, and let $H = A_1\cup\dots\cup A_n$. Then $H$ is a thin module 
%\end{lem}
%\begin{proof}
%By Lemma~\ref{overlapping unions}, $A_1\cup\dots\cup A_n$ is a module. We assume without loss of generality that the modules are ordered such that $A_i$ overlaps or is contained in $A_1\cup A_2\cup\dots \cup A_{i-1}$. Let $\alpha$ be the start state of $H$, and assume w.l.o.g. that $\alpha$ is the start state of $A_1$ and these modules are ordered such that $A_i$ overlaps $A_{i+1}$ for each $i$. The start state of $A_1\cup A_2$ must be $\alpha$, and by Lemma~\ref{unions} all other states are not the start state of some $A_i$ that contains them. This is the base case. If this holds for $A_1\cup\dots\cup A_i$, then again by Lemma~\ref{unions} $\alpha$ is the start state of $A_1\cup\dots\cup A_i\cup A_{i+1}$ and $\alpha$ is the unique state which is the start state of each $A_j$ containing it. By induction, the result holds.
%\end{proof}

\begin{lem} \label{intersections}
Let $A$ and $B$ be overlapping thin modules in an accessible FSM $Z$. Then $A\cap B$ is a module.
\end{lem}
\begin{proof}
Suppose $v\in A\cap B$, and let $\alpha$ and $\beta$ be the start states of $A$ and $B$. By Lemma~\ref{start node in intersection} one of these is in the intersection, so assume w.l.o.g. that $\beta$ is in $A\cap B$. Any arc into $A\cap B$ is in arc into $B$, so must go to $\beta$. Any arc $\arc[x]{A\cap B}{v}$ is either an arc out of $A\cup B$, in which case every state in $A\cup B$ has an $x$-arc and $v$ is the unique $x$-exit, or $v\in B\setminus A$ or $A\setminus B$. In the first case, this is an arc out of $A$, and $A$ is thin, so $A\cap B$ is thin and there is an $x$-arc from every state and $v$ is the only $x$-exit. The second case is identical. Thus $A\cap B$ is a module.
\end{proof}

\begin{thm}[Theorem~\ref{Kq theorem}]
Let $q$ be a state in an FSM $Z$ that is not the start state. Define $\repr_Z(q)$ as the intersection of all thin modules $M$ which contain $q$ but where $q$ is not the start state.
%\[
%\repr_Z(q) := \bigcap_{\substack{M\text{ thin module} \\q\in M\\q\text{ not start node of $M$}}} M
%\]
Then $\repr_Z(q)$ is a basis module, and for each basis module $H$ there exists a $q$ such that $\repr_Z(q)=H$.
\end{thm}
\begin{proof}
We begin by proving the statement for an FSM $Z$, then we extend it to HFSMs. First, observe that for a given $q$, $Q(Z)$ is a module containing $q$ where it is not the start state, so this intersection is always non-empty. For any two modules $A$, $B$ containing $q$ as the start state of neither, by Lemma~\ref{start node in intersection} we know that $q$ is not the start state of $A\cap B$. Thus $\repr(q)$ is a thin module by Lemma~\ref{intersections} and is not a singleton because $q$ is not its start state. 

\noindent\emph{Claim: $\repr(q)$ is indecomposable.}

Suppose for contradiction that $A_1,\dots,A_n$ are overlapping modules with $\repr(q)=A_1\cup\dots\cup A_n$. By Lemma~\ref{start node in intersection}, the start state of $\repr(q)$ is the unique state which is the start state of every $A_i$ which contains it. $q$ is not the start state of $\repr(q)$, so there exists an $A_j$ which contains $q$ with $q$ not its start state, but then by definition of $\repr(q)$, $\repr(q)\subseteq A_j$, which contradicts our assumption that $A_1,\dots,A_n$ are overlapping.

\noindent\emph{Claim: For every basis module $M$, there exists a $q$ such that $\repr(q)=M$.}

Let $\alpha$ be the start state of a basis module $M$, and let $G_1,\dots,G_m$ be those modules whose start state is in $M$ but is not $\alpha$, and let $H_1,\dots,H_n$ be all the indecomposable modules with start state $\alpha$ that do not properly contain $M$ (and are not equal to $M$). If there is a state $q \neq \alpha $ in $M$ but not any $H_i$ or $G_j$, then $M$ must be the smallest indecomposable module containing it where it is not the start state, so $\repr(q)=M$ and we are done. If no such state exists, then
\[
M\subseteq \bigcup_i H_i \cup \bigcup_j G_j
\]
but then
\[
M= \bigcup_i (M\cap H_i) \cup \bigcup_j (M\cap G_j)
\]
Every $M\cap H_i$ contains $\alpha$, so they all overlap each other. However, as $M$ is indecomposable, there is a collection of overlapping modules $M\cap G_{j_1},\dots,M\cap G_{j_m}$ which do not contain $\alpha$ and do not overlap any other $M\cap H_i$ or $M \cap G_j$. The union of these are a module by Lemma~\ref{unions}, and its start state (we call $q$) is the start state of any $M\cap G_{j_i}$ which contains it (Lemma~\ref{start node in intersection}). As $q$ is also contained in no $H_i$ we conclude that $\repr(q)=M$.

For the HFSM case, observe that every HFSM module $M$ corresponds to a module $H$ in some constituent FSM $X_i$. Hence a module in an HFSM $Z$ is a basis module if the associated module is a basis module of $X_i$. Let $H$ be a basis module in $X_i$, with $M$ the associated module of $Z$, and let $q\in Q(X_i)$ be a state with $\repr_{X_i}(q) = H$. If $q$ does not contain a nested FSM in $Z$, then it is a state of $Z$ and any module in $Z$ containing $q$ where it is not the start state either contains all of $Q(X_i)$ or corresponds to a module of $X_i$ which contains $q$ not the start state. Because $\repr_{X_i}(q) = H$ we get $\repr_Z(q) = M$. Otherwise, if an FSM $X_j$ is nested in $X_i$ at $q$. Then $\start(X_j)$ is a state of $Z$, and as the start state of $X_j$ it is the start state of all modules which contain it in $X_j$, and so $\repr_Z(q)$ is equal to the intersection of modules of $X_i$ which contain $q$, so again $\repr_Z(q) = M$.
\end{proof}

\begin{prop} \label{set bases}
Let $F$ be a family of subsets of a finite set $X$. If $F$ is closed under unions of overlapping elements, then a set is in $F$ if and only if it is a union of overlapping indecomposable elements of $F$.
\end{prop}
\begin{proof}
One direction is trivial: by closure, overlapping unions of indecomposable elements are in $F$. For the other direction, we only need to show that decomposable elements can be formed as unions of only indecomposable ones. Suppose that $A$ overlaps $B$ and $A\cup B$ overlaps $C$. Without loss of generality, $A\cap B\neq \emptyset$, so either $A\subseteq C$, in which case $C$ overlaps $B$, or $A$ overlaps $B$. In either case, $\{A,B,C\}$ is an overlapping collection. If a set is decomposable it is a union of strictly smaller sets, and so by induction we can form each $M$ as a union of indecomposable elements. Using the above argument inductively, this collection of sets must be overlapping, completing the proof.
\end{proof}

\begin{prop}[Unique decomposition] \label{decomposition theorem}
Let $M$ be a thin module. Then there exists a unique set of overlapping basis modules $A_1,\dots,A_n$ where $M = A_1\cup \dots\cup A_n$ and $A_i\not\subseteq A_j$ for any $i$ and $j$.
\end{prop}
\begin{proof}
Let $M$ be a thin module. We call a basis module $H$ a \emph{maximal $M$-module} if $H\subseteq M$ and there does not exist a basis module $K$ with $H\subset K \subseteq M$. By Proposition~\ref{set bases} and Lemma~\ref{unions}, $M$ is the union of all the maximal $M$-modules. We show that these are an overlapping collection, and any other overlapping collection whose union is $M$ must include all maximal $M$-modules.
\emph{Claim}: if $K,H_1,\dots,H_n$ are basis modules, and $K\subseteq H_1\cup\dots\cup H_n$, then $K\subseteq H_i$ for some $i$.
Suppose $K\subseteq H_1\cup\dots\cup H_n$.
%We will show that $M$ must be in collection of overlapping modules whose union is $H$. Let $A_1\cup\dots\cup A_h = H$ where $A_i$ are indecomposable modules, where $A_i\not\subseteq A_j$ (Lemma~\ref{set bases}). Let $A_1\cup\dots\cup A_n$ be a subset of overlapping modules such that $M\subseteq A_1\cup\dots\cup A_n$, and we assume this subset is minimal, so that removing any module either results in them being non-overlapping or not covering $M$.
Then $(H_1\cap K)\cup \dots \cup (H_n\cap K) = K$. If $H_i\cap K$ is non-empty, then it is a thin module (Lemma~\ref{intersections}). We show this is an overlapping collection of modules. Let $\alpha$ be the start state of $H_1\cup\dots\cup H_n$ (Lemma~\ref{unions}) and $\mu$ be the start state of $K$. Let $m$ be in $K$, and let $\pth{\alpha}{m}$ be a non-repeating path from $\alpha$ to $m$ within $K$, which must exist by Lemma~\ref{modules connected}. By Lemma~\ref{paths through start node}, there exists a subpath $\pth{\mu}{m}$ within $K$. Consider any arc $\arc{a}{b}$ on this path. By Lemma~\ref{start node in intersection}, there exists a module $H_k$ containing $b$ of which $b$ is not the start state. As there is an arc from $a$ to $b$, $a$ must be in $H_k$. We conclude then that $H_1\cap K,\dots,H_n\cap K$ must be overlapping, as every path from $\mu$ to any state in $K$ passes through only overlapping modules. However, because $K$ is indecomposable, we conclude that this union is trivial, that is $K = H_i$ for some $i$.

\emph{Claim}: If $\{H_1,\dots, H_n, K_1, \dots K_m, M\}$ is an overlapping collection, and $K_i\subseteq M$ for all $i$ and $H_i\not\subseteq M$ for all $i$, then $\{H_1,\dots, H_n,M\}$ is also overlapping. For any $H_i$, $H_j$, there is a sequence $X_i,Y_1,\dots,Y_n,X_j$ of pairwise overlapping sets with $Y_i \in \{H_1,\dots, H_n, K_1, \dots K_m, M\}$. We want to show there exists such a sequence with $Y_i\in \{H_1,\dots, H_n, M\}$. If $K_i$ does not appear, we are done. Otherwise, let $a$ and $b$ be the indices of the first and last sets in the sequence which are contained in $M$. Because $Y_{a-1}\not\subseteq M$ and overlaps $Y_a\subseteq M$, we conclude that $Y_{a-1}$ overlaps $M$ and similarly $Y_{b+1}$ overlaps $M$. Then the sequence $X_i,Y_1,\dots,Y_{a-1},Y_a,M,Y_b,Y_{b+1},\dots,Y_n,X_j$ is pairwise overlapping and doesn't contain any $K_i$. 

Finally, suppose $M = A_1\cup \dots\cup A_n$. By the first clain, if $H$ is a maximal $M$-module, then $H= A_i$ for some $i$, so this union contains all maximal $M$-modules. If it also contains some non-maximal $M$-modules, we can delete them while remaining an overlapping set whose union is $M$, so the set of maximal $M$-modules is the unique minimal overlapping set of basis modules whose union is $M$.
\end{proof}
%\begin{corol} \label{unique nodes}
%For any thin module $M$, there exists a unique minimal set of nodes in $T$ the union of whose descendants are $M$.
%\end{corol}
%\begin{proof}
%If $M$ is indecomposable or strong, then there is a node $m$ representing $M$ in $T$, and so $\{m\}$ is clearly the minimal set of nodes required. If $M$ is decomposable and not strong, then by Theorem~\ref{decomposition theorem} there is a unique minimal collection of indecomposable modules $Q_1,\dots,Q_n$ whose union is $M$. Since these modules are in every decomposition of $M$, and they are indecomposable, there is a corresponding set of nodes $q_1,\dots,q_n$ in $T$ which are in every union making up $M$, and so this must be the unique minimal set of such nodes.
%\end{proof}

Propositions~\ref{decomposition theorem} and~\ref{set bases} together establish the second claim in Theorem~\ref{basis theorem}. Lemma~\ref{linear growth} and Proposition~\ref{linear arcs} establish the first claim, which is that the modular decomposition is small.

%\begin{proof}[Proof of Corollary~\ref{linear growth}]
%Firstly, $Q(Z)$ is always a thin module and it has at least two nodes, and so either there is at least one indecomposable thin module contained in it with at least two nodes, or it is itself indecomposable. Also, each singleton is always trivially indecomposable. This gives the lower bound. By Theorem~\ref{Kq theorem}, each non-singleton indecomposable thin module $H$ has a representative node $q$ where $K_q=H$. This representative is not the start node of $Z$, so there are only $n-1$ possible representatives and so at most $2n-1$ possible distinct indecomposable modules, proving the result.
%\end{proof}

\begin{lem} \label{linear growth}
In any accessible $n$-state FSM with $n > 1$, there are between $n+1$ and $2n-1$ indecomposable modules.
\end{lem}
\begin{proof}
Firstly, $Q(Z)$ is always a thin module and it has at least two nodes, and so either there is at least one basis module contained in it, or it is itself indecomposable. Also, each singleton is always trivially indecomposable. This gives the lower bound. By Theorem~\ref{Kq theorem}, each basis module $H$ has a representative node $q$ where $\repr(q)=H$. This representative is not the start node of $Z$, so there are only $n-1$ possible representatives and so at most $2n-1$ possible distinct indecomposable modules, proving the result.
\end{proof}

\begin{prop} \label{linear arcs}
If $Z$ is an HFSM, then the modular decomposition $\mathcal{T}$ of $Z$ has a linear number of nodes and arcs compared to $Z$.
\end{prop}
\begin{proof}
Let $K$ be an indecomposable module in $Z$, $k$ the node of $T$ corresponding to $K$, and $d$ the in-degree of $k$.

\noindent\emph{Claim: If $d>2$, then there exist a collection of $d$ symbols $x_1,\dots,x_d$ in $\Sigma$ such that every state in $K$ has an $x_1$-arc, an $x_2$-arc, ... , an $x_d$-arc.}

Firstly, let $M_1,\dots,M_d$ be indecomposable modules whose respective nodes in $T$ are predecessors of $k$. By definition of $T$, we know $K\subset M_i$ for each $i$, and so $K\subseteq \bigcap_{i=1}^d M_i$, and these modules are pairwise overlapping because (1) each contains $K$ and (2) $T$ is transitively reduced. By Lemma~\ref{paths through start node}, there is a path 
\begin{tikzcd} g \ar[r,squiggly] & \alpha_1 \ar[r,squiggly] & \mu \end{tikzcd} where $g$ is the start state of $Z$, $\alpha_1$ is the start state of $\bigcup M_i$ and $M_1$ (w.l.o.g) and $\mu$ is in $\bigcap M_i$. The start states $\alpha_2,\dots,\alpha_d$ of $M_2,\dots,M_d$ must be on the path $\pth{\alpha_1}{\mu}$. We show they are all the same state, $\alpha_1$. Because each $M_i$ are basis modules which don't contain each other, for each $i\in\{2,\dots,d\}$ there must be a symbol $x_i$ where the $x_i$-path of $\mu$ goes to a state $m_i$ which is contained in $M_i$ exclusively (and all states on this path are in $M_i$). But since $\mu$ is in $\bigcap M_i$, for any $j$, $k$ in $1,\dots,d$ there is a $x_j$-path $\pth[x_j]{m_j}{m_k}$. $\alpha_k\in M_1$, but if $\alpha_k\neq \alpha_1$, then the path $\pth[x_j]{m_j}{m_k}$ contains $\alpha_1$ (Lemma~\ref{paths through start node}) but $\alpha_1\not\in M_k$, which is a contradiction. Thus $\alpha_1=\alpha_2=\dots=\alpha_d$. Also, as for the other $x_i$, since $\alpha_1\in \bigcap M_i$, there exists a state $m_1$ in $M_1$ exclusively, and a respective symbol $x_1$ where the $x_1$-path out of $\mu$ goes to $m_1$. $K\subseteq \bigcap M_i$, so each state in $K$ has a $x_i$-arc for $i\in \{1,\dots,d\}$. This proves the claim.

We call a module which doesn't overlap any others a \emph{strong} module, and for each state $q$, let $S(q)$ denote the largest non-trivial strong thin module of which $q$ is the start state. This is unique because strong modules do not overlap. Now, to each indecomposable $K$ we will associate a set $\Lambda_K$ of arcs of $Z$, as follows. If $d\leq 2$, then we define $\Lambda_K = \emptyset$. If $d>2$ and $K$ is a singleton $\{r\}$, define $\Lambda_K$ as the $x_1,\dots,x_d$-arcs out of $r$, which exist by the previous claim. Otherwise $d>2$ and $K$ is a basis module, and we define
\[
\Lambda_K := \left\{ \arc[x_i]{a}{b} |\ a\in S(q), b\not\in S(q),i\in\{1,\dots,d\} \right\}
\]
where $q$ is a representative of $K$, so $\repr(q) = K$ (Theorem~\ref{Kq theorem}), and we fix a representative for each basis module. Different choices of representative $q$ may lead to different sets $\Lambda_K$, but this won't matter for the lower bound we seek to establish on the arcs of $Z$. %we will choose any arbitrarily because we only use $\Lambda_K$ to establish a lower bound on the total number of arcs in $Z$, using the fact that $|\Lambda_K| = d$. 
Because $S(q)$ doesn't overlap $K$, by definition, and $q\in S(q)\cap K$, either $K\subseteq S(q)$ or $S(q)\subseteq K$. If $K\subseteq S(q)$, then by Lemma~\ref{unions} $q$ would be the start state of $K$, which it isn't because $\repr(q) = K$ (Theorem~\ref{Kq theorem}). Hence $S(q)\subseteq K$, and because $S(q)$ is a thin module and $q$ has $x_1,\dots,x_d$-arcs, $S(q)$ must have $x_1,\dots,x_d$-exits by the previous claim, so $\Lambda_K$ is well-defined.

We note two important facts from this definition. Firstly, because $S(q)\subseteq K$, the tail of any arc in $\Lambda_K$ is always in $K$. Secondly, because $S(q)\subseteq K\subseteq \bigcap M_i$, and the $x_i$-path out of $q$ goes to a node $m_i\in M_i$ exclusively, the head of the $x_i$-arc in $\Lambda_K$ is always in $M_i$.

%for some $i$, where $M_i$ is defined as above, so is in $S(M_1) = S(M_2) = \dots = S(M_d)$.\todo{this bit doesn't make sense with S(M)}

 \noindent\emph{Claim: If $K$ and $L$ are distinct indecomposable modules, then $\Lambda_K\cap\Lambda_L = \emptyset$.}
 
 This is trivially true if either $\Lambda_K$ or $\Lambda_L$ is empty, so we assume otherwise. If $K$ and $L$ are disjoint the result holds, since the tails of the arcs must be disjoint. Now suppose $K$ and $L$ overlap, which means that both are basis modules. Let $q$ and $\ell$ be representatives of $K$ and $L$ respectively. By definition, $S(q)$ and $S(\ell)$ do not overlap. If $S(q)\subseteq S(\ell)$, then because $q\in S(\ell)$ but not the start state we have $K=\repr(q)\subseteq S(\ell) \subseteq L$, which contradicts the fact that $K$ and $L$ overlap, so we must have $S(q)\cap S(\ell) = \emptyset$% with $S(q)\subset K$ and $S(\ell)\subset L$. If $S(q)\cap S(L)\neq \emptyset$, then both are contained in $K\cap L$, but w.l.o.g. $\ell\in S(q)$ but is not the start node. Because $S(q)\subset L$, this contradicts the fact that $K_\ell = L$. Thus $S(q)\cap S(\ell) = \emptyset$
 , and so $\Lambda_K\cap\Lambda_L = \emptyset$.
 
 If $K$ and $L$ are both singletons, then clearly $\Lambda_K$ and $\Lambda_L$ are disjoint. Now assume w.l.o.g. that $L$ is a basis module with $\repr(\ell) = L$, and $K$ is either a basis module with representative $q$ or $q$ is its sole element. In the latter case we will also write $S(q)$ for the singleton $\{q\}$ to simplify the presentation. Now if $S(q)$ and $S(\ell)$ are disjoint we are done.  By definition, $S(q)$ and $S(\ell)$ do not overlap. If $S(q)\subseteq S(\ell)$, then because $q\in S(\ell)$ but not the start state we have $S(q) \subseteq K=\repr(q)\subseteq S(\ell) \subseteq L$.  Since $d=\indeg(K) > 2$, we know there exists basis modules $M_1,\dots,M_d$ containing $K$. Because these are the smallest basis modules containing $K$, we know that $M_i\subseteq L$ for some $i$. If $\ell\in M_i$, then $\ell=\alpha$ (start state of all $M_i$) because otherwise $L = \repr(\ell) \subseteq M_i$. Because $S(\ell)$ is the largest overlapping module with start state $\ell$, $M_i\subseteq S(\ell)$ for all $i$. If $\ell\not\in M_i$ then because $K\subseteq S(\ell)$ we must again have $M_i\subseteq S(\ell)$ because $S(\ell)$ does not overlap any modules and $\ell\not\in M_i$. However, we established earlier that the head of each $x_i$-arc in $\Lambda_K$ is contained in $M_i$, and so is contained in $S(\ell)$, and hence $\Lambda_K$ and $\Lambda_L$ are disjoint.
 
 Finally we can prove the main claim of the theorem. For any $K$, $d\leq 2+|\Lambda_K|$ because either $d\leq 2$ or $d>2$, in which case $|\Lambda_K| = d$, by definition.
 Then
 \begin{equation*}
 |A(T)| = \sum_{v\in N(T)} \indeg(v) 
  \leq \sum_{M\text{ in basis}} (2 + |\Lambda_M|)
  = 2\dim Z + \sum_{M\text{ in basis}}|\Lambda_M|
  \leq 2\dim Z + |A(Z)|
 \end{equation*} by the fact that all the $\Lambda_M$s are disjoint sets of arcs in $Z$. By Lemma~\ref{linear growth}, the dimension of $Z$ is linear in the number of nodes and hence the number of arcs of $Z$, which completes the proof.
\end{proof}
Finally, Lemma~\ref{representation bijection} establishes the third claim of Theorem~\ref{basis theorem}.

\begin{lem} \label{representation bijection}
    The map $\repr_Z(q) \mapsto \repr_W(q)$ is a bijection between the bases of equivalent HFSMs $Z$ and $W$, and the contracted forms of $\repr_Z(q)$ and $\repr_W(q)$ are equal up to state labels.
\end{lem}
\begin{proof}
    Let $Z$ be an HFSM, and $M$ a module. Let $W$ be the HFSM where $Z[M]$ is nested at $M$ in $Z/M$. We demonstrate that $\repr_Z(q)\mapsto \repr_W(q)$ is a one-to-one correspondence between the bases of $Z$ and $W$. This is sufficient to prove the whole theorem, because the equivalence between $Z$ and $Z^F$ can be broken into a chain of such individual nestings, so by induction we obtain a one-to-one correspondence between $Z$ and $Z^F$, and hence between any two equivalent HFSMs, by Theorem~\ref{flat equivalence}.

The result is easy for basis modules which are subsets of $M$. A set $H\subseteq M$ is a module of $Z$ if and only if it is a module of $Z[M]$ (by Theorem~\ref{gallai}), and hence $W$, giving $\repr_Z(q) = \repr_{Z[M]}(q) = \repr_{W}(q)$. Likewise, if $H\supseteq M$, then by Theorem~\ref{reverse gt}, $H$ is a module of $Z$ if and only if $H/M$ is a module of $Z/M$. As an HFSM module $H$ is a module of $W$ if and only if $H/M$ is a module of $Z/M$. Notice that the start states of these modules are also the same. %Thus the set of modules of $W$ and $Z$ which contain $q$ not as the start state is the same set, and so again $\repr_Z(q) = \repr_W(q)$, by Theorem~\ref{Kq theorem}.
    %if $\repr_Z(q)\supseteq M$, then $\repr_Z(q) = \repr_{Z[M]}(q) = \repr_{W}(q)$ (by Theorem~\ref{gallai}) , and so there is a one-to-one correspondence between these modules. Now we show there is also a one-to-one correspondence between modules that aren't contained in $M$. Observe first that any set $H$ containing $M$ is a module of $W$ if and only if it is a module of $Z$. This is because, as an HFSM module, $H$ is a module of $W$ containing $M$ if and only if $H/M$ is a module of $Z/M$, and by Theorem~\ref{reverse gt}, $H/M$ is a module of $Z/M$ if and only if $H$ .

    (Claim: If $\repr_W(q)$ contains $M$, then $\repr_W(q) = \repr_Z(q) \cup M$.)
    If $\repr_W(q)$ contains $M$, all modules $K$ in $W$ with $q\in K$ but not the start state must contain $M$. Hence, each such $K$ is also a module of $Z$, and again $q$ is not the start state. By Theorem~\ref{Kq theorem} and the claim above, 
    \[\repr_W(q) = \bigcap_{\substack{H\text{ module of $W$} \\ q\in H \\ q\text{ not start state}}} H = \bigcap_{\substack{H\text{ module of $Z$} \\ q\in H \\ H\cap M \neq \emptyset \\ q\text{ not start state}}} (H \cup M) = M \cup \bigcap_{\substack{H\text{ module of $Z$} \\ q\in H \\ q\text{ not start state}}} H = \repr_Z(q) \cup M\]
    %is the intersection of all modules of $Z$ containing $q$ not start node and also $M$. But that is equal to $\repr_W(q) \subseteq \cap_{H in Z} (H\cup M)$ for all $H$ which intersect $M$. Hence $\repr_W(q) = \repr_Z(q) \cup M$.

    To show that $\repr_Z(q)\mapsto\repr_W(q)$ is a bijection, it is sufficient to show it is well-defined as a function in both directions, that is, if $\repr_Z(q)= \repr_Z(h)$ then $\repr_W(q)= \repr_W(h)$, and vice versa. Let $q$ and $h$ be distinct states. First, suppose $\repr_Z(q) = \repr_Z(h)$. If $\repr_Z(q)$ doesn't overlap $M$, then $\repr_Z(q) = \repr_W(q) = \repr_W(h) = \repr_Z(q)$. If $\repr_Z(q)$ overlaps $M$, then $\repr_W(q) = \repr_Z(q)\cup M = \repr_Z(h)\cup M = \repr_W(h)$. For the converse, suppose $\repr_W(q) = \repr_W(h)$. If $\repr_W(q)$ doesn't contain $M$, then $\repr_W(q) = \repr_W(h) = \repr_Z(h) = \repr_Z(q)$. Otherwise, $\repr_Z(q) \cup M = \repr_Z(h) \cup M$. Suppose for contradiction that $\repr_Z(q)\neq \repr_Z(h)$. By Theorem~\ref{basis theorem}, $M = B_1\cup \dots \cup B_n$, where $B_i$ are basis modules, no two of which are contained in each other, by uniqueness. However, then $\repr_Z(q) \cup B_1\cup \dots \cup B_n = \repr_Z(h) \cup B_1\cup \dots \cup B_n$ but this violates uniqueness (Theorem~\ref{basis theorem}) because these are two distinct unions of overlapping basis modules, and no two contain each other.

    Finally, we need to show that the contracted forms of $\repr_Z(q)$ and $\repr_W(q)$ are the same up to state labelling. We denote the contracted form of a module $M$ by $\cf(M)$. As before, this is easy to prove if they are contained in $M$, because then $\repr_Z(q) = \repr_W(q) = \repr_{Z[M]}(q)$, and so $\cf(\repr_Z(q)) = \cf(\repr_W(q)) = \cf(\repr_{Z[M]}(q))$. Similarly, it is trivially true if they are disjoint from $M$. Now assume that $\repr_W(q)$ contains $M$, and denote by $H_1,\dots H_n$ the maximal thin modules contained in $\repr_W(q)$, and these are disjoint, and $\cf(\repr_W(q)) = Z[\repr_W(q)]/H_1,\dots,H_n$. If $\repr_Z(q)$ also contains $M$, then $\repr_Z(q) = \repr_W(q)$ and so $\cf(\repr_W(q)) = Z[\repr_W(q)]/H_1,\dots,H_n = \cf(\repr_Z(q))$ (up to state labels). Suppose instead that $\repr_Z(q)$ overlaps $M$, in which case $\repr_W(q) = \repr_Z(q) \cup M$. However, observe that if $H$ is a module which contains $M$, then $Z[\repr_Z(q)]/H = Z[\repr_Z\cup M]/H$ and so by extension $\cf(\repr_W(q)) = Z[\repr_W(q)]/H_1,\dots,H_n = Z[\repr_Z(q)\cup M]/H_1,\dots,H_n = Z[\repr_Z(q)]/H_1,\dots,H_n = \cf(\repr_Z(q))$ (again, up to state labels).
\end{proof}

\begin{thm}[Theorem~\ref{module characterisation}]
For $q\in Q(Z)$, we write $\uparrow_v\!\! q$ to denote the ancestors of $q$ in $\mathcal{G}_v$. Then $\uparrow_v\!\! q$ is a module, and a subset $H \subset Z$ is a thin $v$-module if and only if there exist states $q_1,\dots,q_m\in Q(Z)$ where $H = \bigcup_{i\in 1,\dots,m}\uparrow_v\!\!q_i$.
\end{thm}
\begin{proof}
\noindent\emph{Claim: for any $v$-module $H$ with $b\in H$, if $\pth{a}{b}$ in $\mathcal{G}_v$ then $a\in H$.}

If $b=v$, then $b$ has no predecessors in $\mathcal{G}_v$, so this is true trivially. Now assume $b\neq v$. We can assume without loss of generality that there is an arc $\arc{a}{b}$ in $\mathcal{G}_v$. As a result, either:
\begin{itemize}
    \item (\textbf{a}) ($\arc[x]{a}{b}$ in $Z$): in this case $a\in H$ because $b$ is not the start state of $H$.
    \item (\textbf{b}) ($\arc[x]{b}{a}$ in $Z$, but there is no path $\pth[x]{v}{a}$): if $a\not\in H$, then $a$ would be the $x$-exit of $H$, but the lack of an $x$-path $\pth[x]{v}{a}$ contradicts thinness. Thus $a\in H$.
    \item (\textbf{c}) ($\arc[x]{b}{q}$ in $Z$, $q\neq v$, and there is a path \begin{tikzcd} v \ar[r,squiggly,"x"] & a \ar[r,"x"] & q \end{tikzcd}): if $q\in H$, then $a\in H$ as $q\neq v$. If $q\not\in H$, then it is the $x$-exit of $H$, and by thinness $v$'s $x$-path goes to $q$ within $H$, and so $a\in H$.
    \item (\textbf{d}) ($a = q_x$ for some $x$, and $b$ has no $x$-arc): $b\in H$, so $H$ must have no $x$-exits. As a result, all states $q$ with an $x$-path $\pth[x]{v}{q}$ must be within $H$.
\end{itemize}
By repeating this argument, we deduce that for any $b\in H$, $\uparrow_v\!\! b\subseteq H$. 
Let $s$ be the start state of $Z$. Using this claim, we see that if $s\neq v$ and there is a path $\pth{s}{b}$, then every $v$-module containing $b$ contains $s$, but none contain $s$ so $b$ is in no $v$-modules. Thus all states removed from $\mathcal{G}_v$ on line~14 of Alg.~\ref{alg:Gv} are in no $v$-modules.

\noindent\emph{Claim: $\uparrow_v q$ is a thin module.} 

First we show $v$ is the only entrance of $\uparrow_v q$. If $v=s$ we are trivially done. Otherwise, suppose for contradiction that $\ell\neq v$ was an entrance of $\uparrow_v q$. Then $\ell$ has a predecessor $p$ in $Z$ that is not in $\uparrow_v q$, but then there is a path to $\ell$ which does not contain $v$, but then $\ell$ is removed from $\mathcal{G}$, so cannot be in $\uparrow_v q$.

then the arc $\arc{p}{\ell}$ still exists in $\mathcal{G}_v$ (as $\ell\neq v$) and so $p\in\uparrow_v q$, which is a contradiction.

Now let $a\in\uparrow_v\!\!q$, and let $\arc[x]{a}{b}$ be an arc in $Z$, with $b\not\in\uparrow_v q$. Clearly $b\neq v$. Because $b\not\in \uparrow_v q$, by Algorithm~\ref{alg:Gv} there must be a path \begin{tikzcd} v \ar[r,squiggly,"x"] & t \ar[r,"x"] & b \end{tikzcd} in $Z$. By (\textbf{c}), either $a=t$ or there is an arc $\arc{t}{a}$ in $\mathcal{G}_v$, so all states on the path $\pth[x]{v}{t}$ are in $\uparrow_v q$. Hence any $x$-exit of $\uparrow_v q$ must be on the $x$-path out of $v$, and have all its predecessors on that path be within $\uparrow_v q$. This proves that $b$ is the unique $x$-exit, and so $\uparrow_v q$ is a thin module.

Since $\uparrow_v q$ is always a thin module, $\bigcup_i \uparrow_v q_i$ is a collection of overlapping thin modules (they all contain $v$) and so is a thin module by Lemma~\ref{unions}. For the converse, suppose $H$ is a thin $v$-module. All states in $H$ must be in $\mathcal{G}_v$ as the removed states are in no $v$-modules. Thus for every $q\in H$, $\uparrow_v q$ is a module contained in $H$, and so $H = \bigcup _{q\in H} \uparrow_v q$.
\end{proof}

\begin{algorithm}
\SetAlgoLined
\SetKwInOut{Input}{Input}
\SetKwInOut{Output}{Output}
\SetKwData{Ap}{apices}
\Input{\ An indecomposable module $K$, and a digraph $\mathcal{T}$.}
\Output{\ Modifies $\mathcal{T}$ to contain a node representing the module $K$.}
Create new node $t_K$, queue $Q$ and set $\Ap \gets \emptyset$\;
\tcp{Here $t_q$ is the node of $\mathcal{T}$ corresponding to state $q$ of $Z$}
$Q \gets [ t_q\ \text{for $q$ in $K$}]$\;
\tcp{$value$ stores an integer for each node of $T$}
$value\gets \{ node \mapsto \text{in-degree}(node)\ \text{for node in $T$} \}$\;
\While{$Q$ is non-empty}{
    $v \gets Q.pop()$, $flag \gets True$\;
    \For{each predecessor $p$ of $v$}{
        $value[p] \gets value[p] - 1$\;
        \If{index = 0}{
            $Q.push(p)$\;
            $flag \gets False$\;
        }
    }
    If $flag$ is true, add $v$ to $\Ap$\;
}
For each $apex$ in $\Ap$, add an arc $\arc{t_K}{apex}$ in $\mathcal{T}$\;
 \caption{AddModule}
 \label{alg:merge module}
\end{algorithm}

Algorithm~\ref{alg:merge module} is responsible for constructing the modular decomposition as a graph from the basis modules, iteratively. It uses a variant of breadth-first-search on the modular decomposition which, given a basis module $K$ to add to $\mathcal{T}$, efficiently locates the nodes $t_H$ representing modules $H$ which are immediate successors of $K$ in the inclusion order on indecomposable modules. It works by maintaining a queue $Q$ with the invariant that each node in $Q$ represents a module wholly contained in $M$. The queue is initialised with the singleton subsets of $K$, and the invariant is maintained by adding nodes to $Q$ only once all of their predecessors have been visited (which guarantees that the modules they represent are subsets of $K$). We store the nodes $t_H$ visited which have no successors added to the queue in a set \texttt{apices}. These $H$ are the immediate successors of $K$ in the inclusion order, so we add arcs $\arc{t_K}{t_H}$ to $\mathcal{T}$.

\begin{lem} \label{merge module correct}
Algorithm~\ref{alg:merge module} works; given $M$, the arcs added out of $t_M$ are exactly those in the modular decomposition of $Z$.
\end{lem}
\begin{proof}
In this proof, we write $t_K,t_H,t_M$ to represent nodes of $\mathcal{T}$ corresponding to indecomposable modules $K,H$ and $M$ respectively. Recall that given indecomposable modules $K$ and $M$, there is an arc $\arc{t_K}{t_M}$ in $\mathcal{T}$ if $K$ \emph{covers} $M$, that is $K\supset M$ and there is no indecomposable module $H$ with $K\supset H\supset M$.

\noindent\emph{Claim: When adding $K$ to $\mathcal{T}$, then for any module $M$ covered by $K$, $t_M$ is already in $\mathcal{T}$.}

This follows from Theorem~\ref{algorithm correct} where we proved that modules are constructed in an order compatible with the inclusion order.

\noindent\noindent\emph{Claim: $t_M$ appears in $Q$ if and only if $M\subset K$.}

We proceed by induction on the length of the longest path from a sink of $\mathcal{T}$ to the node $t_M$. The base case is the nodes $t_v$ corresponding to individual nodes of $Z$, and the claim holds for this case as $Q$ is initialised to contain precisely these nodes. Now assume that for all nodes up to $n$ steps from a sink, a node $t_M$ is added to $Q$ if and only if $M\subset K$. Then let $t_H$ be a node which is $n$ steps from a sink. $H$ is the union of the modules corresponding to successors of $t_H$, which are all at most $n-1$ steps from a sink. If $H\subset K$, then all of its predecessors are added to $Q$, and as each is processed $value[t_H]$ decreases by one, from its initial value which is equal to the number of predecessors of $t_H$. Thus as the last predecessor of $t_H$ is processed, $value[t_H]$ reaches zero and so $t_H$ is added to $Q$. Conversely, if $H\not\subset K$, then at least one of its predecessors is not in $Q$, so $value[t_H] > 0$ for the duration of the algorithm, so $t_H$ is never added to $Q$. This proves the claim.

It follows that every $t_M$ where $K$ covers $M$ is eventually visited, and these are precisely the nodes visited which have no predecessors added to $Q$, and so the $flag$ variable remains $true$ and so these nodes are exactly those added to \texttt{apices}.
\end{proof}

\begin{thm}[Theorem~\ref{algorithm correct}]
Algorithm~\ref{alg:findmodules} works, i.e. the sets $\uparrow_v\!\!q$ added to $\mathcal{T}$ are exactly the basis modules of $Z$.
\end{thm}
\begin{proof}
Let $s$ be the start state of $Z$. Consider $\uparrow_v\!\!q$, for any $v$ and $q$, noting that either $s\not\in \uparrow_v\!\!q$ or $v = s$. Let $M$ be a $v$-module that contains $q$. By Theorem~\ref{module characterisation}, $M$ is the union of $\uparrow_v\!\!q_i$ for some $q_1,\dots,q_n$. However, $q\in M$ implies that there exists an $i$ with $q\in\uparrow_v\!\!q_i$. By transitivity, $\uparrow_v\!\!q\subseteq \uparrow_v\!\!q_i\subseteq M$, and so \[
\uparrow_v\!\!q = \bigcap_{\substack{M\text{ thin module}\\q\in M\\v\text{ start state}}} M
\]
Combining this result with Theorem~\ref{Kq theorem}, we deduce that for any state $w\neq s$, $]\repr(w) = {\uparrow_\sigma\!\! w}$, where $\sigma$ is the start state of $\repr(w)$. Similarly, every basis module $H$ with start state $v$ has a representative $h$ with $H = \repr(h) = \uparrow_v\!\! h$.
%Now assume the current node in the outer loop of the algorithm is $v$. We will show the modules that are output are precisely the indecomposable thin $v$-modules, and that all indecomposable $v$-modules are constructed on this iteration.

Let $v$ be the state on the current iteration of the outer loop of Algorithm~\ref{alg:findmodules}, and let $M \neq \{v\}$ be the strongly connected component of $\mathcal{G}_v$ on the current iteration of the inner loop. We will show by induction that for every $q\in M$, $q\not\in used$ if and only if $\uparrow_v q$ is a basis module. It will follow that the sets we add to the modular decomposition are exactly the basis modules. Assume true for all iterations up to this point.

Firstly, suppose $q\not\in used$. If $\repr(q) \neq \uparrow_v q$, then $\repr(q) = \uparrow_u q$ for some $u\in \uparrow_v q$ by the above. But then by Lemma~\ref{paths through start node} every path from $s$ to $u$ contains $v$, and so $u$ must precede $v$ in every reverse breadth-first-search of $Z$ from $s$. By the inductive hypothesis, the basis module $\uparrow_u q$ has already been constructed, and $q$ is already in $used$. Hence $\repr(q) = \uparrow_v q$, so $\uparrow_v q$ is indecomposable.

For the converse, suppose that $q\in M$ and $\uparrow_v q$ is indecomposable. For any other thin module $H$ containing $q$ where $q$ is not the start state, $\repr(q)\subseteq H$. If $H$ has start state $\alpha \neq v$, then $H$ is constructed later in the reverse breadth-first-search order. If $M$ is a $v$-module, then either $M = \repr(q)$ or $\repr(q) \subset M$, in which case $M = \uparrow_v h \supset \uparrow_v q = \repr(q)$, and so $M$ is constructed after $\repr(q)$ as $q$ precedes $h$ in the topological order. Since $q$ is added to $used$ only when we add a basis module containing it to the modular decomposition, we conclude that $q\not\in used$ on this iteration.
\end{proof}

\end{document}